\documentclass[fleqn,a4paper,usenatbib]{mnras}

\usepackage[T1]{fontenc}
\usepackage{ae,aecompl}

\usepackage{graphicx}	
\usepackage{amsmath}	
\usepackage{amssymb}	
\usepackage{pdflscape}
\usepackage{times,txfonts}
\usepackage{hyperref}
\usepackage[final]{changes}
\definechangesauthor[name={Ted Mackereth}, color=orange]{TM}

\newcommand{\afe}{$\mathrm{[} \alpha \mathrm{/Fe]}$}
\newcommand{\feh}{$\mathrm{[Fe/H]}$}

\title[Age-metallicity structure of the Milky Way disk]{The age-metallicity structure of the Milky Way disk }

\author[J. E. Mackereth et al.]{
J. Ted Mackereth,$^{1}$\thanks{E-mail: J.E.Mackereth@2011.ljmu.ac.uk (LJMU)}
Jo Bovy,$^{2,3}$\thanks{Alfred P. Sloan Fellow}
Ricardo P. Schiavon,$^{1}$
Gail Zasowski,$^{4,5}$\newauthor
Katia Cunha,$^{6}$
Peter M. Frinchaboy,$^{7}$
Ana E. Garc\'{\i}a Perez,$^{8,9}$
Michael R. Hayden,$^{10}$\newauthor
Jon Holtzman,$^{11}$
Steven R. Majewski,$^{12}$
Szabolcs M{\'e}sz{\'a}ros,$^{13}$\thanks{Premium Postdoctoral Fellow of the Hungarian Academy of Sciences}
David L. Nidever,$^{14,15,16}$\newauthor
Marc Pinsonneault$^{17}$
and Matthew D. Shetrone$^{18}$
\\
$^{1}$Astrophysics Research Institute, Liverpool John Moores University, 146 Brownlow Hill, Liverpool L3 5RF, UK\\
$^{2}$Department of Astronomy and Astrophysics, University of Toronto, 50 George Street, Toronto, ON M5S 3H4, Canada\\
$^{3}$Center for Computational Astrophysics, Flatiron Institute, 162 5th Ave NY NY 10010, USA\\
$^{4}$Space Telescope Science Institute, 3700 San Martin Drive, Baltimore, MD 21218, USA\\
$^{5}$Department of Physics and Astronomy, University of Utah, 201 James Fletcher Bldg., 115 South 1400 East, Salt Lake City, UT 84112-0830, USA\\
$^{6}$Observat\'{o}rio Nacional, S\~{a}o Crist\'{o}v\~{a}ao, Rio de Janeiro, Brazil\\
$^{7}$Department of Physics \& Astronomy, Texas Christian University, TCU Box 298840, Fort Worth, TX 76129, USA\\
$^{8}$Instituto de Astrofísica de Canarias (IAC), Vía Lactea S/N, E-38205, La Laguna, Tenerife, Spain\\
$^{9}$Departamento de Astrofísica, Universidad de La Laguna (ULL), E-38206, La Laguna, Tenerife, Spain\\
$^{10}$Laboratoire Lagrange (UMR7293), Universite de Nice Sophia Antipolis, CNRS, Observatoire de la Cote dAzur, BP 4229, F-06304 Nice Cedex 4, France\\
$^{11}$Department of Astronomy, New Mexico State University, P.O. Box 30001, MSC 4500, Las Cruces, NM 88003, USA\\
$^{12}$Dept. of Astronomy, University of Virginia, Charlottesville, VA 22904-4325, USA\\
$^{13}$ELTE Gothard Astrophysical Observatory, H-9700 Szombathely, Szent Imre Herceg st. 112, Hungary\\
$^{14}$Department of Astronomy, University of Michigan, Ann Arbor, MI, 48104, USA\\
$^{15}$Large Synoptic Survey Telescope, 950 North Cherry Avenue, Tucson, AZ 85719, USA\\
$^{16}$Steward Observatory, 933 North Cherry Avenue, Tucson, AZ 85719, USA\\
$^{17}$Department of Astronomy, The Ohio State University, Columbus, OH 43210, USA\\
$^{18}$University of Texas at Austin, McDonald Observatory, USA\\}

\date{Accepted XXX. Received YYY; in original form ZZZ}

\pubyear{2017}

\begin{document}
\label{firstpage}
\pagerange{\pageref{firstpage}--\pageref{lastpage}}
\maketitle

\begin{abstract}
The measurement of the structure of stellar populations in the Milky Way disk places fundamental constraints on models of galaxy formation and evolution. Previously, the disk's structure has been studied in terms of populations defined geometrically and/or chemically, but a decomposition based on stellar ages provides a more direct connection to the history of the disk, and stronger constraint on theory. Here, we use positions, abundances and ages for 31,244 red giant branch stars from the SDSS-APOGEE survey, spanning $3 < R_{\mathrm{gc}} < 15$ kpc, to dissect the disk into mono-age and mono-\feh{} populations at low and high \afe{}. For each population, with $\Delta \mathrm{age} < 2$ Gyr and $\Delta \mathrm{[Fe/H]} < 0.1$ dex, we measure the structure and surface-mass density contribution. We find that low \afe{} mono-age populations are fit well by a broken exponential, which increases to a peak radius and decreases thereafter. We show that this profile becomes broader with age, interpreted here as a new signal of disk heating and radial migration. High \afe{} populations are well fit as single exponentials within the radial range considered, with an average scale length of $1.9\pm 0.1$ kpc. We find that the relative contribution of high to low \afe{} populations at $R_0$ is $f_\Sigma = 18\% \pm 5\%$; high \afe{} contributes most of the mass at old ages, and low \afe{} at young ages. The low and high \afe{} populations overlap in age at intermediate \feh{}, although both contribute mass at $R_{0}$ across the full range of \feh{}. The mass weighted scale height $h_Z$ distribution is a smoothly declining exponential function. High \afe{} populations are thicker than low \afe{}, and the average $h_Z$ increases steadily with age, between 200 and 600 pc. 
\end{abstract}




\section{Introduction}


The understanding of the present day spatial, kinematic and chemical configuration of the stars of the Milky Way disk is a cornerstone of Galactic archaeology, placing key constraints on models of galaxy disk formation and evolution. Much of our understanding of the time evolution of galaxy disks like that of the Milky Way has arisen from studies which match galaxies of a given stellar mass at $z=0$ to their progenitors at higher $z$ \citep[and therefore, lookback time, e.g. ][]{2013ApJ...771L..35V,2015ApJ...803...26P,2016MNRAS.462.4495H}. However, the Sun's position in the Milky Way presents a high-fidelity insight into the structure of a Galactic disk on a star by star basis, which has provided a great many insights into the problem \citep[e.g.][]{1962ApJ...136..748E,1993A&A...275..101E,2013A&A...560A.109H}. Data for large numbers of disk stars over a wide range of Galactocentric distances, including positions, chemical abundances and stellar ages are now becoming readily available, due to the advent of modern spectroscopic surveys such as APOGEE \citep{2015arXiv150905420M}, Gaia-ESO \citep{2012Msngr.147...25G} and GALAH \citep{2016arXiv160902822M}, with future instruments aiming to bolster the ESA-\emph{Gaia} data releases \citep{2016A&A...595A...1G}, such as WEAVE \citep{2014SPIE.9147E..0LD} and MOONS \citep{2012SPIE.8446E..0SC}.

Galaxy disks are commonly considered to have stellar density distributions described by exponential laws of some form \citep[e.g.][]{1959HDP....53..311D,1970ApJ...160..811F,1983MNRAS.202.1025G,2006A&A...454..759P}, assumed classically as the result of gas collapse with angular momentum conservation \citep[e.g.][]{1980MNRAS.193..189F}, and more recently, the redistribution or 'scrambling' of the angular momentum of individual stars \citep[e.g.][]{2013ApJ...775L..35E,2016ApJ...830..115E,2016arXiv161203171H}. External disks have relatively well constrained photometric scale lengths \citep[e.g.][]{2010MNRAS.406.1595F}, but estimates of that of the Milky Way vary greatly \citep[see][for a review]{2016ARA&A..54..529B}, and seem to be in tension with those in external galaxies which are assumed to be similar to the Milky Way. This suggests a discord between internal and external scale length measurement methods, or that the Milky Way has a structure distinct from the bulk of disk galaxies. 

In the Milky Way, the definition of the measured population appears to have a great effect on this estimate, with thicker, geometrically defined populations having generally flatter radial profiles \citep[e.g.][]{2008ApJ...673..864J} than, for example, the populations enhanced in $\alpha$-element abundances \citep[e.g.][]{2012ApJ...752...51C,2012ApJ...753..148B,2016ApJ...823...30B}. Theoretical results have suggested that these \emph{geometric} thick disks are formed from embedded flaring of co-eval populations  \citep{2015ApJ...804L...9M}. The discrepant scale lengths between geometric and abundance selected thick disks was framed most recently by \citet{2016arXiv160901168M} as the result of a hitherto unaccounted-for radial age gradient in the disk.

When considered \emph{in toto}, nearby galaxy disks, and the Milky Way disk, are observed to have a two-component vertical spatial structure, commonly referred to as the \emph{geometric} `thick' and `thin' disks \citep{1979ApJ...234..842T,1979ApJ...234..829B,1982PASJ...34..365Y,1983MNRAS.202.1025G,2008ApJ...673..864J}. Classically, the `thick' components have been characterised by older, kinematically hotter stellar populations, enriched in \afe{}, whereas the `thin' populations assume lower, near solar \afe{} and are kinematically cooler \citep[e.g.][]{2005A&A...433..185B}. It has, however, also been posited that these populations are in fact composed of multiple sub-populations that smoothly span this range of properties \citep[e.g.][]{1987ApJ...314L..39N,1991PASP..103...95N,2012ApJ...755..115B,2012ApJ...753..148B,2016ApJ...823...30B}, thus, in terms of structural parameters, the disk cannot be characterised as the superposition of two distinct structures with different scale heights \citep{2012ApJ...751..131B}. It has been shown  that the total vertical stellar spatial distribution resulting from the overlap of such sub-populations is consistent with a double exponential \citep[see, e.g., Figure 14 of][]{2013A&ARv..21...61R}. It is difficult to explain the presence of a continuity in structure alongside the discontinuity in chemistry seen in the Milky Way.

The Milky Way's disk has a complex chemical structure, with a bimodality in \afe{} seen at fixed \feh{} across many of the observable regions of the disk \citep{2003A&A...410..527B,2005A&A...433..185B,2014ApJ...796...38N,2015ApJ...808..132H}. This characteristic is difficult to explain using one-zone Galactic chemical evolution (GCE) models \citep[most recently shown by][]{2016arXiv160408613A}, giving rise to attempts to explain it by means other than pure chemical evolution. Examples of such models include the heating of the old disk by high redshift mergers \citep[e.g.][]{2004ApJ...612..894B,2008MNRAS.391.1806V,2009ApJ...700.1896K,2013A&A...558A...9M} and the formation of a dual disk by gradual accretion of stars into disk orbits \citep[e.g.][]{2003ApJ...597...21A}. More recent work has also framed this bimodality as a consequence of discontinuous radial migration of stars in the disk \citep{2016arXiv161009869T}. 

However, such chemical structure can be replicated in part by invoking various Galactic chemical evolution models that do not rely on a `one-zone' approximation \citep[e.g.][]{1997ApJ...477..765C,2000A&A...355..929P,2016arXiv160408613A,2016arXiv160407435W}. For example, \citet{2016arXiv160408613A} showed that a combination of GCE models with varying outflow mass loading parameters and inflow timescales (intended to represent enrichment histories at varying Galactocentric radii) could make a roughly bimodal \afe{} distribution. The same models were shown to present a good explanation of the APOGEE \afe{}-\feh{} plane by \citet{2014ApJ...796...38N}. A deeper understanding of the connection between spatial structure in \afe{} and stellar age selected populations in the Milky Way is necessary to link these results. 

In this paper, we present the first dissection of radially extended samples of Milky Way disk stars in age, \feh{} and \afe{}. A strong correlation is observed between stellar age and \afe{} in the solar vicinity \citep{2013A&A...560A.109H}. On the other hand, but also in the solar vicinity, no correlation is found between age and [Fe/H] \citep[e.g.][]{1993A&A...275..101E,2004A&A...418..989N}, which may be explained by the occurrence of radial migrations. Thickening of the Galactic disk has been invoked as a consequence of outward stellar radial migration \citep[e.g.][]{2009MNRAS.399.1145S}. However, \citet{2012A&A...548A.127M} argue that the effect is small and that such migration in fact only makes disks flare by a small amount. Similarly, \citet{2016ApJ...823...30B} measured the flaring profile of low \afe{} stars to only slowly exponentially increase with Galactocentric radius, and suggest that radial migration is likely not a viable mechanism for forming thickened disk components. Flaring has also been shown to arise as a result of satellite infall, which can be a stronger flaring agent than migrations \citep[e.g.][]{2009ApJ...707L...1B}. The understanding of flaring and its connection to the evolution of the Galactic disk is essential, but as yet incomplete. In this paper, we present new constraints on models of radial migration in the disk by studying its effects on the Milky Way's mono-age stellar populations. 

Many theoretical studies have attempted to understand the observed structure of mono-abundance populations (MAPs) through the use of hydrodynamics and N-body simulations. Few reproduce the observed bimodality in \afe{} at fixed \feh{}, and so an understanding of this has so far proved difficult. However, certain characteristics of the Milky Way \afe{} distribution are beginning to emerge in the most recent cosmological simulations  \citep{2016arXiv160804133M}. Structurally, the mono-age populations of simulated galaxies show good agreement with the Milky Way \citep[e.g.][]{2013MNRAS.436..625S,2013ApJ...773...43B,2014MNRAS.442.2474M,2014MNRAS.443.2452M}. More recent work has brought into question the applicability of MAPs as a proxy for mono-age populations \citep{2017ApJ...834...27M}, showing that, particularly at low \afe{}, MAPs may have significant age spreads due to the differential nature of star formation in the disk. We show in this paper that the structures of mono-age and mono-abundance populations in the Milky Way disk differ, but present complementary insights into the temporal and chemical evolution processes in the disk.

Previous work has studied MAPs in the Milky Way by analysing samples of SEGUE G-dwarfs \citep{2012ApJ...755..115B,2012ApJ...753..148B,2012ApJ...751..131B} and APOGEE red-clump (RC) giants \citep{2016ApJ...823...30B}. In this work we map the spatial distribution of mono-age, mono-\feh{} populations at low and high \afe{} using a catalogue containing \feh{} and \afe{} from the APOGEE survey \citep{2015arXiv150905420M} and ages from \citet{2016MNRAS.456.3655M} for 31,244 red giant stars. We complement earlier work by adapting the method developed by \citet{2016ApJ...823...30B} for RC stars, to enable its application to the full red giant branch (RGB) sample from APOGEE. Stars in the RGB are better tracers of the underlying stellar population than their RC counterparts because of reduced uncertainties in the stellar evolution models, and because they are generally brighter, and so are observed at greater distances. On the other hand, this means that the method developed by  \citet{2016ApJ...823...30B} must be adapted to account for the spread in absolute magnitude over the RGB (whereas RC stars can be considered as a near-standard candle). On the basis of these measurements, we establish the local mass-weighted age-\feh{} distribution, showing the contributions from both low and high \afe{} stellar populations.

In Section \ref{sec:dataa}, we discuss the APOGEE data and the distance and age catalogues used for this work. Section \ref{sec:methoda} describes the stellar density fitting method, drawing greatly on work by \citet{2016ApJ...818..130B,2016ApJ...823...30B}. Specifically, we describe the generalities of the maximum likelihood fitting procedure and the calculation of the effective survey selection function for RGB stars in Section \ref{sec:densfit}, the adopted parametric stellar density model in Section \ref{sec:densitymodel}, and the method for calculating stellar surface-mass densities in Section \ref{sec:surfmasscalc}. We present the fits in Section \ref{sec:resultsa}, along with the calculated surface-mass density contributions for each mono-age, mono-\feh{} population. In Section \ref{sec:discussiona} we compare our findings to those in the literature and discuss possible scenarios for the formation of the Milky Way's disk in light of our findings. Section \ref{sec:conclusionsa} summarises our results and conclusions.

\section{Data}

\label{sec:dataa}
\subsection{The APOGEE Catalogue}
\label{sec:APOGEE}
We use data from the twelfth data release \citep[DR12,][]{2015ApJS..219...12A} of the SDSS-III APOGEE survey \citep[][]{2015arXiv150905420M}, a high signal-to-noise ratio (SNR$>100\ \mathrm{pixel}^{-1}$), high resolution ($R \sim 22,500$), spectroscopic survey of over 150,000 Milky Way stars in the near-infrared $H$ Band ($1.5 - 1.7 \mathrm{\mu m}$). Stars were observed during bright time with the APOGEE spectrograph \citep{2010SPIE.7735E..1CW} on the 2.5m Sloan Foundation Telescope \citep{2006AJ....131.2332G} at Apache Point Observatory. Targets were selected in general from the 2MASS point-source catalog, employing a dereddened $(J-K_S)_0 \geq 0.5$ colour cut (in the fields which are of interest here) in up to three apparent $H$ magnitude bins \citep[for a full description of the APOGEE target selection, see][]{2013AJ....146...81Z}. Reddening corrections were determined for the colour cut via the Rayleigh-Jeans Colour Excess method \citep[RJCE,][]{2011ApJ...739...25M}. Corrections are found by applying the method to 2MASS \citep{2006AJ....131.1163S} and mid-IR data from \emph{Spitzer}-IRAC GLIMPSE-I, -II, and -3D \citep{2009PASP..121..213C} when available and from WISE \citep{2010AJ....140.1868W} otherwise. For this work, we use distance moduli (which make use of the aforementioned reddening corrections) from the \citet{2015ApJ...808..132H} distance catalogue for DR12 (see Section \ref{sec:distances}). 

All APOGEE data products employed in this paper are those output by the standard data reduction and analysis pipeline used for DR12. The data were processed \citep{2015AJ....150..173N}, then fed into the APOGEE Stellar Parameters and Chemical Abundances Pipeline \citep[ASPCAP,][]{2016AJ....151..144G}, which makes use of a specifically computed spectral library \citep{2015AJ....149..181Z}, calculated using a customised $H$-band line-list \citep{2015ApJS..221...24S}. Outputs from ASPCAP are analysed, calibrated and tabulated \citep{2015AJ....150..148H}. The output \afe{} and \feh{} abundances for DR12 have been shown to have a high degree of precision (at least between $4500 \lesssim \mathrm{T_{\mathrm{eff}}} \lesssim 5200$ K), such that $\sigma_{\mathrm{[Fe/H]}} = 0.05\ \mathrm{dex}$ and $\sigma_{\mathrm{[\alpha/Fe]}} = 0.02\ \mathrm{dex}$ \citep{2016ApJ...823...30B}. We apply here the same external calibrations to \afe{} and \feh{} as \citet{2016ApJ...823...30B}, constant offsets of -0.05 dex and -0.1 dex, respectively. We use the tabulated \feh{} value rather than the globally fit $\mathrm{[M/H]}$ which is included in the table.

We select stars from the DR12 catalogue which were targeted as part of the main disk survey (i.e were subject to the $(J-K_S)_0 \geq 0.5$ cut), have reliably measured abundances (i.e. no warning or error bits set in the ASPCAPFLAG field) and have a well defined distance modulus and age measurement (see Sections \ref{sec:distances} and \ref{sec:ages}). We apply a secondary cut at $1.8 < \log{g} < 3.0$ to restrict the sample to stars on the red giant branch (RGB), removing most contaminating dwarfs, and very evolved stars near the tip of the RGB. The high end of our $\log{g}$ cut is more conservative than other studies in this regime, however we find this gives the best agreement between the data and stellar evolution models, without significantly reducing the sample size or introducing unwanted bias. These cuts give a final sample of 31,244 stars, spanning $4200 \lesssim \mathrm{T_{\mathrm{eff}}} \lesssim  5050\ \mathrm{K}$ for which the effective survey selection function can be reconstructed.

\begin{figure}
 \centering
	\includegraphics[width=\columnwidth]{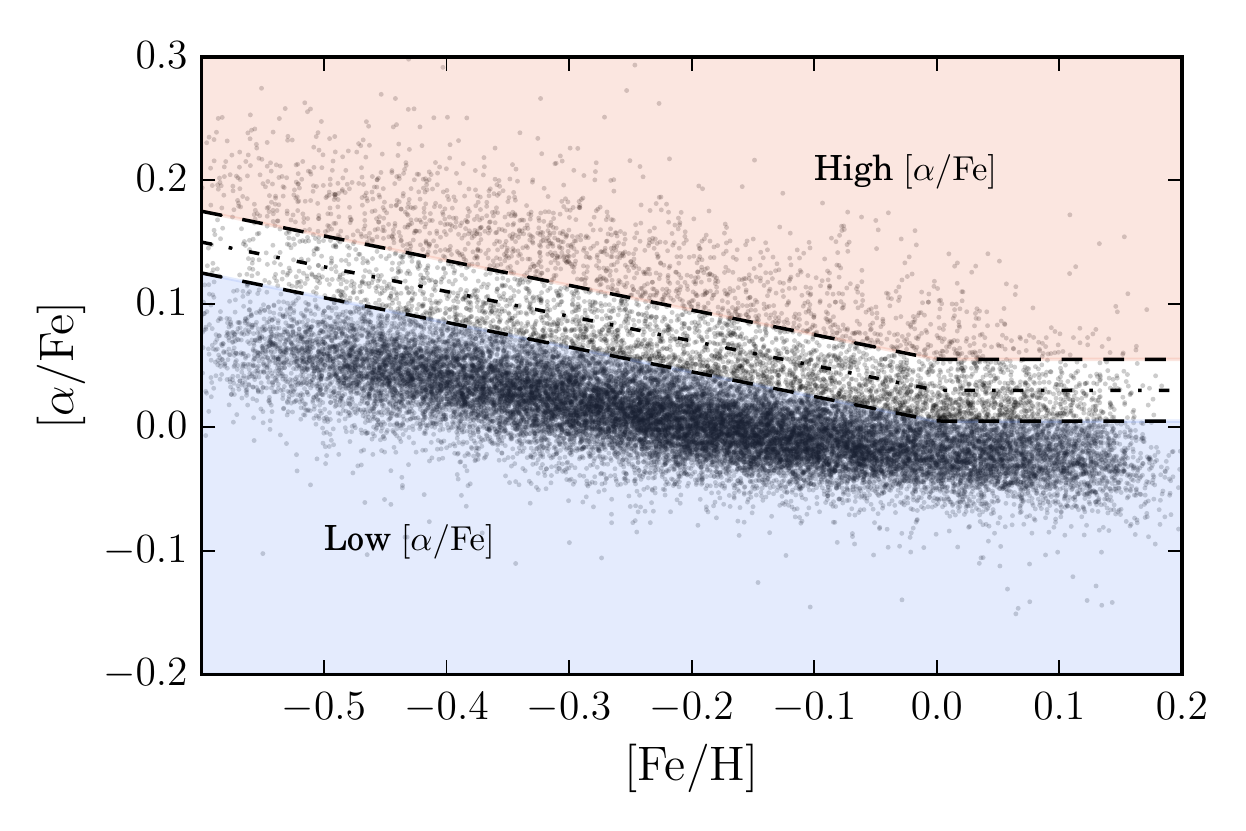}
    \caption{The full RGB sample in \afe{}--\feh{} space. The coloured regions show our division between low and high \afe{} subsamples. At each \feh{} the division between the samples is \afe{} $= 0.05 \ \rm{dex}$, roughly twice the mean uncertainty on \afe{} abundance determinations in APOGEE DR12. The bimodality in \afe{} at fixed \feh{} is visible across many \feh{}, and the lower number of stars in the high \afe{} sample is clear from this plot.}
    \label{fig:afe_feh}
\end{figure}

We further divide the sample into low and high \afe{} sub-samples, as it has been shown in previous work that the two populations have quite different structural parameters \citep{2016ApJ...823...30B}, and as such it makes sense to fit their mono-age sub-populations separately. We separate visually the low and high \afe{} populations, leaving a gap between the two samples of 0.05 dex in \afe{} at each \feh{} (our separation is shown in Figure \ref{fig:afe_feh}), to minimise contamination between the subsamples, particularly at the high \feh{} end, where the two populations partially overlap. The final density fits are performed on finer bins in age and \feh{}, which we define in Section \ref{sec:ages}. As the adopted separation in \afe{} removes 6532 stars from the full count, when calculating the surface-mass density contributions from the stellar number counts in each age-\feh{} bin, we remove the separation in \afe{}, using the star counts as if the populations were separated along the midpoint of the division (as shown by the dot-dashed line in Figure \ref{fig:afe_feh}). 

\begin{figure}
\centering
	\includegraphics[width=\columnwidth]{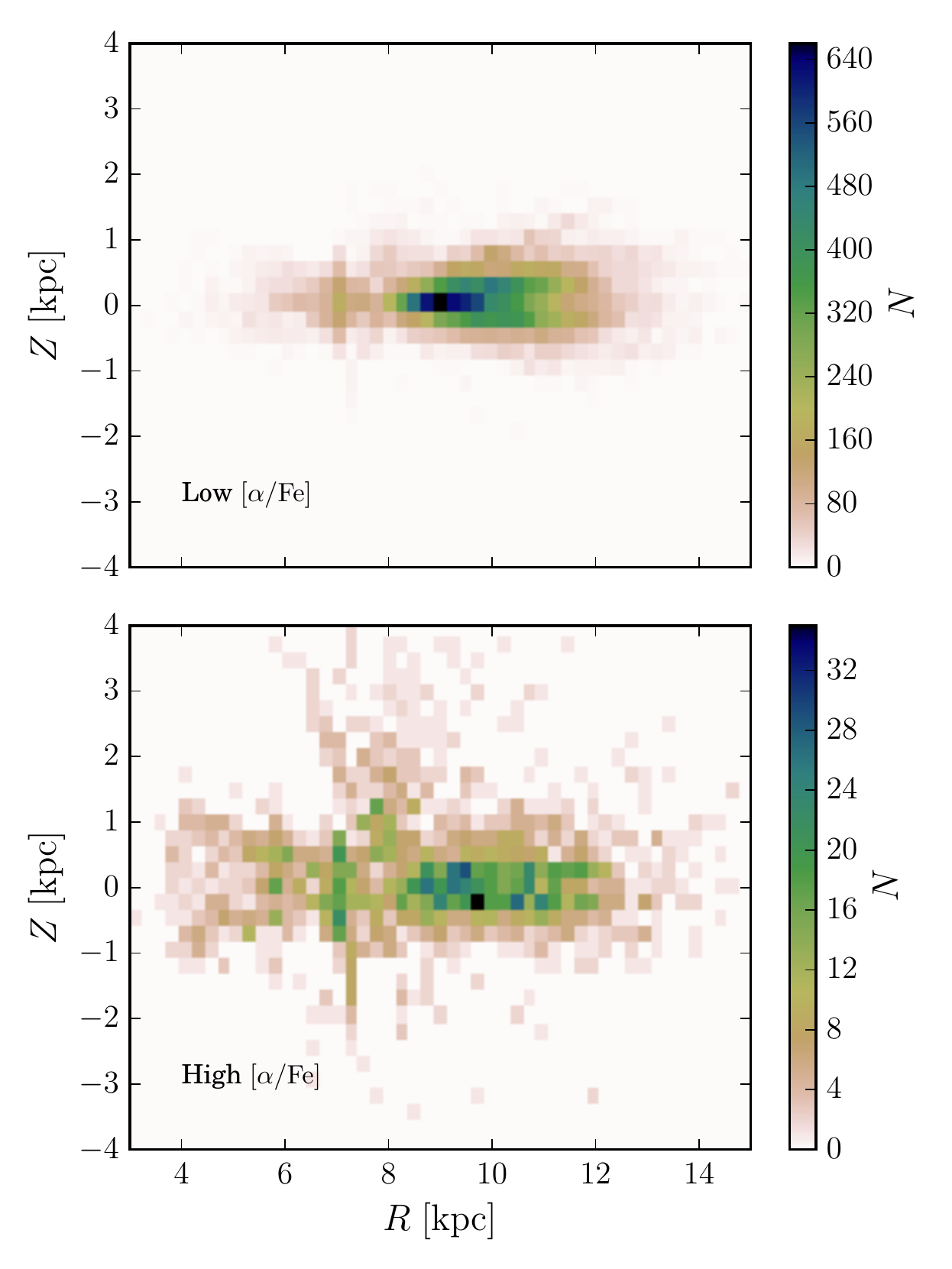}
    \caption{2D histograms of the spatial distribution (in Galactocentric $R$ and $Z$) of the high and low \afe{} subsamples shown in Figure \ref{fig:afe_feh}. The high \afe{} sample appears more diffuse and extended in height even before selection effects are accounted for. Readers should notice the different colour scale adopted for each panel, due to the much lower number of stars in the high \afe{} sample.}
    \label{fig:spatial}
\end{figure}

Our method, discussed in Section \ref{sec:methoda}, corrects for selection effects induced by interstellar extinction (in addition to the RJCE reddening corrections) using 3D dust maps for the Milky Way derived by \citet{2006A&A...453..635M} for the inner disk plane, combined with those for a large majority of the APOGEE footprint by \citep{2015ApJ...810...25G}, adopting conversions $A_H/A_{K_S}=1.48$ and $A_H/E(B-V) = 0.46$ \citep{2011ApJ...737..103S,2013MNRAS.430.2188Y}.  Fields with no dust data (of which there are $\sim 10$) are removed from the analysis. \citet{2016ApJ...823...30B} discuss the relative merits and limitations of these dust maps as opposed to others which are available, and determine that this combination of dust maps provides the best density fits.

\subsection{Distance estimates}
\label{sec:distances}
We use distance estimates from \citet{2014AJ....147..116H} \citep[But see also][for further description]{2015ApJ...808..132H}. Distances are estimated by computing the probability distribution function (PDF) of all distance moduli to a given star using a Bayesian method applied to the spectroscopic and photometric parameters from the DR12 catalogue and the PARSEC isochrones \citep{2012MNRAS.427..127B}. The distance estimates are found to have accuracy at the $15-20\%$ level, upon comparison with cluster members of well known distance observed by APOGEE. 

\begin{figure}
	\includegraphics[width=\columnwidth]{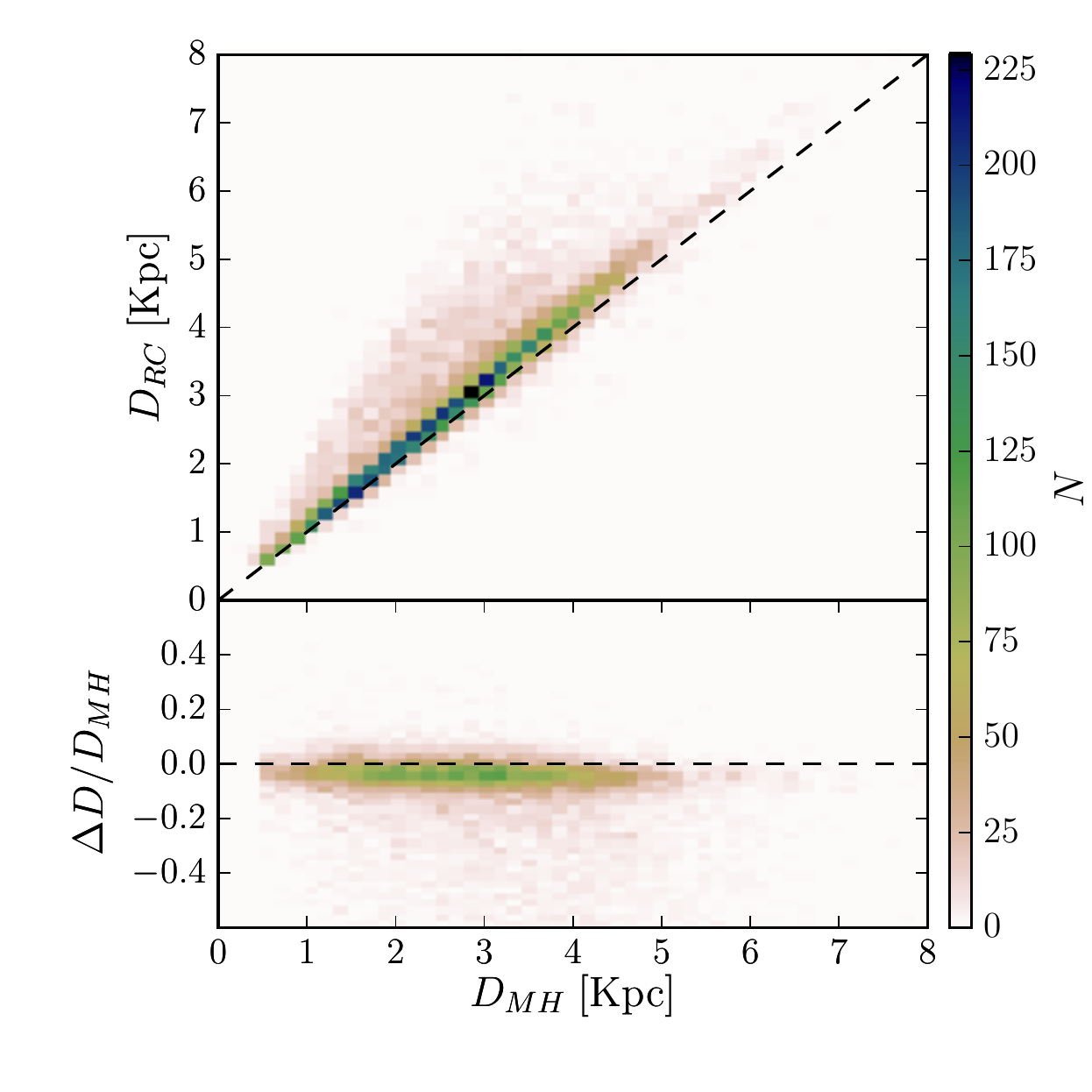}
    \caption{Comparison of APOGEE red clump catalogue (APOGEE-RC) distances $D_{RC}$ with distances derived by \citet{2015ApJ...808..132H}, $D_{MH}$. The top panel directly compares the distances, where the bottom panel shows the difference as a fraction of $D_{MH}$, as a function of that distance.  There are many stars with good agreement, but a distinct fraction of MH distances are underestimated compared to RC ($\sim 20\%$ with distances underestimated by more than $\sim 10\%$). As the variations in the density occur on scales which are, in general, far larger than these discrepancies, these are not problematic in our analysis. The two distance scales differ systematically by a factor of $\sim 5\%$, but we do not correct for this in the following discussion and it does not impact any of our results.}
    \label{fig:distcomp}
\end{figure}

We use the median of the posterior PDF (which is given in the output catalogue) as the estimate for the distance modulus, and compute the Galactocentric cylindrical coordinates, $R, \phi\ \rm{and}$  $Z$ for each star using the $l,b$ coordinates provided in the APOGEE-DR12 catalogue. The spatial distribution of the two \afe{} sub-samples is shown in Figure \ref{fig:spatial}. We perform a simple cross match between our sample and the APOGEE red clump (RC) value added catalogue (VAC) \citep{2014ApJ...790..127B} and plot the red-clump derived distance ($D_{RC}$) against the estimate from \citet{2014AJ....147..116H} ($D_{MH}$) in Figure \ref{fig:distcomp}. The red clump catalogue has precise distance estimates, which can be determined due to the red-clump having a near constant absolute magnitude. The majority of the \citet{2014AJ....147..116H}  distances compare well to the RC distances, but there are notable differences. The \citet{2014AJ....147..116H} distances can be underestimated by as much as 50\%, and we find that $\sim20\%$ of our sample have distances underestimated by more than 10\%. Our density fitting method is insensitive to uncertainties on the distances at these scales, as the scale of any variations in the density distribution can be assumed to be far greater than the distance uncertainties. 

Figure \ref{fig:distcomp} also shows that there is a systematic offset between the RC and \citet{2014AJ....147..116H} distances of the order $\sim 5\%$ across the full range of distances. We find that adopting this offset as a correction to the distances makes little impact on the final results, merely broadening fitted density profiles slightly, spreading the star counts over a wider Galactocentric distance, meaning that the final stellar surface-mass density estimates are unchanged.

\subsection{Age estimates}
\label{sec:ages}
We use age estimates for APOGEE DR12 catalogued by \citet{2016MNRAS.456.3655M}, who derive an empirical model for the $\mathrm{[C/N]} -M_*$ relation using asteroseismic masses from \emph{Kepler} and abundances from APOGEE for their overlapping samples \citep[APOKASC ][]{2014ApJS..215...19P}. Masses are predicted for DR12 stars which meet quality and stellar parameter criteria outlined in \citet{2016MNRAS.456.3655M}, and ages are estimated from that mass using the PARSEC isochrones with the nearest metallicity to that of the given star. \citet{2016MNRAS.456.3655M} use this empirical relation to build a model which predicts mass and age as a function of $[\text{[M/H], [C/M], [N/M], [(C+N)/M],}\log{g},T_{\mathrm{eff}}]$. It is important to note here that \citet{2016MNRAS.456.3655M} derive a model and fit for the ages in DR12 using the uncalibrated, raw stellar parameters, found in the FPARAM arrays in the APOGEE catalogue. This is difficult to account for when using the age catalogue alongside the calibrated parameters, and must be borne in mind in future comparisons of this work with models and observational results. \added[id=TM]{In addition to this, \citet{2016MNRAS.456.3655M} also mention that care should be taken when applying these ages to regions of the Milky Way where the chemical evolution may have been complex (e.g. the Bulge/Bar region). However, in their Figure 12, they compare the [C/N] ratio as a function of [M/H] in a sample of pre-dredge-up giants in the inner and outer disk, showing that the shapes of the distributions are similar. This suggests that differences in chemical evolution do not affect the $\mathrm{[C/N]}$-age relation within a wide range of galactocentric distances. Therefore, the assumption that it is safe to adopt the \citet{2016MNRAS.456.3655M} ages over the extent of the disk covered by our sample is robust, regardless of the fact that they are trained on the \emph{Kepler} sample, which is limited in its spatial extent.}

Although individual uncertainties on ages are not given in the catalogue, \citet{2016MNRAS.456.3655M} state that the model predicts ages with r.m.s errors of $\sim 40\%$. Although uncertainties are potentially very large at high age, our sample is binned with $\Delta \text{age}= 2$ Gyr in order to gauge general trends with age. It should be understood that such trends are smoothed by the age uncertainties\added[id=TM]{, particularly at high age}, and detailed comparisons to models should take the age uncertainty into account. \added[id=TM]{We discuss the effect of these uncertainties on our recovered trends with age in Appendix \ref{sec:ageerror}, showing that our methodology can still reliably recover such trends, even though mixing between bins may be present in the data.}

\begin{figure}
	\includegraphics[width=\columnwidth]{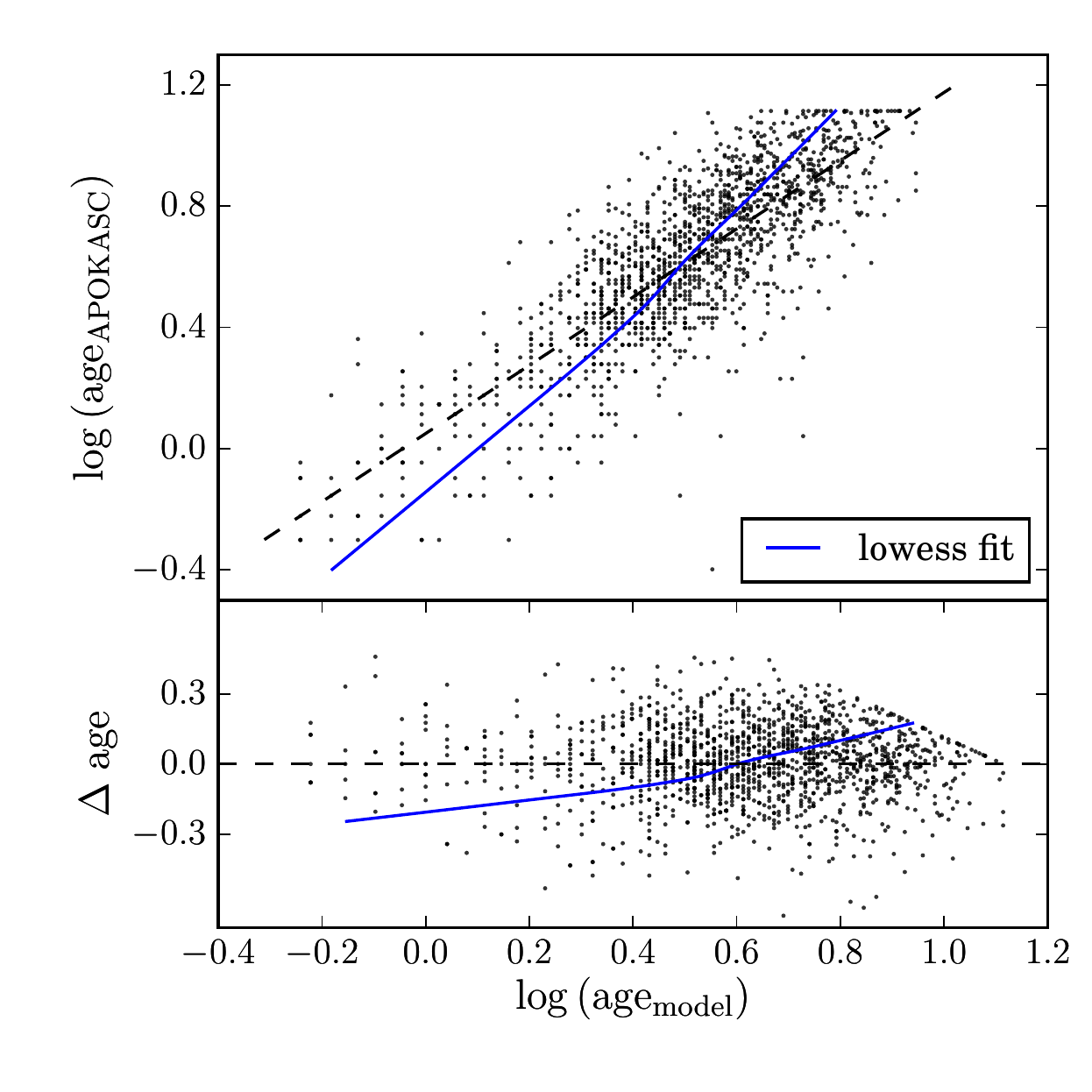}
	\centering
    \caption{Asteroseismically determined ages from APOKASC against the $\mathrm{[C/M]}$ and $\mathrm{[N/M]}$ based ages from \citet{2016MNRAS.456.3655M}. The line gives the fitted correction for ages from \citet{2016MNRAS.456.3655M}, based on the values for the APOKASC training set, given in their Table 1. We fit the data using a non-parametric lowess fit. Before corrections, older ages are under-predicted, and young ages are over-predicted. The corrections mainly change the scaling of the ages, such that the high \afe{} sample occupies an age range more in-line with the existing literature \citep[e.g.][]{2016arXiv160407771A,2013A&A...560A.109H} }
    \label{fig:correction}
\end{figure}

Figure 11 of \citet{2016MNRAS.456.3655M} shows that there is a significant bias in the ages returned by the model, such that ages are underpredicted at high age when compared to the training set. For this reason, we fit for and apply a correction to the catalogued ages before performing the density fitting. Using Table 1 of \citet{2016MNRAS.456.3655M}, we perform a non-parametric lowess fit to the predicted age--true age distribution. This fit is then used to derive age corrections as a function of predicted age. We show the fitted correction in Figure \ref{fig:correction} in both predicted vs. true age space and also in $\Delta$ age against the predicted age. The correction as a function of predicted age is then applied to each of the ages in the DR12 catalogue. In all further analysis, we refer only to the corrected ages. \added[id=TM]{The main effect of this correction is to make the high-\afe{} stars older. Consequently, our surface-mass density estimates (presented in Section \ref{sec:surfmassdens}) become more conservative, as the mass contribution per star in older bins is lower (as discussed in Section \ref{sec:discrepant}). \citet{2016MNRAS.456.3655M} comment on the bias in the context that the ages returned for high \afe{} stars appear younger than previous estimates \citep[from][]{2013A&A...560A.109H,2014A&A...562A..71B,2014A&A...565A..89B}. Our correction brings these data more in line with those estimates.}  

It is also important to account for all other cuts made by \citet{2016MNRAS.456.3655M} on the stellar parameters in the APOGEE catalogue, outlined in full at the beginning of their Section 6.2. While we account for cuts in $T_{\mathrm{eff}}$ and $\log{g}$ by applying the same cuts to the isochrone grid when calculating surface-mass density contributions, it is not possible to properly account for cuts made on the stellar abundances in this way. We find that 9,041 stars are removed from the 40,285 star catalogue (those with distances, after the $\log{g} $cut mentioned above) by these abundance cuts, to give our final catalogue size of 31,244. This means that $\sim 25\%$ of star counts are missing from the age catalogue, and therefore unaccounted for by our analysis. With no robust method for determining the age distribution of these missing stars, we are forced to make the assumption that these star counts can be added uniformly to each age-\feh{} bin. We make this correction by simply increasing the counts in each bin by $25\%$ when calculating the surface-mass density. This correction simply increases the final surface-mass density values systematically by $25\%$.

\added[id=TM]{Another important consideration when using this set of ages regards stars whose chemical compositions are such that the ages fit from the model (after making the corrections) would be higher than 13 Gyr and left out of our analysis. As the age estimates are computed based on the surface parameters and abundances of the stars using a fitting function, many stars with strongly outlying abundances and parameters can be assigned ages which are greater than 13 Gyr. We find that the 3020 such stars (in the final sample) have an average [C/Fe] which is lower than the general sample, and an average [N/Fe] which is enhanced with respect to the stars with reliably measured ages. We also find that at fixed \feh{} and $\log{g}$ these stars have warmer $T_{\mathrm{eff}}$. While these properties are expected given their age measurements, we believe that there is a distinct possibility of some peculiarity of these stars. For example, if such stars were early AGB stars \citep[having gone through the second dredge-up, reducing the surface C abundance, e.g.][]{1999ApJ...510..232B}, which had been fit as RGB stars, their actual age may be considerably younger, and their counts missed in the younger bins (where mass contribution per star is higher). As the nature of these stars is debatable and a correction cannot be made confidently before carrying out the full analysis, we regard the missing counts from these stars as a contribution to the systematic error budget, which we discuss fully in Section \ref{sec:discrepant}.}

\begin{figure}
	\includegraphics[width=\columnwidth]{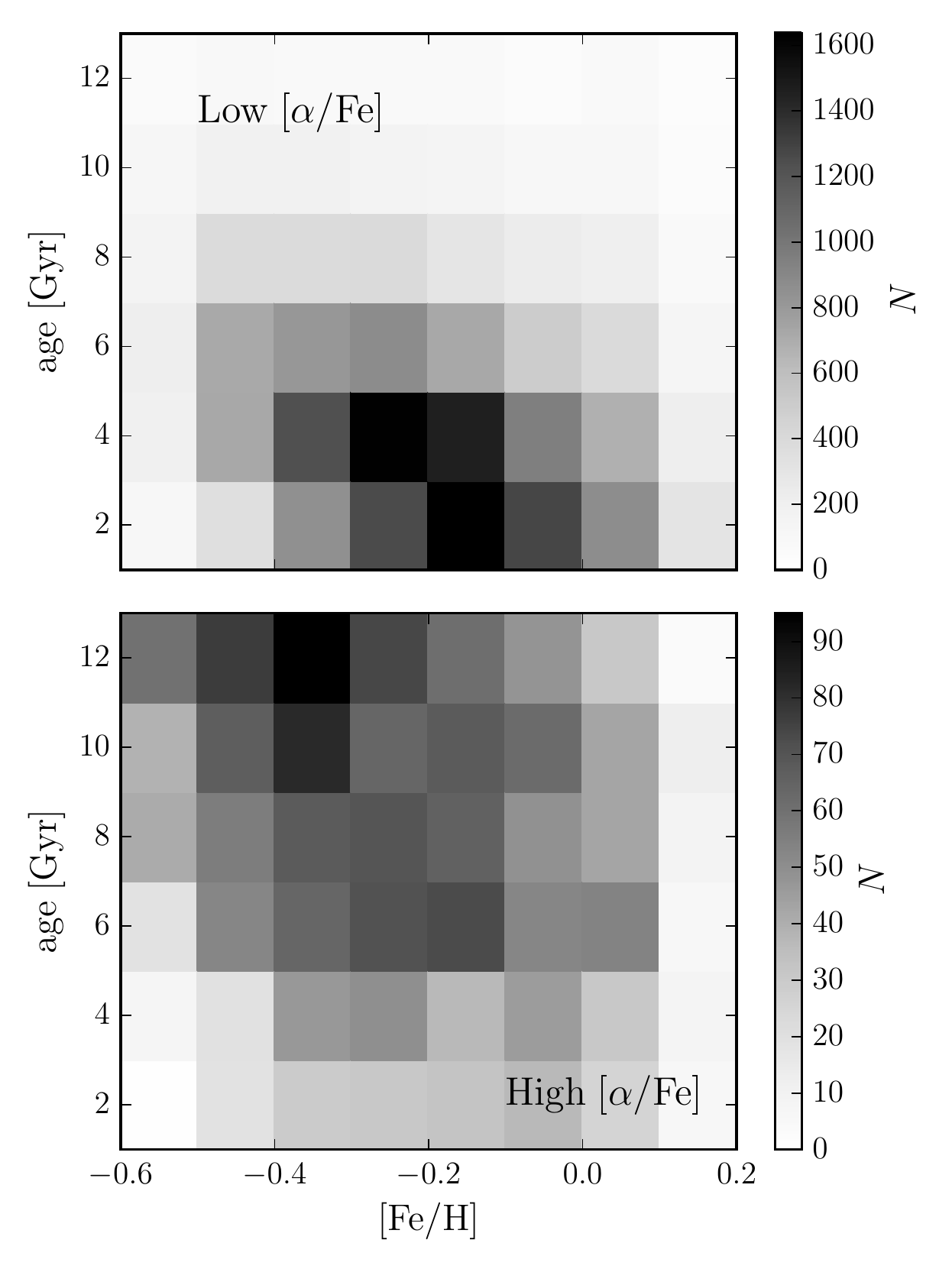}
    \caption{2D Histograms showing the raw number of stars in each (age,\feh{}) bin of the low (\emph{left}) and high (\emph{right}) \afe{} sub-samples. We draw the reader's attention to the difference in amplitude between the two sub-samples (and the associated difference in colour scale normalisation). Although the majority of bins are well sampled ($\gtrsim 30$ stars), there are some greatly undersampled bins, for which well-defined fits are not possible. }
    \label{fig:numbins}
\end{figure}

We demonstrate the adopted binning in (age,[Fe/H]) space in Figure \ref{fig:numbins}, showing also the number of stars which fall in each bin and the general distribution of stars in (age,[Fe/H]) space. There is a notable separation in age between the high and low \afe{} subsamples. 

\section{Method}
\label{sec:methoda}
In this section we describe the method for fitting the underlying number density of stars in the Milky Way from APOGEE observations, which we represent here as $\nu_*(X,Y,Z|\theta)$, in units of stars kpc$^{-3}$. The calculation of this quantity requires allowances to be made for the survey selection function, which is non-trivial due to the presence of inhomogeneous dust extinction along lines of sight observed by APOGEE, the target selection invoking different $H$ magnitude limits, and the use of RGB stars as a tracer, which cannot be considered as standard candles. The quantity which we are ultimately interested in is the surface-mass density of stars at the solar radius, $\Sigma_{R_0}$, in units of $\mathrm{M_{\odot}}\ \mathrm{pc^{-2}}$, which we infer from the number of stars in the APOGEE sample as a function of position. We describe the method for this calculation in Section \ref{sec:surfmasscalc}. Our methodology consists of an adaptation of that used by \citet{2016ApJ...823...30B}, employing a modified version of their publicly available code\footnote{Available at https://github.com/jobovy/apogee-maps}. Although the general method is identical, we describe again the key components for clarity and completeness.

 As some readers may find it unnecessary to read in full the details of the methodology (which are described in the following sections), we summarise the procedure as follows:
 \begin{itemize}
 \item We fit parametric density models to the APOGEE star counts using a maximum likelihood fitting procedure, based on the assumption that star counts are well modelled as an inhomogeneous Poisson point process. The density models which we assume throughout the paper are described by radially broken exponentials, with scale lengths $h_{R,\mathrm{[in,out]}}$ either side of a break radius $R_{\mathrm{peak}}$ (where $h_{R,\mathrm{in}}$ denotes the scale length of the inner profile and vice versa), and a vertical distribution which is a single exponential with scale height $h_Z$, which is modified as a function of $R$ by an exponential flaring term with scale length $R_{\mathrm{flare}}$. We show that, in general, if the density is better fit by a single exponential, it is recovered as so by our procedure.
 \item We obtain a best fit density model for every bin in age and \feh{}, at high and low \afe{}. This best fit model is then used to initiate an MCMC sampling of the posterior PDF. We then use the median and standard deviation of one dimensional projections of the MCMC chain as our adopted parameter values and uncertainties.
 \item As the fitting procedure does not fit for the normalisation of the density $N_{R_0}$, the number surface density of stars at the solar radius in stars pc$^{-2}$, we calculate this value by comparing the observed number of stars in each bin to that which would be observed in APOGEE for the fitted density model if $N_{R_0} = 1 $ star pc$^{-2}$. We then convert $N_{R_0}$ for each bin into the surface-mass density in visible stars at the solar radius $\Sigma_{R_0}$ by converting the mass in RGB stars observed to the total mass using stellar evolution models.
\end{itemize}
 Readers can then pick up the results in Section \ref{sec:resultsa}.

\subsection{Density Fitting Procedure}
\label{sec:densfit}
We first fit for the number density of stars for each sub-population as defined by Figure \ref{fig:numbins}. The following discussion describes the general procedure used for fitting density models with a generic set of parameters $\theta$. The actual stellar number density model adopted is discussed in Section \ref{sec:densitymodel}. \citet{2016ApJ...823...30B,2012ApJ...753..148B,2013A&ARv..21...61R} have shown that the observed rate of stars as a function of position, magnitude, colour and metallicity can be modelled as an inhomogeneous Poisson point process. Stars are distributed in the space defined by $O = [l,b,D,H,\mathrm{[J-K_S]_0}, \mathrm{[Fe/H]}]$ -- position, magnitude, colour and metallicity -- with an expected rate $\lambda(O|\theta)$ (which has units of stars per arbitrary volume in O), parameterised by a set of parameters $\theta$ (which are, in this particular case, the parameters describing an adopted density profile). This rate function is written fully as
\begin{multline}
\lambda(O|\theta) = \nu_*(X,Y,Z|\theta) \times |J(X,Y,Z;l,b,D)| \\ \times \rho(H, [J-K_S]_0, \mathrm{[Fe/H]}|X,Y,Z) \times S(l,b,H) 
\label{eq:rate}
\end{multline}
where $\nu_*(X,Y,Z|\theta)$ is the quantity we aim to estimate, which is defined as the stellar number density in rectangular coordinates, in units of stars kpc$^{-3}$. $ |J(X,Y,Z;l,b,D)|$ is the Jacobian of the transformation from rectangular $(X,Y,Z)$ to Galactic $(l,b,D)$ coordinates and $ \rho(H, [J-K_S]_0, \mathrm{[Fe/H]}|X,Y,Z) $ denotes the density of stars in magnitude, colour and metallicity space given a spatial position $(X,Y,Z)$, in units of stars per arbitrary volume in magnitude, colour and metallicity space. $S(l,b,H)$ is the survey selection function (the fraction of stars observed in the survey) which includes dust extinction effects, which we discuss in the following. When expressed in this way, fitting the density model parameters $\theta$ becomes a maximum likelihood problem.

The likelihood is a sum over all data-points considered in a given age-\feh{} bin, and gives the likelihood of the parameters $\theta$ given the data. For this application, it is written as
\begin{multline}
\ln \mathcal{L}(\theta) = \sum_{i} \left[ \ln \nu_{*}(X_i, Y_i, Z_i|\theta)- \ln \int dO \lambda(O|\theta) \right]
\label{eq:likelihood}
\end{multline}
where second term on the right hand side of the equation, $ \int dO \lambda(O|\theta)$, describes the effective volume of the survey. we drop the other factors in the rate in Equation (\ref{eq:rate}) in the argument of the logarithm, because the other factors do not depend on the model parameters $\theta$. The effective volume is independent of the data-point considered, and is an intrinsic property of the survey for a given $\theta$. It provides the normalisation for the rate likelihood, and is non-trivial to evaluate due to the presence of patchy dust extinction along lines of sight in the survey.

The effective volume is written generally as
 \begin{multline}
\label{eq:effvol}
\int dO \lambda(O|\theta) = \sum_{\text{fields}} \Omega_f \int dD D^2 \nu_*([X,Y,Z](D,\text{field})|\theta)\\ \times \mathfrak{S}(\text{field},D)
\end{multline}
which is a sum over all APOGEE fields, where $\Omega_f$ is the solid angle of the field considered. The integrand $ \nu_*([X,Y,Z](D,\text{field})|\theta)$ is the density at each point along a line of sight, assumed to be constant over the angular size of the field. $\mathfrak{S}(\text{field},D)$ represents the effective survey selection function, which is given by the integration of the survey selection function over the area of the field and is written, in this case, as
\begin{multline}
\mathfrak{S}(\text{field}, D) = \sum_k S(\text{field},k) \\ \int dM_H \frac{\Omega_k(H_{\text{[min,max]},k}, M_H, A_H[l,b,D], D)}{\Omega_f}.
\label{eq:effsel}
\end{multline}
This is a sum over the apparent magnitude bins, $k$, in the APOGEE target selection, with the integral representing the fractional area of the APOGEE field where stars are observable, given the distance modulus and extinction at a given position. The term describing this area is $\Omega_k$, which is the observable area of the field at a given distance and absolute magnitude, written as
\begin{multline}
\Omega_k(H_{\text{[min,max]},k}, M_H, A_H[l,b,D], D) =\\ \Omega(H_{\text{min},k} - [M_H - \mu(D)] < A_H(l,b,D) < H_{\text{max},k} - [M_H - \mu(D)])
\end{multline}
where $H_{\text{[min,max]},k}$ denotes the minimum and maximum $H$ for an apparent magnitude bin $k$ in the APOGEE target selection and $\mu(D)$ is the distance modulus at $D$.  $A_H(l,b,D)$ is the $H$ band extinction at a given position, which we obtain from the 3D dust maps described in Section \ref{sec:APOGEE}. This area is integrated (in Equation (\ref{eq:effsel})) over the full absolute $H$-band magnitude, $M_H$, distribution in an (age,\feh{}) bin. We find the $M_H$ distribution for each (age,\feh{}) bin using the PARSEC isochrones \citep{2012MNRAS.427..127B} within that bin, weighted with a \citet{2003PASP..115..763C} IMF. We apply the same cuts in $\log{g}$ and $(J-K_S)_0$ colour to the isochrone points as are imposed on the data, and perform a Monte Carlo integration using the resulting $M_H$ distribution to evaluate the integral in Equation (\ref{eq:effsel}). S(\text{field},k) in Equation (\ref{eq:effsel}) denotes the 'raw' APOGEE selection function, which gives the fraction of the stars in the photometric catalogue that were observed spectroscopically \citep[see][for details]{2013AJ....146...81Z}. This number is constant within an apparent magnitude bin and within an APOGEE field, which is why $S$ is cast as a function of field and magnitude bin in Equation (\ref{eq:effsel}). The values of S(\text{field},k) (and $\mathfrak{S}(\text{field}, D)$) are evaluated using the \texttt{apogee} python package \footnote{Available at {https://github.com/jobovy/apogee}}.

We evaluate $\mathfrak{S}(\text{field}, D)$ on a grid of distances for each APOGEE field for simple computation of $\int dO \lambda(O|\theta)$. We then optimise the likelihood function in Equation (\ref{eq:likelihood}) for a given density model and data-set using a downhill-simplex algorithm, to obtain the best fitting set of parameters $\theta$. A Markov Chain Monte Carlo (MCMC) sampling of the posterior PDF is then initiated using this optimal solution. This is implemented with an affine-invariant ensemble MCMC sampler  \citep{goodmanweare2010,2013PASP..125..306F}. All parameter values and associated uncertainties for individiual (age,\feh{}) bins which are reported in the following sections represent the median and standard deviation $\sigma$, respectively, of one dimensional projections of the MCMC chain. 

\subsection{Adopted stellar number density models}
\label{sec:densitymodel}
It was shown in \citet{2016ApJ...823...30B} that density profiles of MAPs are well represented by axisymmetric profiles that can be written as 
\begin{equation}
\nu_*(R, \phi, Z) = \Sigma(R)\zeta(Z|R) \quad \text{where } \int dZ \zeta(Z|R) = 1.
\end{equation}
Furthermore, the exact form of the best fitting profile is that of a radially broken exponential, with a vertical profile that is an exponential with a scale height which varies exponentially with R (a flaring profile), such that
\begin{equation}
\ln \Sigma(R) \propto  \begin{cases}
    -h_{R,\text{in}}^{-1}(R-R_0)    & \quad \text{where } R \leq R_{\text{peak}}\\
   -h_{R,\text{out}}^{-1}(R-R_0)  &\quad \text{where } R > R_{\text{peak}}\\
  \end{cases}
\end{equation}
and
\begin{equation}
\ln \zeta(Z|R) \propto h_Z^{-1} \exp{(R_{\mathrm{flare}}^{-1}[R-R_0])} |Z|-\ln{h_Z(R)}.
\end{equation}
$R_0$ denotes the solar radius, which we assume here to be 8 kpc. This number only sets the radius at which the profiles are normalised, and so does not have any effect on the fitting procedure. We use the same general set of density profiles to describe the mono-age, mono-metallicity populations which are studied here. We bin in \feh{} to account for the observed \feh{} spread at fixed age \citep[e.g.][and our Figure \ref{fig:numbins}]{1993A&A...275..101E}. \citet{2016ApJ...823...30B} also showed that when mock data were fit using the procedure in Section \ref{sec:densfit} and the density profile above, the input parameters were always recovered within acceptable uncertainty ranges. In particular, mock data generated from a single exponential profile was still recovered as such (i.e. with $R_{\mathrm{peak}} = 0$) even when fit assuming a broken exponential profile.

We also note here that our sample is not limited to stars which are members of any specific Galactic component, and as such, may include small numbers of halo stars in the very high \afe{} and low \feh{} regimes. However, our fitting procedure is agnostic to these contaminants, which would only cause the fits to have larger uncertainty (from the MCMC exploration) about the best fit from the dominant population in a given bin.

\subsection{Stellar surface-mass densities}
\label{sec:surfmasscalc}
We compute the surface-mass density in visible stars for each of our age and \feh{} populations using the method originally outlined in \citet{2012ApJ...751..131B}. As our fitting procedure does not fit for the normalisation of the density (we normalise to a surface density of 1 at $R_0$), we first compute the normalisation $N_{R_0}$, which represents the number density of stars at the solar radius in units of stars pc$^{-2}$ in an (age,\feh{}) bin. $N_{R_0}$ is given by the relation 
\begin{equation}
N_{R_0} = \frac{N_{*,\text{observed}}}{\int dO \lambda(O|\theta)}
\end{equation}
where $N_{*,\text{observed}}$ is the number of stars observed in the survey for a given (age,\feh{}) and $\int dO \lambda(O|\theta)$ is the usual definition of the effective volume (given by Equation (\ref{eq:effvol})) for a given set of parameters $\theta$, found using the method in Section \ref{sec:densfit}. 

We find the contribution to stellar surface-mass density by first multiplying $N_{R_0}$ by the average mass of a red giant star in the same range of age and \feh{}, given the selection criteria on $\log{g}$ given in Section \ref{sec:APOGEE} (which picks out the RGB) and  $(J-K_S)_0 \geq 0.5$ (given that we only use fields in which this cut was applied).  We then correct this value to represent the total stellar population by dividing by the fractional contribution of the red giants to the total underlying population. These values are found using PARSEC isochrones \citep{2012MNRAS.427..127B}, weighted with a Log-normal \citet{2001ApJ...554.1274C} IMF, as described in the calculation of the effective volume in Section \ref{sec:densfit}. This then leads us to a stellar surface-mass density $\Sigma_{R_0}$ as a function of age and \feh{}. This conversion can be expressed as
\begin{equation}
\Sigma_{R_0}(\mathrm{age,[Fe/H]}) = N_{R_0} \frac{\langle M_{\text{RGB}} \rangle (\mathrm{age,[Fe/H]})}{\omega(\mathrm{age,[Fe/H]})}
\end{equation}
Where $\langle M_{\text{RGB}} \rangle (\mathrm{age,[Fe/H]})$ is the mean stellar mass in an (age,\feh{}) bin, and $\omega(\mathrm{age,[Fe/H]})$ is the fraction of stars in the total stellar population in an (age,\feh{}) bin which are within the $\log{g}$ and $(J-K_S)_0$ cuts in APOGEE. The stellar surface-mass density contributions of each bin can then be summed to give the total stellar surface-mass density at the solar radius $\Sigma_{R_0, \text{tot}}$.

Our final surface-mass density estimate is strongly dependent on the conversion factors in the above equations, the average RGB star mass, $\langle M_{\mathrm{RGB}} \rangle$, and the fractional contribution from giants, $\omega$. We find that the average giant masses in our range of ages and \feh{} span $0.9 \lesssim \langle M_{\mathrm{RGB}} \rangle \lesssim 2.1 \mathrm{M_{\odot}}$. The most metal poor and oldest populations have the lowest average mass, and the youngest, most metal rich populations have the highest. The fractional contribution from giants in this regime ranges between $0.002 \lesssim \omega \lesssim 0.02 $. The oldest and most metal poor populations have the least giants, whereas the youngest, metal rich populations have the most. These values appear to sit well with recent inventories of the solar neighbourhood, which suggest giants should make up of the order of a few percent of the mass \citep{2015ApJ...814...13M}. We discuss the potential systematics introduced by the use of stellar evolution models in Section \ref{sec:discrepant}.

\section{Results}
\label{sec:resultsa}
We now present results from the density fitting procedure, and the subsequent calculation of the surface-mass density contribution of each mono-age, mono-\feh{} population in \autoref{fig:numbins}. Density fitting is performed on all populations, but we only display the fits for populations with $> 30$ stars, as as data below this level become too noisy to render reliable fits. Although the remaining fits can be noisy when star counts are near this limit, this is reflected in the error analysis arising from the MCMC exploration of the posterior PDF of the fitted parameters. We refer the reader to Appendix \ref{sec:densityfits} for a comparison between the data and the fitted models for each mono-age, mono-\feh{} bin, and a qualitative discussion regarding the rationale behind the decision to discuss fits to only the broken exponential density profile.

\subsection{The radial profile of mono-age, mono-\feh{} populations}
We first show the fits to the surface density in the low and high \afe{} sub-samples in Figure \ref{fig:surfdens}. We display fits for all age and \feh{} bins with $> 30$ stars. By shading the profiles by their surface-mass density contribution (as shown in Section \ref{sec:surfmassdens}), we intend to draw the eye to the profiles which contribute most to the mass of the Milky Way disk. We defer a discussion of the individual mass contributions of each bin to Section \ref{sec:surfmassdens}, concentrating in this section on trends in the shapes of the density profiles.

\begin{landscape}
\begin{figure}
     \includegraphics[width=1.\textwidth]{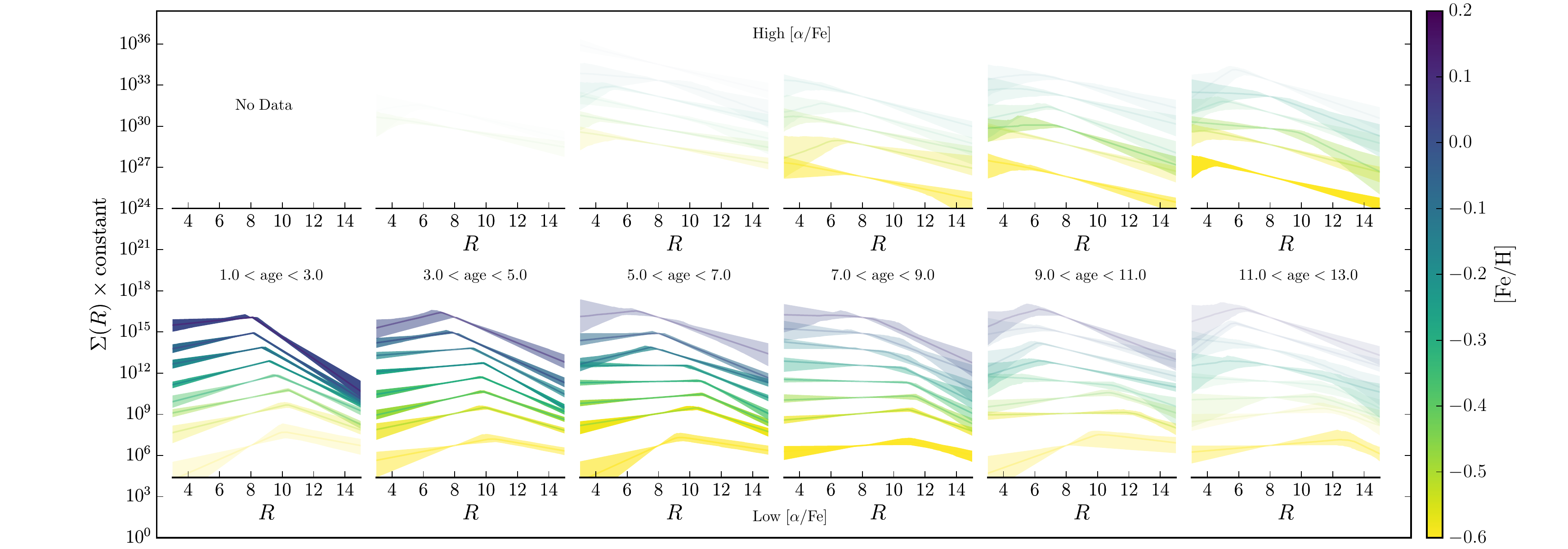}
     \centering
    \caption{The fitted surface density profiles for the high \afe{} (\emph{top}) and low \afe{} (\emph{bottom}) sub-samples as a function of \feh{} (colour) and age (increasing from left to right). The coloured bands represent the 95\% uncertainty range.  Only profiles for bins containing$ > 30$ stars are shown. The profiles have a transparency according to the surface-mass density calculated for each bin in Section \ref{sec:surfmassdens}, normalised separately for each row (i.e. in each \feh{} bin), to draw the eye to those profiles which contribute most to the Milky Way surface-mass density. High \afe{} profiles are described well by a single exponential, whereas young, low \afe{} profiles are broken exponentials with a peak density which varies in radius in the disk.}
    \label{fig:surfdens}
\end{figure}

\begin{figure}
	\includegraphics[width=1.\textwidth]{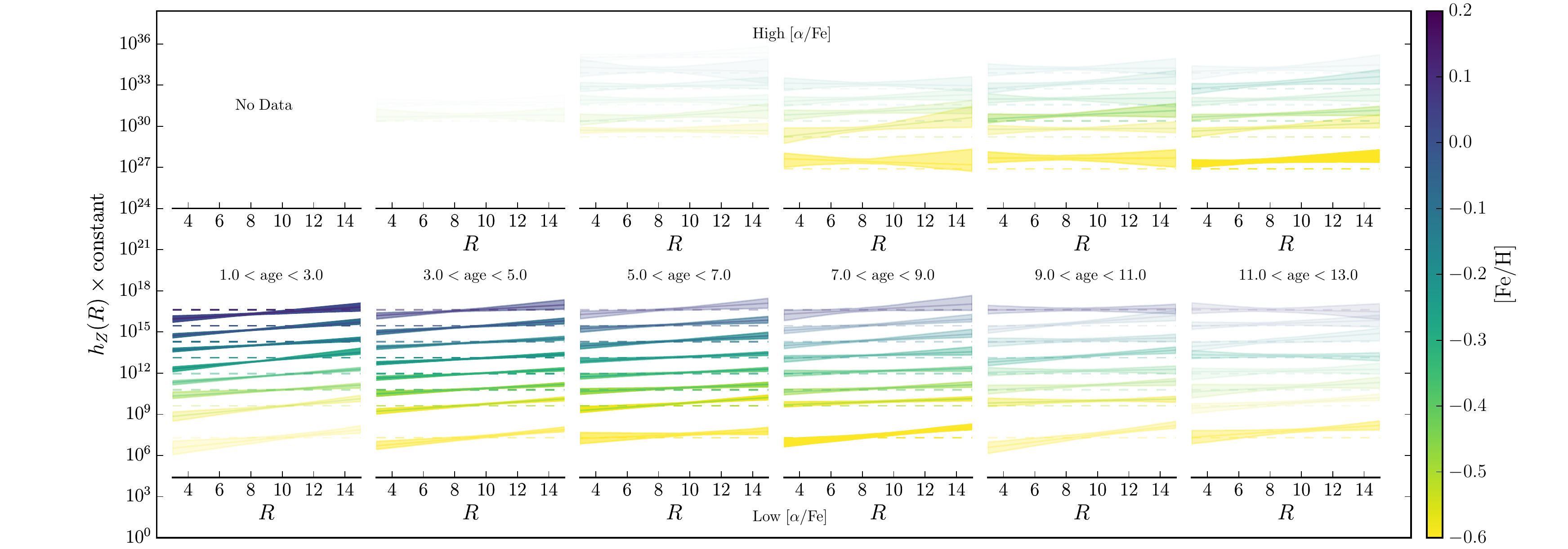}
	\centering
    \caption{Vertical profiles for the high \afe{} (\emph{top}) and low \afe{} (\emph{bottom}) sub-samples as a function of \feh{} (colour) and age (increasing from left to right). The coloured bands represent the 95\% uncertainty range. Only profiles for bins with $> 30$ stars are shown, with profiles shaded according to their surface-mass density contributions (discussed in Section \ref{sec:surfmassdens}).  The dashed lines represent $h_Z = 0.3$ kpc, for reference.}
    \label{fig:hzprofile}
\end{figure}
\end{landscape}
Although fit with the broken exponential, high \afe{} profiles are generally better described by near-single exponentials, either showing no break in the radial range, or being fit by a profile with a break at low significance (i.e. a single line could be drawn through the coloured band). Many of the outer profiles (after $R_{\mathrm{peak}}$) in the high \afe{} sub-sample appear to have a similar slope, suggesting that they may all be represented by the same exponential. The mean outer scale length for the high \afe{} populations is $h_{R,\text{out}} = 1.9\pm 0.1$ kpc. The picture is noticeably different in the low \afe{} sub-sample, with profiles showing clear breaks, at well defined radii. Any trends in break radius in this regime are determined with high significance.
Low \afe{} profiles have a density which increases with radius out to the break radius, and declines outward of this radius. We do not constrain the fits to behave in this way, and this indicates that mono-age, mono-\feh{} populations at low \feh{} are shaped approximately as donut-like annuli. The variation of the break radius then represents the moving peak of stellar density as a function of age and \feh{}. Concentrating on the bins youngest bin ($1 < \mathrm{age} < 3$ Gyr), the break radius is a declining function of metallicity, moving between $R_{\mathrm{peak}}=10$ kpc at $-0.6 < \mathrm{[Fe/H]} < -0.5\ \mathrm{dex}$ down to $R_{\mathrm{peak}} < 8$ kpc at $0.1 < \mathrm{[Fe/H]} < 0.2\ \mathrm{dex}$. This trend is also present in older bins but with decreased amplitude. In a fixed \feh{} bin, $R_{\mathrm{peak}}$ appears to remain roughly constant (within $\sim1$ kpc) at ages between 1 and 6 Gyr.  At ages older than this $R_{\mathrm{peak}}$ varies in unexpected ways, but there is much less mass contribution from these populations, and we attribute much of this behaviour to noise from the narrow age bins. 

\begin{figure}
	\includegraphics[width=\columnwidth]{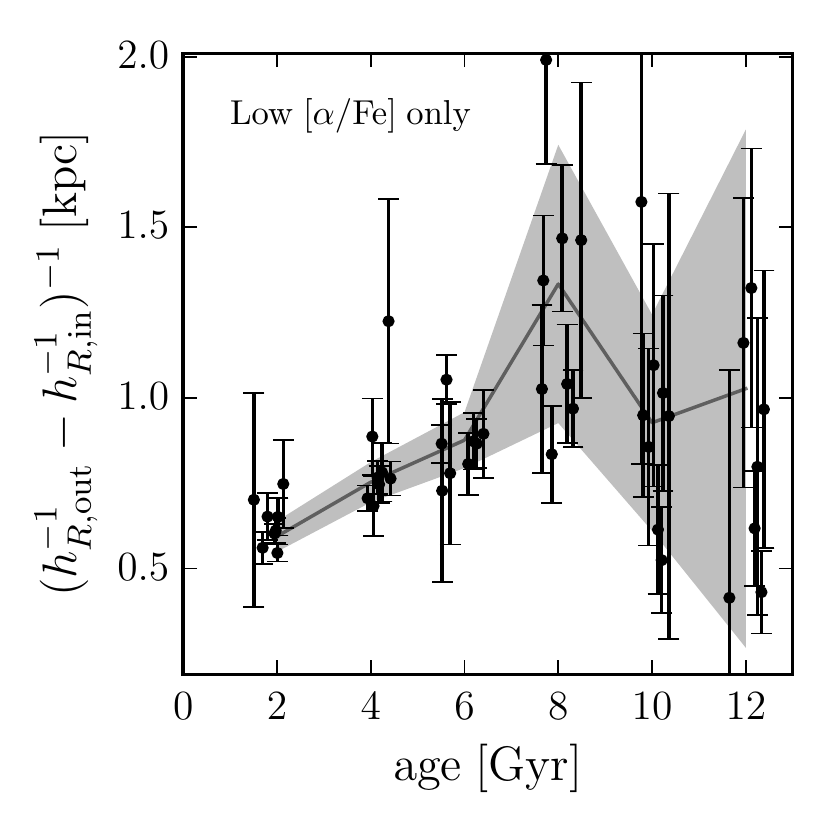}
    \caption{The profile width $(h_{R,\mathrm{out}}^{-1} - h_{R,\mathrm{in}}^{-1})^{-1}$ against age for the low \afe{} populations (this diagnostic is irrelevant for the high \afe{} populations, which are generally fit by single exponentials). We add a small random jitter to the central age of each age bin, to make individual points and their uncertainty clearer. The relations and coloured band shows the running surface-mass density weighted mean and standard deviation in the age bins. The profile width increases with age. A higher value of this diagnostic suggests a broader surface density profile, showing that older populations are flatter and broader around the peak density.}
    \label{fig:agevsbroadening}
\end{figure}

On the other hand, the low \afe{} profiles change \emph{shape} (either side of $R_{\mathrm{peak}}$) with age in a fixed \feh{} bin. The youngest populations show a sharp peak, with a steep increase and decline either side of $R_{\mathrm{peak}}$. As populations grow older, the profile broadens significantly, becoming almost flat in the lowest \feh{} bins. We show this behaviour by finding the inverse of the difference between the inverse outer and inner scale length\footnote{Taking a ratio of the sum of the density at fixed $\Delta R$ either side of $R_{\mathrm{peak}}$ to that at $R_{\mathrm{peak}}$ would give some measure of width. Then, assuming $\Delta R << h_{R,[\text{in,out}]}$, a Taylor expansion of this ratio $\sim \Delta R (h_{R,\text{out}}^{-1}-h_{R,\text{in}}^{-1})$. We then plot the inverse of this factor such that it increases for broader profiles.}, such that a low value denotes a sharper peak, whereas a broader profile has a higher value. We show how this value changes with age for the low \afe{} populations in Figure \ref{fig:agevsbroadening}. The peak is sharpest in the younger populations, and becomes broader with age. Old populations have artificially sharpened peaks in this diagnostic due to their being better described by single exponentials. Notably also, in Figure \ref{fig:surfdens}, at low \feh{} the inner profiles flatten faster than the outer profile, whereas the higher \feh{} populations show the opposite behaviour. For example, in the $-0.3 < \mathrm{[Fe/H]} < -0.2\ \mathrm{dex}$ bin, the outer profile appears to remain roughly constant in slope between 1 and 6 Gyr, while the inner profile flattens significantly. The opposite is seen in the $0.1 < \mathrm{[Fe/H]} < 0.2\ \mathrm{dex}$ bin, where the outer profile flattens considerably with age. 

\subsection{The vertical profile of the disk}
We now examine the variation of $h_Z$ as a function of radius in mono-age, mono-\feh{} populations. Because mono-age, mono-\feh{} populations are well described by a single scale height, which is modified by a flaring term $R_{\text{flare}}$, this means that $h_Z$ is weakly dependent on $R$ for profiles which flare. We show vertical profiles for age-\feh{} bins with $> 30$ stars in Figure \ref{fig:hzprofile}, adopting the same shading as Figure \ref{fig:surfdens} to draw the eye to the profiles with greater mass contribution, and adding a dashed line representing $0.3$ kpc for reference.

\begin{figure*}
	\includegraphics[width=\textwidth]{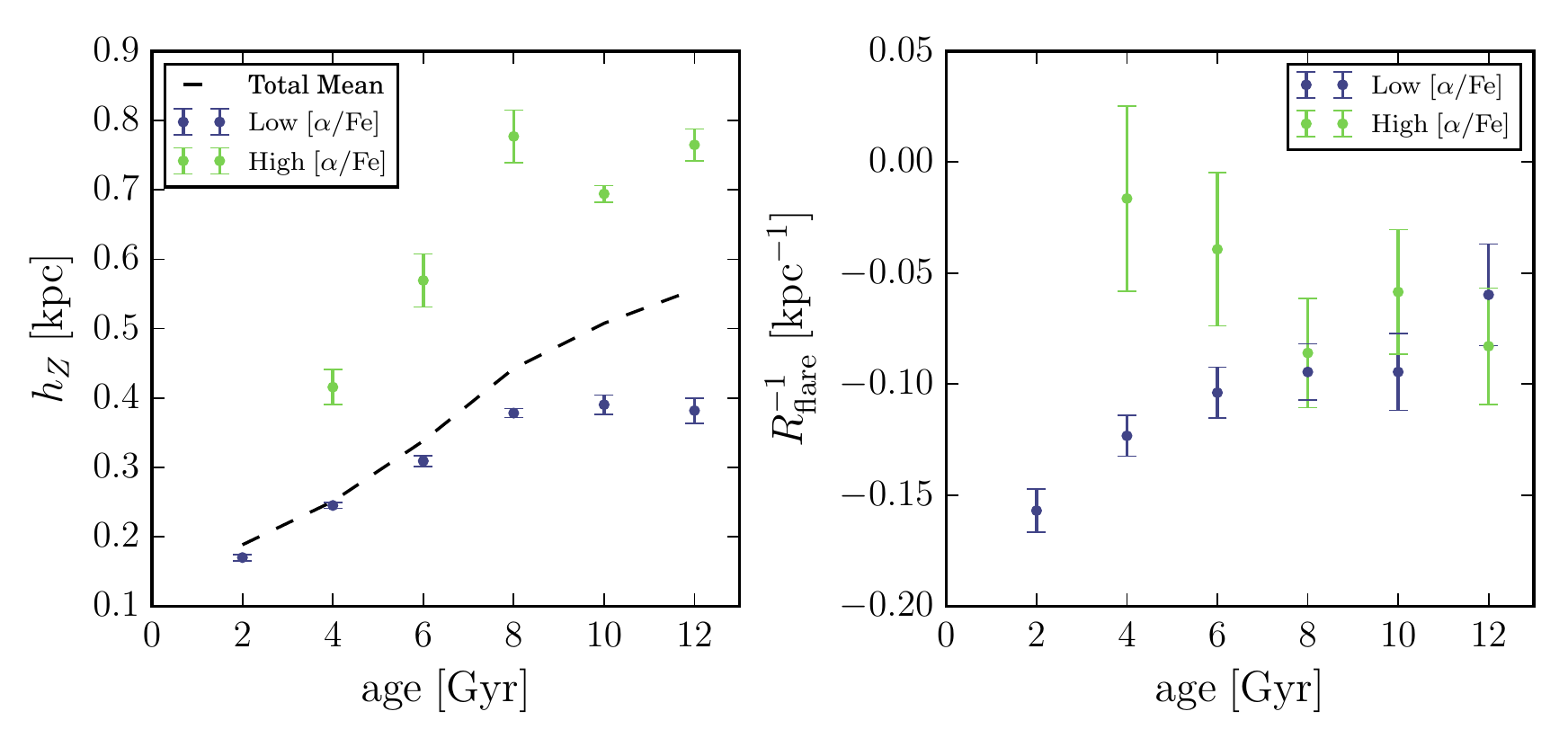}
    \caption{ Mean $h_Z$ at $R_0$ (\emph{left}) and $R_{\mathrm{flare}}^{-1}$ (\emph{right}) against age. The mean value in each age bin is calculated by multiplying together the posterior PDFs of the density fits. The panels show both the low (\emph{purple}) and high (\emph{green}) \afe{} populations. The left panel shows the total surface-mass density weighted mean as a dashed line, which demonstrates that the vertical distribution of the high \afe{} population is only important at the solar radius at old ages due to its low surface-mass density contribution. $h_Z$ increases with age for both low and high \afe{} populations. $R_{\mathrm{flare}}^{-1}$ behaves similarly for the low \afe{} population, meaning flaring decreases with age, but the high \afe{} population shows an opposite behaviour.}
    \label{fig:agevshzrf}
\end{figure*}

Figure \ref{fig:hzprofile} suggests that the disk is thicker as traced by older populations. All \feh{} bins show a thickening as age increases. This is clear in the left panel of Figure \ref{fig:agevshzrf}, which shows the surface-mass density weighted mean variation of $h_Z$ with age. The mean $h_Z$ spans the range between 0.8 and 0.2 kpc. The high \afe{} populations show a bump in the mean $h_Z$ at 8 Gyr, but $h_Z$ generally increases with age, similarly to the low \afe{} populations. The shapes of the profiles of the youngest populations in Figure \ref{fig:hzprofile} in the low \afe{} subsample show little variation with \feh{}, and this trend generally continues to older ages. This is also reflected in the low uncertainties associated with the blue points in the left panel of Figure \ref{fig:agevshzrf}. 

The high \afe{} profiles are generally flat, indicating that these populations show little flaring. By multiplying together the PDFs of the posterior distribution of fits to $R_{\mathrm{flare}}$, we determine that the high \afe{} populations have an average $R_{\mathrm{flare}}^{-1} = -0.06 \pm 0.02$. The low alpha populations flare more strongly, with an average $R_{\mathrm{flare}}^{-1} = -0.12 \pm 0.01$. There is, however, some variation in the flaring as a function of age, so it may not be sensible to ascribe a single $R_{\mathrm{flare}}^{-1}$ to all the populations. We show the variation of $R_{\mathrm{flare}}^{-1}$ with age in the right panel of Figure \ref{fig:agevshzrf}, showing $R_{\mathrm{flare}}^{-1}$ rather than $R_{\mathrm{flare}}$ so that values close to 0 are represented properly. The surface-mass density weighted mean $R_{\mathrm{flare}}^{-1}$ of the low \afe{} populations increases as a function of age, meaning that the most flared populations are the youngest. The behaviour appears opposite for the high \afe{} populations, whose mean $R_{\mathrm{flare}}^{-1}$ seems to decrease with age, but this is determined with low significance as $R_{\mathrm{flare}}^{-1}$ measurements are noisier for these populations.

\subsection{The mass contribution of mono-age, mono-\feh{} populations}
\label{sec:surfmassdens}
We now present the results from the calculation of the surface-mass density at the solar radius using the method described in Section \ref{sec:surfmasscalc}. We compute surface-mass density $\Sigma_{R_0}$ estimates for each age-\feh{} bin, for the high and low \afe{} samples. When quoting the surface-mass densities, we also quote estimates of the systematic uncertainties. We evaluate the sources of these uncertainties in Section \ref{sec:discrepant}.

\begin{figure*}
	\includegraphics[width=\textwidth]{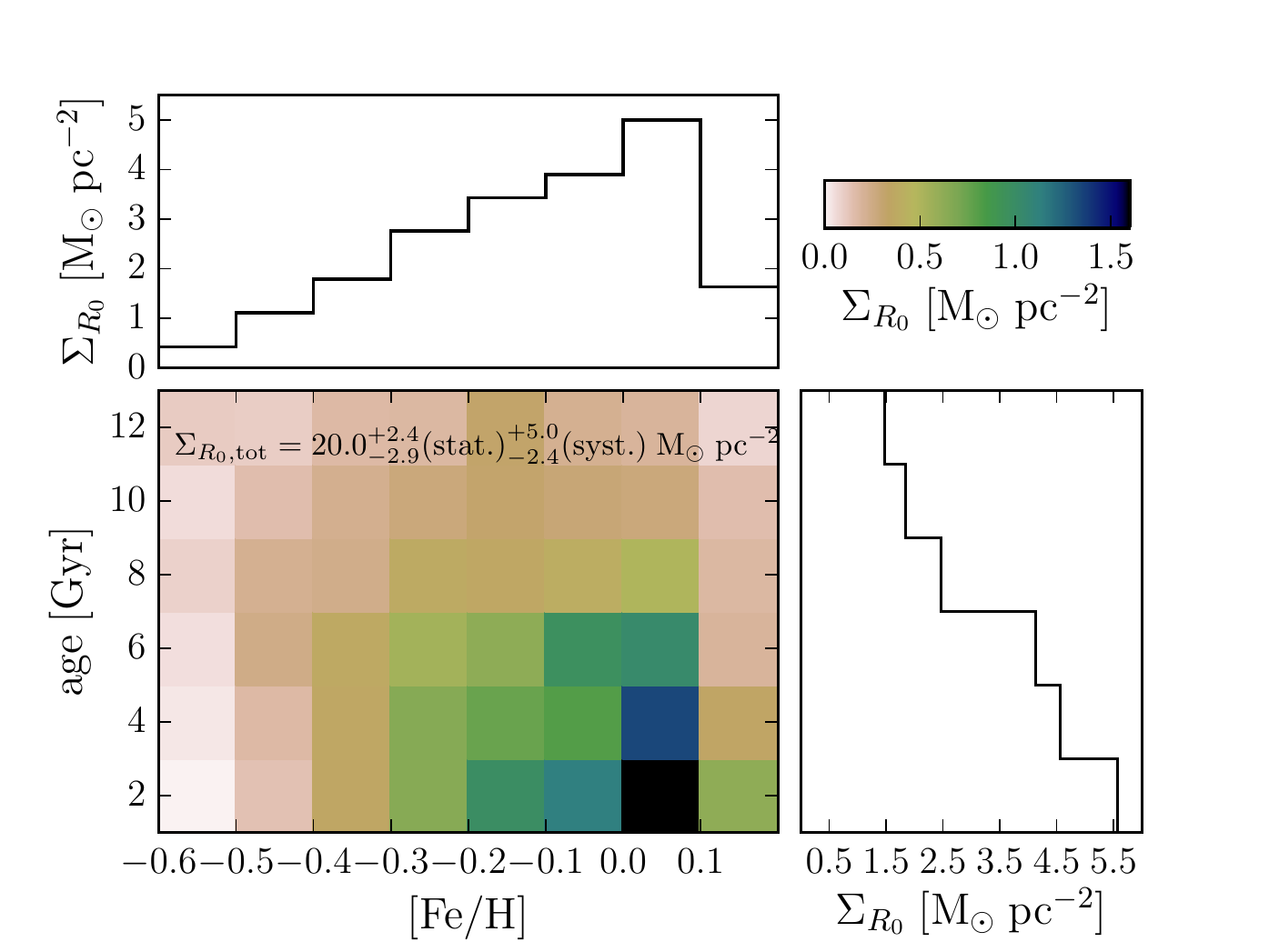}
	\centering
    \caption{The surface-mass density contribution of mono-age, mono-\feh{} populations at $R_0$ (where low and high \afe{} are combined). The total contribution $\Sigma_{R_0,\ \mathrm{tot}}$ is displayed at the top of the main panel. The colour scale is linear and spans the surface-mass density range between $0 < \Sigma_{R_0} < 1.5\ \mathrm{M_{\odot}\ pc^{-2}}$. The marginalised distributions along each axis are shown above and to the right. The mass at the solar radius increases monotonically with both age and \feh{}. }
    \label{fig:margmass}
\end{figure*}

We combine the mass contributions of the high and low \afe{} mono-age, mono-\feh{} populations, and plot the estimates as a function of age and \feh{} in Figure \ref{fig:margmass}. This figure essentially represents the \emph{mass-weighted} age-\feh{} distribution at the solar radius, that is, the probability distribution for age and \feh{} for a randomly selected mass element. The distribution varies smoothly with no sharp peaks, and the surface-mass density increases linearly with both age and \feh{}, peaking at $1 < \mathrm{age} < 3$ Gyr, $0.0 < \mathrm{[Fe/H]} < 0.1$ dex. The mass increases more smoothly with \feh{} than with age, but there is little mass in the highest \feh{} bin, creating a ridge in the marginalised distribution. The marginalised distributions as a function of age and \feh{} show no sign of bimodality, and there is little sign of a bimodality in age at fixed \feh{}. \added[id=TM]{It should be mentioned again here that the age uncertainties may be larger than the bin width, particularly in older bins, which would cause an artificial blurring of a density edge in teh distribution along the age axis. Therefore, we cannot presently determine to high significance that there are no discontinuities in this distribution.}

\begin{figure*}
	\includegraphics[width=\textwidth]{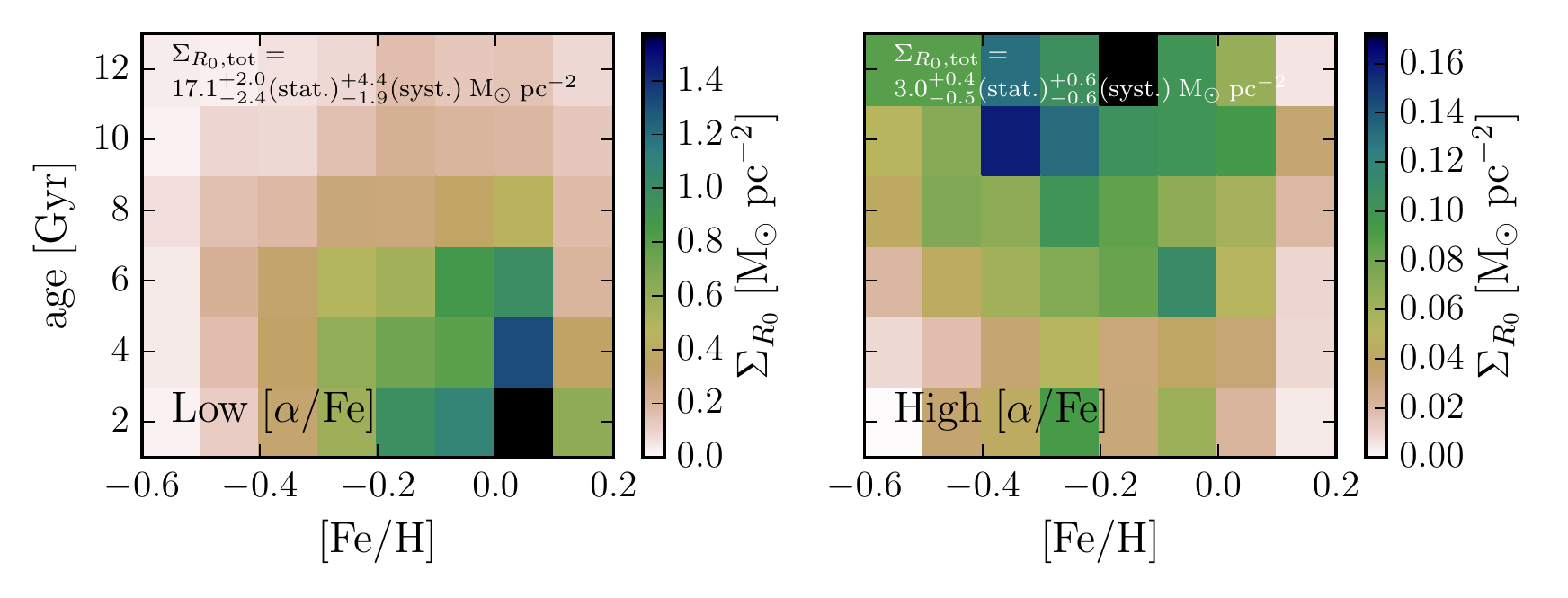}
	\centering
    \caption{The surface-mass density contributions of the low (\emph{left}) and high (\emph{right}) \afe{} sub-samples. The total contributions $\Sigma_{R_0,\ \mathrm{tot}}$ are displayed at the top of each panel. We draw the attention of the reader to the difference in colour scale between the high and low \afe{} panels, which differs by an order of magnitude, and is adopted to better show the behaviour in the high \afe{} sample. The low \afe{} sub-sample has mass at all ages and \feh{} but is concentrated mostly at young ages. The high \afe{} sub-sample contributes far less mass and is concentrated at old age.}
    \label{fig:massafe}
\end{figure*}

An alternative way to look at the surface-mass density distributions is to retain the division in \afe{}. We find that the high \afe{} populations contribute $\Sigma_{R_0,\ \mathrm{tot}} = 3.0_{-0.5}^{+0.4}\mathrm{(stat.)}_{-0.6}^{+0.6}\mathrm{(syst.)}\ \mathrm{M_{\odot}\ pc^{-2}}$ to the total surface-mass density at the solar radius, whereas the low \afe{} populations contribute $\Sigma_{R_0,\ \mathrm{tot}} = 17.1_{-2.4}^{+2.0}\mathrm{(stat.)}_{-1.9}^{+4.4}\mathrm{(syst.)}\ \mathrm{M_{\odot}\ pc^{-2}}$, giving a total surface-mass density in stars at $R_0$ of $\Sigma_{R_0,\ \mathrm{tot}} = 20.0_{-2.9}^{+2.4}\mathrm{(stat.)}_{-2.4}^{+5.0}\mathrm{(syst.)}\ \mathrm{M_{\odot}\ pc^{-2}}$. We plot the individual surface-mass contributions for the separated low and high \afe{} populations in Figure \ref{fig:massafe}, adopting different color scales in each panel, to highlight the behaviour of the high \afe{} populations, which contribute little mass in comparison to the low \afe{}. The low \afe{} mass is mostly concentrated at young age and towards higher \feh{}, although there is mass even at the oldest ages. The high \afe{} mass is concentrated towards older ages, but interestingly the distribution extends to high \feh{}, and we detect mass at some \feh{} in every age bin, but at much lower levels. The tails of the distributions of the low and high \afe{} populations overlap somewhat in age-\feh{} space, around 6 Gyr ago, and there is a hint of a sequence extending from old, low \feh{} and high \afe{} populations, to young, high \feh{} and low \afe{} populations, which is somewhat visible in the combined histogram. There is no clear bimodality in age at fixed \feh{} in the combined histogram, owing to the very low mass contribution of the old, high \afe{} populations. 

\begin{figure}
	\includegraphics[width=\columnwidth]{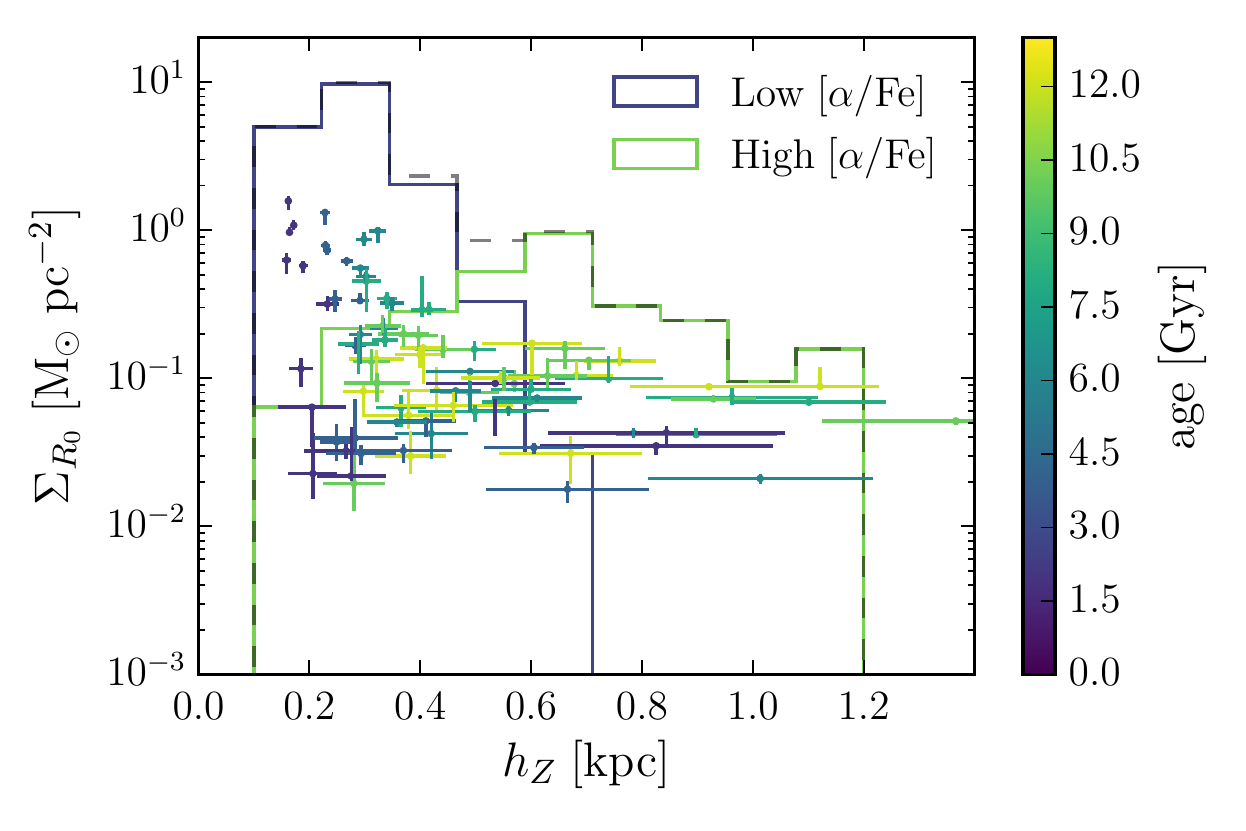}
	\centering
    \caption{The mass weighted vertical scale height $h_Z$ distribution. The individual points represent the $h_Z$ and $\Sigma_{R_0}$ for each mono-age, mono-\feh{} population. We colour the points, which represent both the low and high \afe{} populations, by the central age of the mono-age, mono-\feh{} bin that they represent. The coloured histograms represent the $h_Z$ distributions for the low and high \afe{} populations from the sum of the individual contributions. The dashed histogram represents the total distribution. The total distribution smoothly decreases with $h_Z$, with no hints of bimodality.}
    \label{fig:hzhistogram}
\end{figure}

We have established that the vertical spatial distributions of mono-age, mono-\feh{} populations are well described by single exponentials with characteristic $h_Z$. Next, we use this information to generate the mass-weighted distribution of $h_Z$, which is representative of the probability distribution function for $h_Z$, $p(h_Z)$. For a random stellar mass element, this function gives the probability density for the $h_Z$ of the component to which it belongs. We show this relation in Figure \ref{fig:hzhistogram}, where coloured points represent the individual density contributions of mono-age, mono-\feh{} populations, and the coloured histograms their co-addition within $\sim 0.1$ kpc wide bins in $h_Z$ for the low and high \afe{} populations (purple and green, respectively). The dashed histogram represents the resulting total $p(h_Z)$. Scatter points are coloured by the age of the population they represent. The total distribution is smooth, resulting from the superposition of the low and high \afe{} distributions, which overlap significantly.  The total $\Sigma_{R_0}$ (dashed histogram) declines exponentially with $h_Z$, and is unimodal with no gaps. The trends of both $h_Z$ and $\Sigma_{R_0}$ with age seen in Figures \ref{fig:agevsbroadening} and \ref{fig:margmass} are recovered here, although it is surprising that the trend of $h_Z$ with age at the high $h_Z$ end does not appear as obvious here.

\begin{figure}
	\includegraphics[width=\columnwidth]{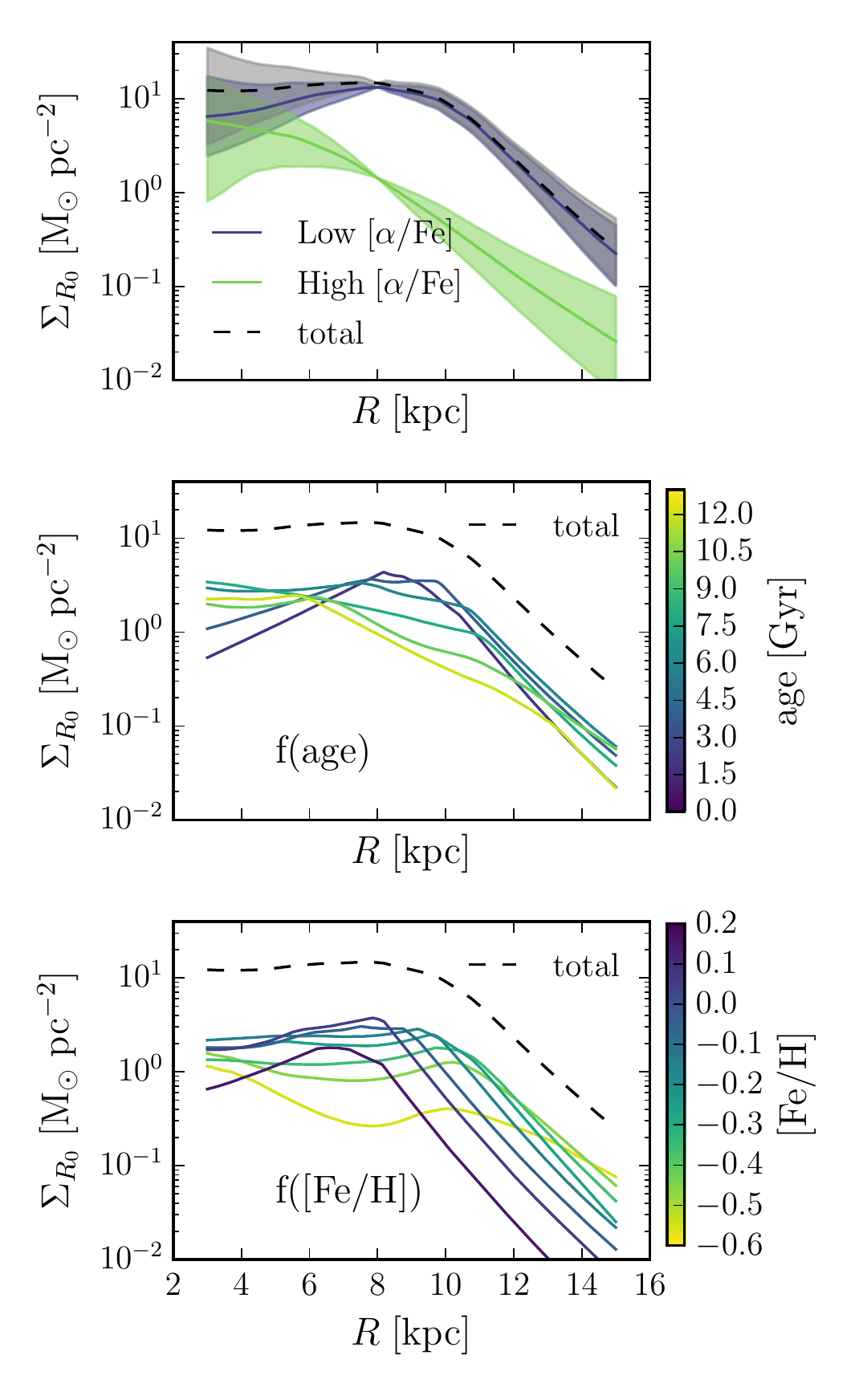}
	\centering
    \caption{The radial surface-mass density profile of the Milky Way, as a function of \afe{} (\emph{top}), age (\emph{middle}) and \feh{} (\emph{bottom}). The profiles are the result of a mass-weighted combination of the fitted density profiles along different axes in age-\feh{} space and (in the top panel) for the combined low and high \afe{} populations. We show the combined uncertainties from the fitting procedure in the top panel, which are sufficient (without addition of the individual statistical and systematic errors on the surface-mass densities, which are substantial) to show that the apparent flattening at $R < R_0$ is not found to high significance. The surface-mass density of low \afe{} stars extends to a higher radius than the high \afe{} stars. The youngest populations show a clearly peaked surface-mass density around the solar radius, whereas the older populations peak more centrally. Behaviour with \feh{} is complex, with flat profiles at low $R$, becoming exponentially decreasing at high $R$.}
    \label{fig:profcombo}
\end{figure}

We can also now mass-weight and combine the fitted density profiles to attain the surface-mass density profile of the Milky Way as a function of age, \feh{} and \afe{}. The resulting profiles are displayed in Figure \ref{fig:profcombo}. The different nature of the low and high \afe{} populations in terms of spatial structure is clear here, with the low \afe{} profile having a clear break between 8 and 10 kpc, and the high \afe{} declining exponentially with $R$. It is interesting to note that extrapolation by eye of the high and low \afe{} profiles to low $R$ would result in the high \afe{} population becoming dominant over the low. The total profile appears roughly flat out to $\sim 10$ kpc. However, we \emph{strongly} emphasize that this is not determined to high significance, as even when only the uncertainties from the fitting procedure are included, one could describe the profile as exponentially declining with $R$ within $R < R_0$. The inclusion of the other sources of uncertainty on the surface-mass density estimates would further decrease the significance of the apparent flattening. For example, the systematic uncertainties (discussed in Section \ref{sec:discrepant}) act to increase the fraction of surface-mass density contributed by the high \afe{} populations, which would only \emph{increase} the slope of the inner exponential. Using dynamcial tracers, \citet{2013ApJ...779..115B} find that the surface density should decline exponentially with R, so it seems logical to assume that the inner profile should not be increasing with R.

As a function of age, the peak in the surface-mass density visible in the youngest population becomes less prominent, and the profile becomes a roughly single exponential at the oldest ages (i.e., it monotonically decreases with $R$).  The behaviour with \feh{} is more complex, but the variation of the peak radius with \feh{} is obvious, and the turnover in the total profile at $\sim 10$ kpc appears to be a result of the outermost breaks.

\section{Discussion}
\label{sec:discussiona}
In the above analysis, we have, for the first time, determined the detailed structure of the Milky Way's disk as a function of stellar age and \feh{}. In our method, we have drawn heavily from previous dissections of the disk into its mono-abundance constituents \citep[MAPs;][]{2012ApJ...753..148B,2016ApJ...823...30B}, and so use these previous findings as a benchmark with which to compare these results. We also show that our results are also broadly consistent with other measurements, whilst shedding new light onto the problem of the formation of the Milky Way disk.

 \subsection{Surface-mass density systematics}
\label{sec:discrepant}
We first adress the sources of systematic uncertainty in our surface-mass density estimates, which are pertinent to the following discussions. Our total local surface-mass density, including the correction for stars missing from the age catalogue, but before accounting for any other systematic uncertainties, is $\Sigma_{R_0, \text{tot}} = 20.0_{-2.9}^{+2.4}\ \mathrm{M_{\odot} \ pc^{-2}}$. This result is roughly two thirds as large as previous estimates, which are of order $\Sigma_{R_0, \text{tot}} \sim 30\ \mathrm{M_{\odot} \ pc^{-2}}$ \citep[e.g.][]{2006MNRAS.372.1149F,2012ApJ...751..131B,2015ApJ...814...13M}. From canonical stellar evolution it is known that giants contribute very little to the total stellar mass in any population. For instance, \citet{2015ApJ...814...13M} find that giants make up $\sim 2\%$ of the local stellar mass. By virtue of this fact, conversions of the stellar mass inferred from giant-star counts to that of the total underlying stellar population require a multiplication of the observed counts by a factor of $\sim 50$, meaning that any uncertainty in the star counts is amplified in the final surface-mass density estimate. Our quoted statistical error estimates, however, which account for Poisson fluctuations in the stellar counts, cannot fully account for the discrepancy. 

We first evaluate whether such a discrepancy may be due to the assumed IMF or stellar evolution model. Tests adopting exponential 
\citet{2003PASP..115..763C} and \citet{2001MNRAS.322..231K} IMFs for the mass calculation resulted in variations of the final $\Sigma_{R_0, \text{tot}}$ estimate of the order $\sim 1\ \mathrm{M_{\odot}\ pc^{-2}}$, which we incorporate into the systematic error budget. We also re-ran our analysis on the basis of the BaSTI stellar evolution models \citep{2004ApJ...612..168P}, for which there also exists calculations for $\alpha$-enhanced stars \citep{2006ApJ...642..797P}, which produced comparable estimates to the PARSEC models (after correcting for the fact that the lowest mass in the BaSTI isochrones is $0.5 M_{\odot}$ as opposed to $0.1 M_{\odot}$ in the PARSEC models). We also compute the mass using only APOGEE fields away from the plane (with $|b|\geq 6^{\circ}$), to test for the effects of extinction on the star counts, but attain results within the Poisson uncertainties of the original estimate.

We also apply our analysis procedure to a basic Monte Carlo mock sample, to check the method for converting observed counts to the real number density $N(R_0)$. We sample stars on a broken exponential density distribution with exponential flare then select points within APOGEE fields out to an imposed distance cut (which allows a simple reconstruction of the selection, and calculation of the effective volume). We calculate $N(R_0)$ analytically, and via our method, and find results which are consistent with the input parameters of the model broken exponential profiles, within the Poisson errors, for a wide variety of input parameters.

\added[id=TM]{As mentioned in Section \ref{sec:ages}, we find that, after making corrections to the ages, that the model returns ages greater than 13 Gyr for a sizeable number of stars \citep[][limit ages to 13 Gyr in their table]{2016MNRAS.456.3655M}. While these stars make up approximately $10\%$ of the final sample (3020 stars), they are not included in the number counts in each mono-age mono-\feh{} bin for calculation of the surface-mass density. Adding an extra $10\%$ of counts to each bin (in the same way as the extra $25\%$ is added in Section \ref{sec:ages}) introduces an extra systematic uncertainty of roughly $1 \mathrm{M_{\odot}\ pc^{-2}}$ in each \afe{} sub-sample. However, readers should take into account that this simple correction does not account for a scenario where the stars with ages fitted $> 13$ Gyr might have a specific distribution in age, casting more counts in some bins (which might have more mass contribution per star) than others. For example, if these stars were all old, then the actual surface mass density in older bins would be higher than that found here, which would increase the total surface-mass density estimate. }

From the above, we conclude that the majority of the systematic discrepancy is likely not due to the assumed IMF, stellar evolution model, dust extinction, or some peculiarity in the age measurements which affects star counts in the bins used. At this stage, it is difficult to understand what is the possible origin of this discrepancy with other works in the literature. Interestingly, our study is the only one employing giant stars as the stellar population tracer, which may point to possible systematics in the theoretical isochrones, or the APOGEE stellar parameters, or a combination thereof. It has recently been demonstrated by \citet{2017A&A...597L...3M} that there may be significant issues with the spectroscopic determination of stellar surface gravity, which is dependent on the star's evolutionary state. We find some discrepancy between the $\log{g}$ of the red clump between the PARSEC isochrones and the data, of the order $\sim 0.2$ to $0.3$ dex \citep[similar to that found by][albeit based on APOGEE-DR13 data]{2017A&A...597L...3M}, which could conceivably lead to problems in our conversion. We test for the effect of systematics in the $\log{g}$ scale, shifting the $\log{g}$ cut of the isochrones to lower and higher $\log{g}$ by $0.3$ dex. We find that shifting the $\log{g}$ cut by $-0.3$ dex increases the surface-mass density estimate by $5.0\ \mathrm{M_{\odot}\ pc^{-2}}$. Increasing the $\log{g}$ cut by $0.3$dex results in a decrease of $ 2.4\ \mathrm{M_{\odot}\ pc^{-2}}$. It therefore seems plausible that the discrepancy results from a systematic difference between the $\log{g}$ scales of the theoretical isochrones and APOGEE, and so we incorporate these shifts into the systematic error estimate. 

Upon inclusion of the systematic uncertainties from IMF variations and differences in the surface gravity scales, we attain a final estimate of the total local surface-mass density in visible stars of $\Sigma_{R_0, \text{tot}} = 20.0_{-2.9}^{+2.4}\mathrm{(stat.)}_{-2.4}^{+5.0}\mathrm{(syst.)}\ \mathrm{M_{\odot} \ pc^{-2}}$, from the addition of the low \afe{} surface-mass density of $\Sigma_{R_0, \text{tot}} = 17.1_{-2.4}^{+2.0}\mathrm{(stat.)}_{-1.9}^{+4.4}\mathrm{(syst.)}\ \mathrm{M_{\odot} \ pc^{-2}}$, and the high \afe{} value of $\Sigma_{R_0, \text{tot}} = 3.0_{-0.5}^{+0.4}\mathrm{(stat.)}_{-0.6}^{+0.6}\mathrm{(syst.)}\ \mathrm{M_{\odot} \ pc^{-2}}$.  If the $\log{g}$ systematics are as large as $-0.3$ dex, then our result is in agreement with the recent estimate from \citet{2015ApJ...814...13M}, of $27\pm 2.7\ \mathrm{M_{\odot} \ pc^{-2}}$.

A recent compilation of measurements of the thick and thin disks found that the thick-thin disk surface density ratio at $R_0$ is $f_{\Sigma} = 15\% \pm 6\%$ \citep{2016ARA&A..54..529B}. Our results find $f_{\Sigma} = 18\% \pm 5\%$ for high-low \afe{} disk surface-mass density ratio, consistent with that estimate. While a better understanding of the possible systematics between the theoretical isochrones and APOGEE is beyond the scope of this paper, we have shown that even a slight difference in the $\log{g}$ scale can bring our results in line with existing estimates. This suggests that our surface-mass density measurement discrepancy is indeed systematic, and that the high and low \afe{} disks may still have some relation to the thick and thin components measured by these studies, which are mainly based on geometric decompositions of the disk. 

\subsection{Comparison with MAPs results}
\label{sec:bovycomparison}
We first discuss our density fits in comparison to the MAP measurements of \citet{2012ApJ...753..148B,2016ApJ...823...30B}. Such a comparison is important because our method is based on an extension of that developed by \citet{2016ApJ...823...30B} to the case of RGB stars, whose distances are far more uncertain than those of RC stars. \citet{2016ApJ...823...30B} used the APOGEE RC sample, to find the structure of populations in narrow bins of \afe{} and \feh{}, or MAPs. These MAPs represent stellar populations with a distribution of ages, but their interpretation assumes a significant relationship between age, \afe{} and \feh{}. We can now compare results when the third parameter, age, is known.  

\citet{2016ApJ...823...30B} showed that the radial distribution of low \afe{} MAPs is well described by a broken exponential, and we confirm this result, showing that each of the low \afe{}, mono-age, mono-metallicity populations is also described by a radially broken exponential. We also show that the older, high \afe{} populations are instead described by a single exponential, which is in good agreement with the findings of \citet{2016ApJ...823...30B}.

The dependence of the radial distribution of mono-\feh{} populations on age is interesting in this regard.  The low \afe{} population, for which our sample covers a wide range of ages with high signal-to-noise, shows a broadening of the profile around a density peak towards older populations, at all \feh{}. This effect does not appear to be present in the high \afe{} population (although some populations have slight evidence of a break at low significance), which suggests that it was formed and evolved differently. We discuss the implications of these findings in Section \ref{sec:implications}.

 We also confirm the results of \citet{2016ApJ...823...30B} which showed that the break radius, $R_{\mathrm{peak}}$ is a declining function of \feh{}. We show that, in the low \afe{} population, at fixed age, $R_{\mathrm{peak}}$ moves to smaller radii as \feh{} increases. As a function of age, the amplitude of this variation increases. The difference in $R_{\mathrm{peak}}$ at the highest and lowest \feh{} in the 6 Gyr bin is $\sim 6$ kpc, which is identical with that of the profiles shown in Figure 11 of \citet{2016ApJ...823...30B}.
 
\citet{2016ApJ...823...30B} found that low \afe{} MAPs were fit well with a $h_Z(R)$ which was slowly exponentially flaring with $R_{\mathrm{flare}}^{-1} = -0.12 \pm 0.01 \ \mathrm{kpc^{-1}}$. We confirm this result, finding also that low \afe{} populations have, on average, $R_{\mathrm{flare}}^{-1} = -0.12 \pm 0.01 \ \mathrm{kpc^{-1}}$, but we also find that $R_{\mathrm{flare}}^{-1}$ shows considerable variation with age (around this mean) in the low \afe{} populations. While \citet{2016ApJ...823...30B} found that high \afe{} populations were consistent with a having $R_{\mathrm{flare}}^{-1} = 0.0 \pm 0.02\ \mathrm{kpc^{-1}}$, we find that these populations have $R_{\mathrm{flare}}^{-1} = -0.06 \pm 0.02\ \mathrm{kpc^{-1}}$, showing some evidence of flaring, albeit at a lower level and lower significance than the low \afe{} populations.

It was also shown by \citet{2012ApJ...753..148B,2016ApJ...823...30B} that the $h_Z$(\afe{},\feh{}) of MAPs smoothly spans the range between 0.2 and 1 kpc. We also confirm and extend this result, showing that $h_Z$(age,\afe{},\feh{}) varies smoothly between a maximum $h_Z$ of $\sim 1.2$ kpc in the high \afe{}, low \feh{}, older populations, down to a minimum of $\sim 0.2$ kpc in the youngest, low \afe{}, \feh{} rich populations. We can also directly compare our Figure \ref{fig:hzhistogram} with Figure 2 of \citet{2012ApJ...751..131B}, which showed that the mass-weighted $h_Z$ distribution is not bimodal but smoothly declines with $h_Z$. We confirm that result, showing that for mono-age, mono-metallicity populations, the mass weighted $h_Z$ distribution shows no sign of bimodality; The low and high \afe{} populations' $h_Z$ distributions are distinct but overlap significantly, generating a smooth distribution. This presents an interesting new look at the interplay between spatial and chemical structure in the disk, as there is a clear mixing spatially of the two chemically separated populations. The implications of this finding are intriguing, and we discuss them further in Section \ref{sec:implications}.

\citet{2012ApJ...751..131B} also made a measurement of the local surface-mass density, finding (in the original application of the method used here) that SEGUE G-type dwarfs yield an estimate of $\Sigma_{R_0,\text{tot}} = 30 \pm 1\ \mathrm{M_{\odot}\ pc^{-2}}$, which is in good agreement with other studies based on different samples \citep[e.g.][]{2006MNRAS.372.1149F,2015ApJ...814...13M}.  Comparatively, our result is somewhat smaller, even when systematics are taken into account. There are a number of differences between this study and that of \citet{2012ApJ...751..131B}. For example, the increased radial coverage, adoption of RGB stars as a tracer, and the fits based on mono-age, mono-metallicity (rather than mono-abundance) populations. However, as mentioned above in Section \ref{sec:discrepant}, our results, after accounting for systematic uncertainties, appear in good agreement with other, more recent estimates \citep{2015ApJ...814...13M}. 

 \subsection{Comparison with other Milky Way disk studies}
We now compare qualitatively the findings of our analysis with the broader body of knowledge regarding the Milky Way disk structure \citep[see, e.g.][for recent reviews]{2013A&ARv..21...61R,2016ARA&A..54..529B} . In comparing our results with those from previous studies, we are constrained to making mostly qualitative considerations, as previous work is based on fits of single exponentials to the radial component of the stellar density distribution.

This work strongly constrains the structure of both the low and high \afe{} components in the Galactic disk, which are commonly considered to be interchangeable with the thin and thick components \citep[as asserted by, e.g.][]{2004A&A...415..155B,2012A&A...545A..32A,1998A&A...338..161F}. We have shown that the \afe{} rich component, while corresponding to a thicker configuration in general, is the product of individual mono-age populations of varying thickness. We find that the \afe{} rich populations span the range  $ 0.4 < h_Z < 1$ kpc, with $h_Z$ increasing with age. Studies of the vertical disk structure which fit a double exponential find a thick disk scale height of $\sim 1$ kpc  \citep[e.g.][]{1983MNRAS.202.1025G,2008ApJ...673..864J}, which is fully consistent with measurements of the thickest high \afe{}, old, mono-age populations in our analysis. \added[id=TM]{However, we again stress here that the age uncertainties at old ages may be significantly larger than the bin size, which may cause a blurring of these trends, and should be accounted for when comparing these results to models. As an example, in the \emph{worst case} scenario, assuming gaussian errors, the oldest bins (between 7 and 13 Gyr) may be contaminated by up to $50\%$ of the stars which should be assigned to neighbouring bins, at the oldest end, with the fraction dropping off quickly at younger ages.  We briefly discuss the implications of the worst case blurring on our interpretation of these trends in Appendix \ref{sec:ageerror}}

Regarding the radial scale length of the thick component, we find an obvious discrepancy with literature values, whereby our thick, high \afe{}, populations have an average $h_{R, \text{out}} = 1.9 \pm 0.1  $ kpc, while the aforementioned studies, who define the thick disk geometrically, find values of the order $\sim 4$ kpc \citep{2001MNRAS.322..426O,2008ApJ...673..864J}. This discrepancy appears to arise in the choice of definition of the measured population between a geometric or chemical abundance selection, with many studies finding a scale length for the abundance-selected $\alpha$-rich disk in the range $h_R = 2.0 \pm 0.2$ kpc \citep{2016ARA&A..54..529B,2016ApJ...823...30B,2012ApJ...753..148B,2012ApJ...752...51C}. It should be noted here that our $h_{R, \text{out}}$ would likely be in even better agreement with this value, had we accounted for the 5\% systematic discrepancy in the distances (shown in Figure \ref{fig:distcomp}). \citet{2016arXiv160901168M} recently showed evidence for a radial age gradient in the Milky Way, suggesting this as a source of disagreement between abundance-selected and geometric studies of the thick disk components, where the geometric selected studies see an extended thick disk which is made up of flared low \afe{} populations. 

We also examine claims of a sharp decline in the stellar density at $R\sim 13.5$ kpc \citep[e.g.][]{2009A&A...495..819R,2010MNRAS.402..713S} in light of our results. \citet{2010MNRAS.402..713S} fit a single exponential density profile with scale-length $\sim 3$ kpc to A stars (which preferentially selects stars younger than $\sim 100$ Myr old) and found that after $R\sim13$ kpc, a model with shorter scale-length was necessary to explain the increased rate of decline in stellar density. Our total profile in Figure \ref{fig:profcombo} begins to decline after $R\sim10$ kpc. The uncertainties in the measurement of the mass contribution of each profile may cause some discrepancy here, as implying a higher mass on older or more metal poor populations would shift this turn-over to higher radii. We also fit older populations than \citet{2010MNRAS.402..713S}, which, under the inside-out formation paradigm, might suggest another reason for such a discrepancy, as older populations would be more centrally concentrated. We confirm the assertion of \citet{2016ApJ...823...30B} that this break, clearly visible in the total stellar distribution, is attributable to the outermost break of the mono-\feh{} profiles - which are shown in the bottom panel of Figure \ref{fig:profcombo}. External disk galaxies are also observed to have such a truncation in their stellar density profiles \citep[e.g.][]{2006A&A...454..759P}.

\subsection{Implications for the formation of the Galactic disk}
\label{sec:implications}

In light of the above discussion, we now present the implications of our results for the formation of the Galactic disk. In this paper, we present a detailed dissection of the disk by age, \feh{},and \afe{}, and as such, present a previously unseen picture of the dominant structure of the Milky Way. In studying mono-age populations, we can perform a more direct comparison than previously possible with numerical simulations of Milky Way type galaxies, which tend to use age information in the absence of detailed chemical modelling.

\subsubsection{Disk flaring, profile broadening and radial migration}

By estimating the density profiles of mono-age populations, we place novel constraints on radial migration and its effects on the structure and evolution of the disk. We have two key observables which provide this insight: the flaring of the disk, which has been considered as an effect of vertical action conserving radial migration \citep[where stars have greater vertical excursions as they migrate outward, e.g.][]{2012A&A...548A.127M}, and the broadening of the density profiles around the peak with time, which we discuss as a potential new indicator of radial migration. The right panel of Figure \ref{fig:agevshzrf} shows a clear trend of increasing $R_{\mathrm{flare}}^{-1}$ with age such that the youngest populations flare most. This behaviour is distinct from the results of \citet{2016ApJ...823...30B}, which found that low \afe{} populations were described by a single $R_{\mathrm{flare}}^{-1}$. It is, however, conceivable that if low \afe{} populations of all ages are combined, the resulting population may have a similar behaviour as that in \citet{2016ApJ...823...30B}. Indeed, we find  an average $R_{\mathrm{flare}}^{-1} = -0.12\pm 0.01\ \mathrm{kpc^{-1}}$, which is in good agreement with the value from \citet{2016ApJ...823...30B}. If populations become more flared as radial migration proceeds, then it becomes difficult to reconcile our result with that of a disk whose stars continually underwent radial migration, largely unperturbed by any mergers that might cause structural discontinuity \citep[e.g.][]{2014MNRAS.442.2474M}, especially under suggestions that mergers actually reduce flaring from radial migration \citep[e.g.][]{2014A&A...572A..92M}. In this context, this result is indicative of an old population in the Milky Way which has undergone some mergers, reducing the flaring in the oldest populations. It should be noted here, however, that the age uncertainties (which can be as large as $40\%$) could effectively artificially increase the age bin size, super-imposing populations with different scale heights and flare, and reducing the overall flaring profile.

We showed in Figures \ref{fig:agevsbroadening} and \ref{fig:surfdens}  that the radial surface density profiles of low \afe{} populations become smoother with age. Interestingly, the position of the break radius does not vary monotonically with age. Assuming that the peak radius is at the equilibrium point of chemical evolution for a given population  \citep[where the consumption of gas and its dilution are balanced, as discussed in][]{2016ApJ...823...30B}, then one might consider that such a broadening would occur if stars that formed near the equilibrium point migrated inwards and outwards over time.  If these assumptions are correct, then the specific surface density profile shapes might provide insights into how radial migration has proceeded in the disk. For example, if the slope of the inner or outer profile change slope differently, this might suggest that migration has been asymmetric (i.e. more mass has moved in than out or vice versa). Comparing the $-0.2 <\mathrm{[Fe/H]}<-0.1$ dex and $0.0 <\mathrm{[Fe/H]}<0.1$ dex bins in Figure \ref{fig:surfdens}, it seems that the \emph{inner} slope of the former profile decreases more with age, whereas the \emph{outer} slope of the latter decreases more strongly, suggesting that the former population has preferentially migrated in, whereas the latter migrated out.  it is important to point out that under the interpretation that the increasing profile width is due to migration efficiency, we make an assumption that stars of a given \feh{} and \afe{} must have been born with the same width throughout cosmic time \citep[as discussed by, e.g.][]{2017ApJ...834...27M}.

\begin{figure}
	\includegraphics[width=\columnwidth]{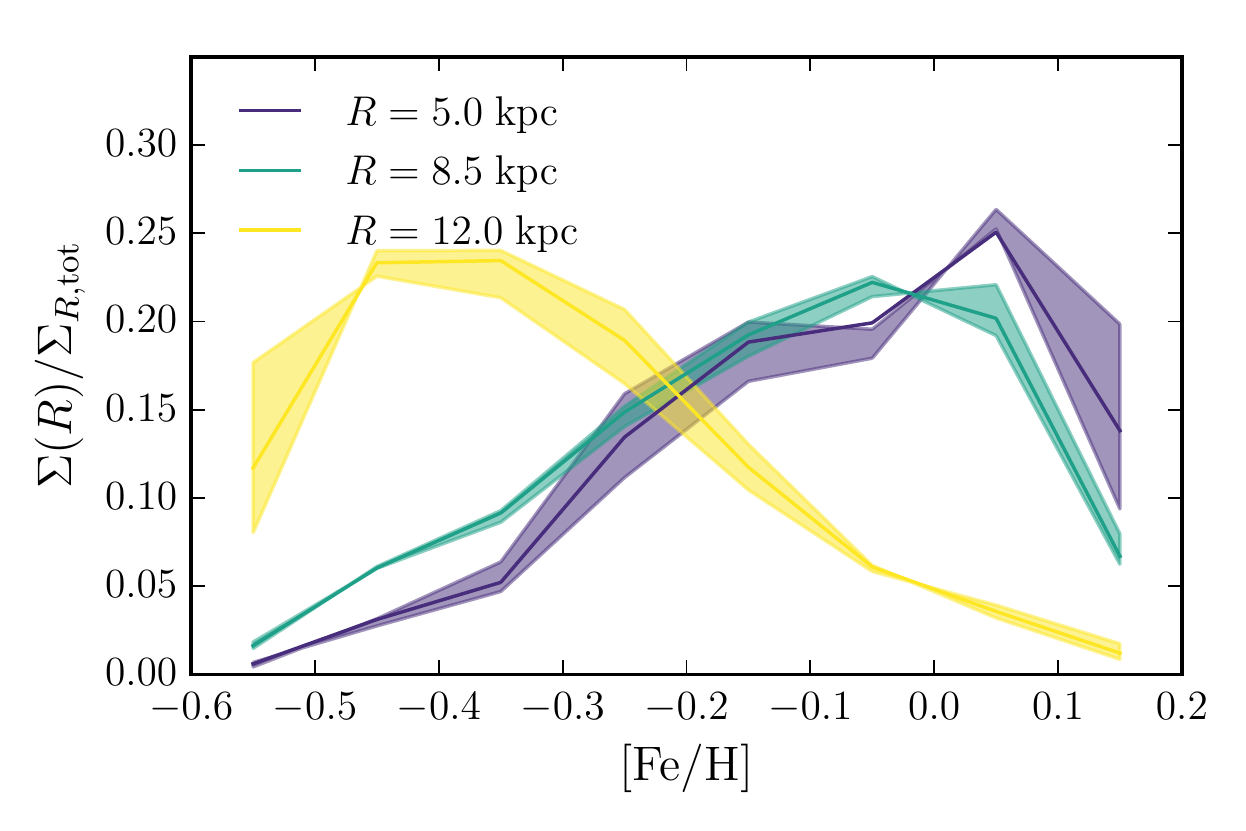}
	\centering
   \caption{The surface-mass density weighted \feh{} distribution (MDF) at 3 radii for profiles fit to the low \afe{} populations. The distribution shown is marginalised in age across all our age bins. The coloured bands give the 95\% uncertainty ranges, where uncertainties are dominated by those on the fitted density profiles. The mean \feh{} is lower at greater $R$. Qualitatively, the skew of the MDF's changes with $R$, such that the innermost $R$ has a tail going to low \feh{}, and the outermost $R$ has a tail going to high \feh{}}
    \label{fig:mdf}
\end{figure}

This picture is also consistent with the suggestion by, e.g, \citet[][]{2015ApJ...808..132H,2011ApJ...737....8L}, that the changing skew in the MDF as a function of Galactocentric radius is caused by such a mechanism. We show the mass-weighted \feh{} distribution for the low \afe{} fits at different Galactocentric radii in Figure \ref{fig:mdf}, showing that our results both find a radial metallicity gradient, and qualitatively reproduce the skew found by \citet[][]{2015ApJ...808..132H}. Unlike \citet[][]{2015ApJ...808..132H}, our analysis fully corrects for sample-selection and stellar-population biases in reconstructing the MDF. \citet{2016MNRAS.460L..94G} also found similar behaviour in a simulated galaxy, finding that spiral structure induces different migration patterns, dependent on birth radius. 

Interestingly, we detect the possible flaring of old, high \afe{} populations, with average $R_{\mathrm{flare}}^{-1} = -0.06 \pm 0.02\ \mathrm{kpc^{-1}}$. However, the right panel of Figure \ref{fig:agevshzrf} shows that the flaring of the high \afe{} populations does not vary as strongly with age as the low \afe{} populations (although it should be noted that most of the mass in the high \afe{} populations is concentrated at older age anyway). The detection of even a slight flare in these populations is surprising, as \citet{2016ApJ...823...30B} found that the high \afe{} MAPs did not have flare. Again, this may be an effect of the superposition of multiple mono-age populations within the MAPs. \citet{2015ApJ...804L...9M}, for example, found that co-eval populations in simulated galaxies always flare, and suggested that the superposition of such flares might be an explanation for thickened disk components. \citet{2017ApJ...834...27M} showed that the superposition of mono-age populations within MAPs can introduce decreased flaring in high \afe{} populations, whilst the mono-age populations themselves still flare. Comparison of these results with \citet{2016ApJ...823...30B} seems to present a consistent scenario. It should be noted here however, that \citet{2013MNRAS.436..625S} found that MAPs in their simulation were coeval in general.

Our results show that the oldest populations are thicker, centrally concentrated, and display the least flaring, whilst the youngest populations, which show the most flaring, have the thinnest vertical distribution (smallest $h_Z$). Between these extremes, consecutive populations in age form a continuum, when the combined low and high \afe{} structure is considered (see Figure \ref{fig:agevshzrf}). It is clearly conceivable then, that the \emph{geometrically} defined thick disk, found to have large scale-length \citep[e.g.,][]{2013MNRAS.431..930J,2008ApJ...673..864J}, may be the superposition of these flared (young) and naturally thick (old) components. An obvious consequence of this scenario would be an age-gradient at high $Z$ above the disk plane, which has been recently shown to be present in the APOGEE data by \citet{2016arXiv160901168M}. It was also recently shown that in the Gaia-ESO survey data, the mean structural characteristics of the abundance selected thick and thin disks appear to overlap at $\mathrm{[M/H]}\sim-0.25$ dex and \afe{} $\sim 0.1$ dex \citep{2014A&A...567A...5R}, further presenting a scenario where the thick and thin disk components are not necessarily separable from one another, or at least not in abundance space. 

\subsubsection{Inside out formation, and the overall vertical disk structure}
\begin{figure}
	\includegraphics[width=\columnwidth]{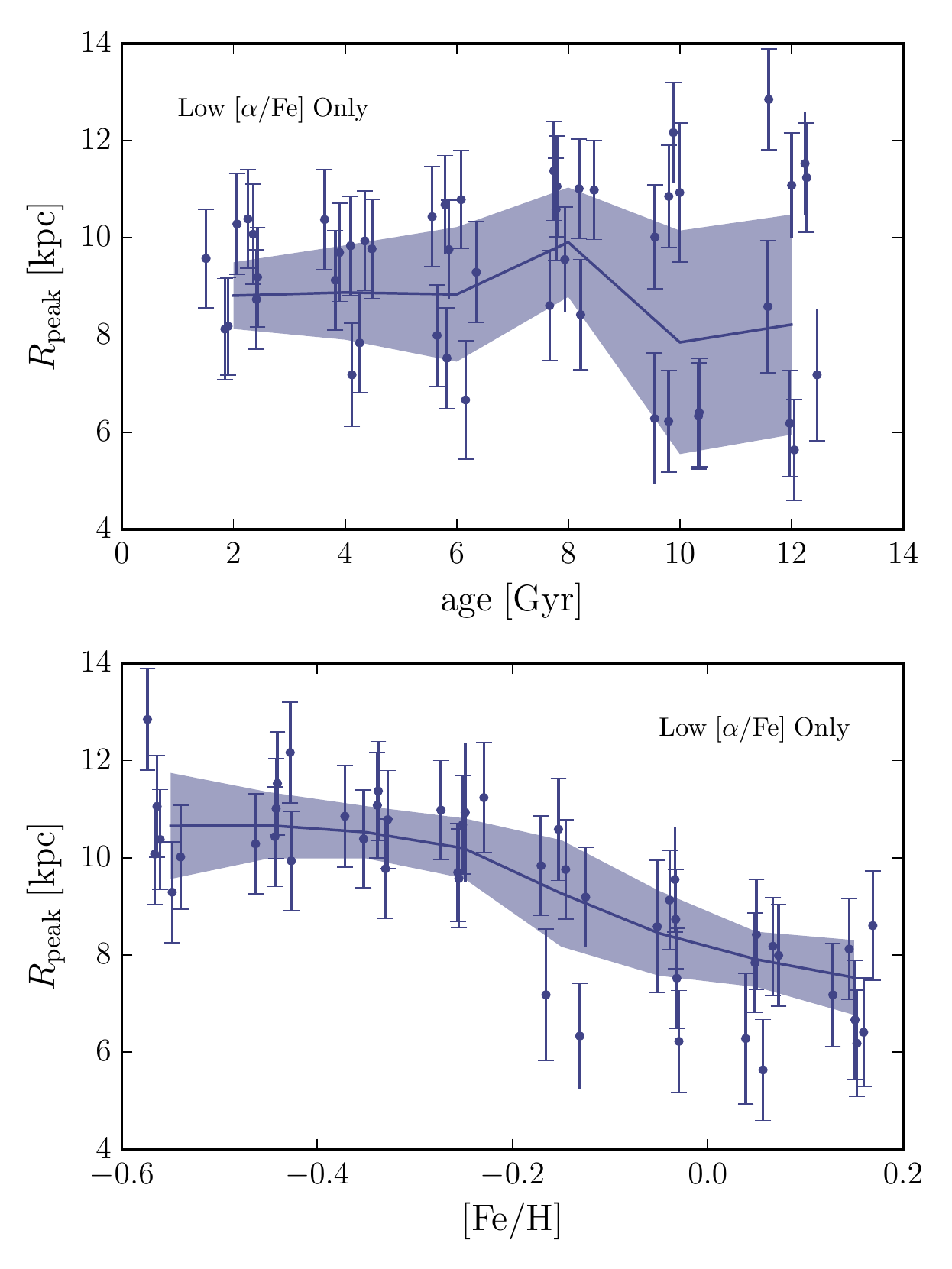}
	\centering
   \caption{The behaviour of $R_{\mathrm{peak}}$ with age and \feh{} for the low \afe{} populations. The high \afe{} populations are better fit by single exponentials, and so $R_{\mathrm{peak}}$ is not an informative diagnostic of these populations. The coloured lines and bands give the surface-mass density weighted mean and standard deviation within an age or \feh{} bin. The mean $R_{\mathrm{peak}}$ does not vary significantly with age, whereas it shows a clear decrease with \feh{}. However, the dispersion in $R_{\mathrm{peak}}$ does increase with age for low \afe{} populations. High \afe{} populations show a slight increasing trend for increasing age and \feh{}, albeit at low significance. }
    \label{fig:rpeakvsage}
\end{figure}

The formation of the Galactic disk is commonly framed in the paradigm of inside-out formation \citep[e.g.,][]{2013ApJ...773...43B,2011ApJ...729...16K,1976MNRAS.176...31L,1989MNRAS.239..885M}. More recently, the effects of radial migration  \citep[e.g.][]{2002MNRAS.336..785S} were added, in order to produce models that agree better with the observations \citep[e.g.][]{2011ApJ...737....8L,2009MNRAS.396..203S,2015ApJ...802..129S,2015A&A...580A.126K} . Our measurements of the peak radius of mono-age populations place strong empirical constraints on the evolution of the Milky Way disk over time. The behaviour of $R_{\mathrm{peak}}$ with age and \feh{} is shown in Figure \ref{fig:rpeakvsage}. We find that the surface-mass density weighted mean $R_{\mathrm{peak}}$ of low \afe{} populations remains roughly constant with age, whilst the dispersion about the mean increases with age. This finding is qualitatively consistent with that of e.g. \citet{2016arXiv160804951A}, who show that the radial metallicity gradient decreases with the age of the population considered. Our results show that the density peaks of mono-\feh{} populations become more separated with age. As the mean \feh{} at a given $R$ is dictated by the dominant population in stellar density at that radius, this indicates that we also see a shallowing gradient in \feh{} with age. This is reinforced by our finding (also shown in Figure \ref{fig:rpeakvsage}) that the mean $R_{\mathrm{peak}}$ decreases with \feh{}.

Our results also place strong constraints on models for the formation of the vertical disk structure. We have already discussed that the vertical structure of the disk is commonly framed as having two geometrically distinct (but overlapping) vertical components \citep[e.g.][]{1983MNRAS.202.1025G}. Our results confirm previous work \citep[e.g.][]{2012ApJ...751..131B} that shows that this picture, while providing an acceptable description of the data when analysing the whole population, is not complete when individual populations, either abundance- or age-selected, are considered. We have shown (in Figure \ref{fig:hzhistogram}) that for a random mass element, the probability that it belongs to a population of a given $h_Z$ exponentially declines as $h_Z$ increases. There are no apparent breaks in this relation at the resolution which we measure it, strongly suggesting that the spatial vertical disk structure is continuous. This finding, in stark contrast to the distinct discontinuities seen in the chemical structure of the entire disk \citep[e.g][]{2014ApJ...796...38N,2015ApJ...808..132H}, presents an interesting conundrum for galaxy formation theory. How is it possible that the \emph{spatial} structure of the disk be smooth and continuous, whilst the \emph{chemical} structure portrays a clear discontinuity?

 Theoretical studies have, thus far, presented some clues as to how galaxy disks such as that of the Milky Way might form. As most studies fit single exponential radial profiles to simulated disks, quantitative comparisons are difficult, yet qualitative considerations can be made. \citet{2013MNRAS.436..625S} used a maximum likelihood method similar to \citet{2012ApJ...753..148B} to fit density profiles to MAPs in a simulated galaxy and found a continuous distribution of scale heights, but also found that their simulation showed a strongly geometrically distinct thick disk component. \citet{2013ApJ...773...43B} made detailed measurements of the mono-age populations in a high-resolution hydrodynamic Milky Way-like galaxy simulation, and found, similarly to our results, that their scale heights gradually decreased with time, while the scale lengths increased, with populations forming thick and retaining that thickness in an 'upside-down and inside-out' disk formation. Our results also find evidence of flaring in the thick components (as discussed in Section \ref{sec:bovycomparison}), which may point towards some structural evolution (via a process such as radial migration) after their formation. However, work on simulations by \citet{2009ApJ...707L...1B}  suggests early, turbulent gas as the origin for thicker disk components. A flare in the gas disk of the Milky Way, associated with the stellar component, is also observed in numerous studies \citep[e.g.][]{2014Natur.509..342F,2014ApJ...794...90K,1963SvA.....6..658L}, suggesting a formation of the disk with structural parameters similar to its progenitor gas disk.
 
\subsubsection{The age-\feh{} distribution and the evolution of the disk}
We now discuss how the present day structural parameters, in combination with the emergent picture of the mass distribution in age-\feh{} space at the solar radius, might offer deeper insights into the formation and evolution of the disk. 

We find, as may be expected, that the high \afe{} populations contribute the majority of their mass at the solar radius at ages older than $\sim 6$ Gyr, although the mass contribution by old stars is extremely low compared to the younger populations. The middle panel of Figure \ref{fig:profcombo} shows that, if the populations follow the density models that we fit, then the older stars become more dominant closer to the Galactic centre, which is suggestive of a weak mean radial age gradient, in qualitative agreement with theoretical predictions \citep[e.g.][]{2015ApJ...804L...9M} and observations of the thick disk \citep[e.g.][]{2016arXiv160901168M}. It is therefore not surprising that the bottom panel of Figure \ref{fig:profcombo} shows a clear variation in mean metallicity, in agreement with findings in other works \citep[e.g.][]{2012ApJ...746..149C,2016arXiv160804951A,2015ApJ...808..132H}. Only with high resolution  hydrodynamical simulations, which accurately reproduce the stellar populations in galaxies, will it be possible to reconstruct the right combination of star formation history and radial mixing that led to these age and metallicity gradients, to gain a better understanding of the details of their formation.

In Figure \ref{fig:massafe}, an overlap in age-\feh{} space is visible between the high and low \afe{} populations. While there appears to be mass at many \feh{} bins in the old populations, the overlap in age occurs at intermediate \feh{} at the solar radius. Previous studies have found that the youngest stars in the high \afe{} sequence overlap in age with the oldest and most \feh{} poor stars in the low \afe{} population \citep[e.g.][]{2013A&A...560A.109H}, and our findings appear to be consistent with that result. It should however, be noted, that at least some of this overlap is likely caused by the age uncertainties, which can be as high as 40\%. If the low \afe{} population emerged from the remnants of the high \afe{}, then it is likely that some sort of infall event must have occurred to return the ISM to low \feh{} and low \afe{} before forming those stars \citep[as expressed by, e.g.,][]{1997ApJ...477..765C}. These scenarios are also discussed in the context of the APOGEE results by \citet{2014ApJ...796...38N}. To fully understand this, however, we will likely require a chemodynamical model which reproduces the bimodality in \afe{} at fixed \feh{}.

\section{Conclusions}
\label{sec:conclusionsa}

We have performed the first detailed dissection of the stellar populations of the Milky Way disk in age, \feh{} and \afe{} space, bridging the gap between the detailed observational understanding of mono-abundance populations \citep[e.g.][]{2012ApJ...753..148B,2016ApJ...823...30B} and the plethora of studies of co-eval stellar populations in simulated galaxies \citep[e.g][]{2013ApJ...773...43B,2014MNRAS.442.2474M,2013MNRAS.436..625S}. We have placed novel constraints on models for the formation of the Milky Way disk by combining detailed density models fit to the mono-age, mono-\feh{} populations of the low and high \afe{} disk, with surface mass density contributions calculated on the basis of these density fits and stellar evolution models. We summarise our key results as follows:
\begin{itemize}
\item \textbf{Radial and vertical profiles} The mono-age, mono-\feh{} populations of the \afe{} poor disk are well fit by a radially broken exponential, with a peak radius, $R_{\mathrm{peak}}$, that varies as a function of age and \feh{}. We find that the distance between $R_{\mathrm{peak}}$'s of the low and high \feh{} populations increases with age, which we interpret as evidence for a decreasing \feh{} gradient with time \citep[e.g][]{2016arXiv160804951A}. The radial variation of the stellar surface density of the high \afe{} mono-age populations is found to have insignificant breaks, and they are better fit by a single exponential in this disk region. As these populations are the oldest, this may be a sign of the disk evolution washing out the density peak over time, or may point to a different formation scenario for high \afe{} stars, where no density peak ever existed. These findings are in good agreement with earlier studies of MAPs \citep{2016ApJ...823...30B}. We measure an average high \afe{} population scale length of $h_{R,\text{in}} = 1.9 \pm 0.1$ kpc, and find scale heights between 600 and 1000 pc, in good agreement with current measures of the \afe{} rich disk scale length and height \citep[e.g. those outlined in][]{2016ARA&A..54..529B}. 
\item \textbf{Profile Broadening} We show that the radial surface density profile of the low \afe{} populations broadens with age in a given \feh{} bin, which we interpret as evidence of the gradual dispersal of mono-\feh{} populations, presumably due to radial migration and radial heating. The variation in shape of the broken exponential profile changes differently depending on the population \feh{}, with \emph{low} \feh{} populations \emph{inner} profiles flattening faster, whereas the \emph{high} \feh{} \emph{outer} profiles flatten faster. We interpret this effect as tentative evidence for \feh{} dependent radial migration arising from pre-existing \feh{} gradients in the star forming disk. We showed that our results qualitatively reproduce those of \citet{2015ApJ...808..132H}, finding a skewed MDF that varies as a function of R.
\item \textbf{Flaring} We find that flaring seems to be present in almost all mono-age populations, at differing levels. We have shown that the inverse flaring scale length $R_{\mathrm{flare}}^{-1}$ increases with age, meaning that the youngest populations flare most strongly. This finding appears inconsistent with that above, under the assumption that flaring is the result of radial migration. However, these results may be reconciled by invoking a more active accretion history in the early life of the disk, which could have suppressed flaring \citep[e.g.][]{2014ApJ...781L..20M}.

\item \textbf{The surface-mass density at $R_0$} We have measured the surface mass-density at the solar radius for each mono-age, mono-\feh{} population, finding a total surface mass density of  $\Sigma_{R_0, \text{tot}} = 20.0_{-2.9}^{+2.4}\mathrm{(stat.)}_{-2.4}^{+5.0}\mathrm{(syst.)}\ \mathrm{M_{\odot} \ pc^{-2}}$. Before allowing for systematics, this value is less than current estimates \citep[e.g.][]{2012ApJ...751..131B,2006MNRAS.372.1149F,2015ApJ...814...13M}, however, the systematic uncertainties are large, mainly due to a mismatch between the $\log{g}$ scales in APOGEE and the PARSEC models, and as such, we find our value to be consistent within the uncertainties. The relative contribution of high to low \afe{} populations, $f_\Sigma$, is $18\% \pm 5\%$, which is consistent with existing measurements \citep[e.g.][]{2016ARA&A..54..529B}. 
\item \textbf{The $h_Z$ distribution at $R_0$} The shape of the mass-weighted $h_Z$ distribution found by this study is in good agreement with that of \citet{2012ApJ...751..131B}, calling into question the existence of a vertical structural discontinuity in the Milky Way disk. The reconciliation of this finding with the discontinuity in chemical space \citep[e.g. the bimodality in \afe{} at fixed \feh{}:][]{2015ApJ...808..132H,2014ApJ...796...38N} may shed new light on our understanding of the formation of the Galactic disk.

\item \textbf{The surface-mass density profile of the Milky Way} We have found the combined (from mono-age, mono-\feh{} populations at low and high \afe{}) surface-mass density weighted profiles of the Milky Way disk as a function of \afe{}, age and \feh{}, and found that the total surface density is also described by a broken exponential. We find that our results fail to determine the sign of the inner exponential to high significance out to $\sim 10$ kpc, but detect a turnover to a declining exponential, at high significance, thereafter. We find evidence of a radial mean age and \feh{} gradient driven by the changing dominant population as a function of radius. A detailed comparison of these findings with numerical simulations is necessary for a proper interpretation. Our finding of a decline in stellar density may be consistent with that found in other studies \citep[e.g.][]{2009A&A...495..819R,2010MNRAS.402..713S}, albeit at shorter radii.
\end{itemize}

These findings are strongly constraining to future theoretical work. With the recent \citep{2016arXiv160904303L} and future releases of \emph{Gaia} data, and the ongoing APOGEE-2 survey \citep{2014AAS...22344006S}, which will include an updated APOKASC sample \citep{2014ApJS..215...19P}, access to improved positions, abundances and age estimates is within reach. \added[id=TM]{We again stress here that the age uncertainties in this data set can be as large as $40\%$, and so until more precise ages are attained for similarly sized samples, our conclusions must be considered under the caveat that the mono-age populations at old age are likely mixed to some extent. It will be possible to investigate this issue better once better ages for a larger sample are released by APOGEE and APOKASC \citep{2014ApJS..215...19P}.}

Future studies of simulations which accurately track chemical evolution, gas and stellar dynamics, and the feedback processes which are dominant in galaxies will no doubt lead to a deeper insight into the physical processes leading to the present day structure of the Milky Way. The understanding of discontinuity in chemical space, namely the bimodality in \afe{} at fixed \feh{}, and how this can be reconciled with the apparent structural continuity which we find here poses an interesting challenge to models of the formation of the Milky Way disk. By performing a first mapping of the 3D distribution of stellar populations as a function of age, metallicity, and [a/Fe], we hope that this work provides the kind of data needed for a comparison with numerical simulations that is unencumbered by the complexities associated with corrections for the survey selection function.

\section*{Acknowledgements}
It is a pleasure to thank Marie Martig for helpful comments and discussion, and also for providing the catalogue of ages for APOGEE DR12 in digital format. We also thank Ivan Minchev, Misha Haywood and Paola Di Matteo for insightful discussion and comments which improved the clarity of the manuscript. The analysis and plots in the paper used \texttt{iPython}, and packages in the \texttt{SciPy} ecosystem \citep{Jones:2001aa,4160265,4160251,5725236}. JB received support from the Natural Sciences and Engineering Research Council of Canada, an Alfred P. Sloan Fellowship, and from the Simons Foundation. SM has been supported by the Premium Postdoctoral
Research Program of the Hungarian Academy of Sciences, and by the Hungarian
NKFI Grants K-119517 of the Hungarian National Research, Development and Innovation Office.

Funding for SDSS-III was provided by the Alfred P. Sloan Foundation, the Participating Institutions, the National Science Foundation, and the U.S. Department of Energy Office of Science. The SDSS-III web site is http://www.sdss3.org/.




\bibliographystyle{mnras}
\bibliography{bib} 

\begin{thebibliography}{}
\makeatletter
\relax
\def\mn@urlcharsother{\let\do\@makeother \do\$\do\&\do\#\do\^\do\_\do\%\do\~}
\def\mn@doi{\begingroup\mn@urlcharsother \@ifnextchar [ {\mn@doi@}
  {\mn@doi@[]}}
\def\mn@doi@[#1]#2{\def\@tempa{#1}\ifx\@tempa\@empty \href
  {http://dx.doi.org/#2} {doi:#2}\else \href {http://dx.doi.org/#2} {#1}\fi
  \endgroup}
\def\mn@eprint#1#2{\mn@eprint@#1:#2::\@nil}
\def\mn@eprint@arXiv#1{\href {http://arxiv.org/abs/#1} {{\tt arXiv:#1}}}
\def\mn@eprint@dblp#1{\href {http://dblp.uni-trier.de/rec/bibtex/#1.xml}
  {dblp:#1}}
\def\mn@eprint@#1:#2:#3:#4\@nil{\def\@tempa {#1}\def\@tempb {#2}\def\@tempc
  {#3}\ifx \@tempc \@empty \let \@tempc \@tempb \let \@tempb \@tempa \fi \ifx
  \@tempb \@empty \def\@tempb {arXiv}\fi \@ifundefined
  {mn@eprint@\@tempb}{\@tempb:\@tempc}{\expandafter \expandafter \csname
  mn@eprint@\@tempb\endcsname \expandafter{\@tempc}}}

\bibitem[\protect\citeauthoryear{{Abadi}, {Navarro}, {Steinmetz}  \&
  {Eke}}{{Abadi} et~al.}{2003}]{2003ApJ...597...21A}
{Abadi} M.~G.,  {Navarro} J.~F.,  {Steinmetz} M.,   {Eke} V.~R.,  2003, \mn@doi
  [\apj] {10.1086/378316}, \href
  {http://adsabs.harvard.edu/abs/2003ApJ...597...21A} {597, 21}

\bibitem[\protect\citeauthoryear{{Adibekyan}, {Sousa}, {Santos}, {Delgado
  Mena}, {Gonz{\'a}lez Hern{\'a}ndez}, {Israelian}, {Mayor}  \&
  {Khachatryan}}{{Adibekyan} et~al.}{2012}]{2012A&A...545A..32A}
{Adibekyan} V.~Z.,  {Sousa} S.~G.,  {Santos} N.~C.,  {Delgado Mena} E.,
  {Gonz{\'a}lez Hern{\'a}ndez} J.~I.,  {Israelian} G.,  {Mayor} M.,
  {Khachatryan} G.,  2012, \mn@doi [\aap] {10.1051/0004-6361/201219401}, \href
  {http://adsabs.harvard.edu/abs/2012A%26A...545A..32A} {545, A32}

\bibitem[\protect\citeauthoryear{{Alam} et~al.,}{{Alam}
  et~al.}{2015}]{2015ApJS..219...12A}
{Alam} S.,  et~al., 2015, \mn@doi [\apjs] {10.1088/0067-0049/219/1/12}, \href
  {http://adsabs.harvard.edu/abs/2015ApJS..219...12A} {219, 12}

\bibitem[\protect\citeauthoryear{{Anders} et~al.,}{{Anders}
  et~al.}{2016a}]{2016arXiv160407771A}
{Anders} F.,  et~al., 2016a, preprint, \href
  {http://adsabs.harvard.edu/abs/2016arXiv160407771A} {} (\mn@eprint {arXiv}
  {1604.07771})

\bibitem[\protect\citeauthoryear{{Anders} et~al.,}{{Anders}
  et~al.}{2016b}]{2016arXiv160804951A}
{Anders} F.,  et~al., 2016b, preprint, \href
  {http://adsabs.harvard.edu/abs/2016arXiv160804951A} {} (\mn@eprint {arXiv}
  {1608.04951})

\bibitem[\protect\citeauthoryear{{Andrews}, {Weinberg}, {Sch{\"o}nrich}  \&
  {Johnson}}{{Andrews} et~al.}{2017}]{2016arXiv160408613A}
{Andrews} B.~H.,  {Weinberg} D.~H.,  {Sch{\"o}nrich} R.,   {Johnson} J.~A.,
  2017, \mn@doi [\apj] {10.3847/1538-4357/835/2/224}, \href
  {http://adsabs.harvard.edu/abs/2017ApJ...835..224A} {835, 224}

\bibitem[\protect\citeauthoryear{{Bensby}, {Feltzing}  \&
  {Lundstr{\"o}m}}{{Bensby} et~al.}{2003}]{2003A&A...410..527B}
{Bensby} T.,  {Feltzing} S.,   {Lundstr{\"o}m} I.,  2003, \mn@doi [\aap]
  {10.1051/0004-6361:20031213}, \href
  {http://adsabs.harvard.edu/abs/2003A%26A...410..527B} {410, 527}

\bibitem[\protect\citeauthoryear{{Bensby}, {Feltzing}  \&
  {Lundstr{\"o}m}}{{Bensby} et~al.}{2004}]{2004A&A...415..155B}
{Bensby} T.,  {Feltzing} S.,   {Lundstr{\"o}m} I.,  2004, \mn@doi [\aap]
  {10.1051/0004-6361:20031655}, \href
  {http://adsabs.harvard.edu/abs/2004A%26A...415..155B} {415, 155}

\bibitem[\protect\citeauthoryear{{Bensby}, {Feltzing}, {Lundstr{\"o}m}  \&
  {Ilyin}}{{Bensby} et~al.}{2005}]{2005A&A...433..185B}
{Bensby} T.,  {Feltzing} S.,  {Lundstr{\"o}m} I.,   {Ilyin} I.,  2005, \mn@doi
  [\aap] {10.1051/0004-6361:20040332}, \href
  {http://adsabs.harvard.edu/abs/2005A%26A...433..185B} {433, 185}

\bibitem[\protect\citeauthoryear{{Bensby}, {Feltzing}  \& {Oey}}{{Bensby}
  et~al.}{2014}]{2014A&A...562A..71B}
{Bensby} T.,  {Feltzing} S.,   {Oey} M.~S.,  2014, \mn@doi [\aap]
  {10.1051/0004-6361/201322631}, \href
  {http://adsabs.harvard.edu/abs/2014A%26A...562A..71B} {562, A71}

\bibitem[\protect\citeauthoryear{{Bergemann} et~al.,}{{Bergemann}
  et~al.}{2014}]{2014A&A...565A..89B}
{Bergemann} M.,  et~al., 2014, \mn@doi [\aap] {10.1051/0004-6361/201423456},
  \href {http://adsabs.harvard.edu/abs/2014A%26A...565A..89B} {565, A89}

\bibitem[\protect\citeauthoryear{{Bird}, {Kazantzidis}, {Weinberg}, {Guedes},
  {Callegari}, {Mayer}  \& {Madau}}{{Bird} et~al.}{2013}]{2013ApJ...773...43B}
{Bird} J.~C.,  {Kazantzidis} S.,  {Weinberg} D.~H.,  {Guedes} J.,  {Callegari}
  S.,  {Mayer} L.,   {Madau} P.,  2013, \mn@doi [\apj]
  {10.1088/0004-637X/773/1/43}, \href
  {http://adsabs.harvard.edu/abs/2013ApJ...773...43B} {773, 43}

\bibitem[\protect\citeauthoryear{{Bland-Hawthorn} \&
  {Gerhard}}{{Bland-Hawthorn} \& {Gerhard}}{2016}]{2016ARA&A..54..529B}
{Bland-Hawthorn} J.,  {Gerhard} O.,  2016, \mn@doi [\araa]
  {10.1146/annurev-astro-081915-023441}, \href
  {http://ukads.nottingham.ac.uk/abs/2016ARA%26A..54..529B} {54, 529}

\bibitem[\protect\citeauthoryear{{Boothroyd} \& {Sackmann}}{{Boothroyd} \&
  {Sackmann}}{1999}]{1999ApJ...510..232B}
{Boothroyd} A.~I.,  {Sackmann} I.-J.,  1999, \mn@doi [\apj] {10.1086/306546},
  \href {http://adsabs.harvard.edu/abs/1999ApJ...510..232B} {510, 232}

\bibitem[\protect\citeauthoryear{{Bournaud}, {Elmegreen}  \&
  {Martig}}{{Bournaud} et~al.}{2009}]{2009ApJ...707L...1B}
{Bournaud} F.,  {Elmegreen} B.~G.,   {Martig} M.,  2009, \mn@doi [\apjl]
  {10.1088/0004-637X/707/1/L1}, \href
  {http://adsabs.harvard.edu/abs/2009ApJ...707L...1B} {707, L1}

\bibitem[\protect\citeauthoryear{{Bovy} \& {Rix}}{{Bovy} \&
  {Rix}}{2013}]{2013ApJ...779..115B}
{Bovy} J.,  {Rix} H.-W.,  2013, \mn@doi [\apj] {10.1088/0004-637X/779/2/115},
  \href {http://adsabs.harvard.edu/abs/2013ApJ...779..115B} {779, 115}

\bibitem[\protect\citeauthoryear{{Bovy}, {Rix}  \& {Hogg}}{{Bovy}
  et~al.}{2012a}]{2012ApJ...751..131B}
{Bovy} J.,  {Rix} H.-W.,   {Hogg} D.~W.,  2012a, \mn@doi [\apj]
  {10.1088/0004-637X/751/2/131}, \href
  {http://adsabs.harvard.edu/abs/2012ApJ...751..131B} {751, 131}

\bibitem[\protect\citeauthoryear{{Bovy}, {Rix}, {Liu}, {Hogg}, {Beers}  \&
  {Lee}}{{Bovy} et~al.}{2012b}]{2012ApJ...753..148B}
{Bovy} J.,  {Rix} H.-W.,  {Liu} C.,  {Hogg} D.~W.,  {Beers} T.~C.,   {Lee}
  Y.~S.,  2012b, \mn@doi [\apj] {10.1088/0004-637X/753/2/148}, \href
  {http://adsabs.harvard.edu/abs/2012ApJ...753..148B} {753, 148}

\bibitem[\protect\citeauthoryear{{Bovy}, {Rix}, {Hogg}, {Beers}, {Lee}  \&
  {Zhang}}{{Bovy} et~al.}{2012c}]{2012ApJ...755..115B}
{Bovy} J.,  {Rix} H.-W.,  {Hogg} D.~W.,  {Beers} T.~C.,  {Lee} Y.~S.,   {Zhang}
  L.,  2012c, \mn@doi [\apj] {10.1088/0004-637X/755/2/115}, \href
  {http://adsabs.harvard.edu/abs/2012ApJ...755..115B} {755, 115}

\bibitem[\protect\citeauthoryear{{Bovy} et~al.,}{{Bovy}
  et~al.}{2014}]{2014ApJ...790..127B}
{Bovy} J.,  et~al., 2014, \mn@doi [\apj] {10.1088/0004-637X/790/2/127}, \href
  {http://adsabs.harvard.edu/abs/2014ApJ...790..127B} {790, 127}

\bibitem[\protect\citeauthoryear{{Bovy}, {Rix}, {Green}, {Schlafly}  \&
  {Finkbeiner}}{{Bovy} et~al.}{2016a}]{2016ApJ...818..130B}
{Bovy} J.,  {Rix} H.-W.,  {Green} G.~M.,  {Schlafly} E.~F.,   {Finkbeiner}
  D.~P.,  2016a, \mn@doi [\apj] {10.3847/0004-637X/818/2/130}, \href
  {http://adsabs.harvard.edu/abs/2016ApJ...818..130B} {818, 130}

\bibitem[\protect\citeauthoryear{{Bovy}, {Rix}, {Schlafly}, {Nidever},
  {Holtzman}, {Shetrone}  \& {Beers}}{{Bovy}
  et~al.}{2016b}]{2016ApJ...823...30B}
{Bovy} J.,  {Rix} H.-W.,  {Schlafly} E.~F.,  {Nidever} D.~L.,  {Holtzman}
  J.~A.,  {Shetrone} M.,   {Beers} T.~C.,  2016b, \mn@doi [\apj]
  {10.3847/0004-637X/823/1/30}, \href
  {http://adsabs.harvard.edu/abs/2016ApJ...823...30B} {823, 30}

\bibitem[\protect\citeauthoryear{{Bressan}, {Marigo}, {Girardi}, {Salasnich},
  {Dal Cero}, {Rubele}  \& {Nanni}}{{Bressan}
  et~al.}{2012}]{2012MNRAS.427..127B}
{Bressan} A.,  {Marigo} P.,  {Girardi} L.,  {Salasnich} B.,  {Dal Cero} C.,
  {Rubele} S.,   {Nanni} A.,  2012, \mn@doi [\mnras]
  {10.1111/j.1365-2966.2012.21948.x}, \href
  {http://adsabs.harvard.edu/abs/2012MNRAS.427..127B} {427, 127}

\bibitem[\protect\citeauthoryear{{Brook}, {Kawata}, {Gibson}  \&
  {Freeman}}{{Brook} et~al.}{2004}]{2004ApJ...612..894B}
{Brook} C.~B.,  {Kawata} D.,  {Gibson} B.~K.,   {Freeman} K.~C.,  2004, \mn@doi
  [\apj] {10.1086/422709}, \href
  {http://adsabs.harvard.edu/abs/2004ApJ...612..894B} {612, 894}

\bibitem[\protect\citeauthoryear{{Burstein}}{{Burstein}}{1979}]{1979ApJ...234..829B}
{Burstein} D.,  1979, \mn@doi [\apj] {10.1086/157563}, \href
  {http://adsabs.harvard.edu/abs/1979ApJ...234..829B} {234, 829}

\bibitem[\protect\citeauthoryear{{Chabrier}}{{Chabrier}}{2001}]{2001ApJ...554.1274C}
{Chabrier} G.,  2001, \mn@doi [\apj] {10.1086/321401}, \href
  {http://adsabs.harvard.edu/abs/2001ApJ...554.1274C} {554, 1274}

\bibitem[\protect\citeauthoryear{{Chabrier}}{{Chabrier}}{2003}]{2003PASP..115..763C}
{Chabrier} G.,  2003, \mn@doi [\pasp] {10.1086/376392}, \href
  {http://adsabs.harvard.edu/abs/2003PASP..115..763C} {115, 763}

\bibitem[\protect\citeauthoryear{{Cheng} et~al.,}{{Cheng}
  et~al.}{2012a}]{2012ApJ...746..149C}
{Cheng} J.~Y.,  et~al., 2012a, \mn@doi [\apj] {10.1088/0004-637X/746/2/149},
  \href {http://adsabs.harvard.edu/abs/2012ApJ...746..149C} {746, 149}

\bibitem[\protect\citeauthoryear{{Cheng} et~al.,}{{Cheng}
  et~al.}{2012b}]{2012ApJ...752...51C}
{Cheng} J.~Y.,  et~al., 2012b, \mn@doi [\apj] {10.1088/0004-637X/752/1/51},
  \href {http://ukads.nottingham.ac.uk/abs/2012ApJ...752...51C} {752, 51}

\bibitem[\protect\citeauthoryear{{Chiappini}, {Matteucci}  \&
  {Gratton}}{{Chiappini} et~al.}{1997}]{1997ApJ...477..765C}
{Chiappini} C.,  {Matteucci} F.,   {Gratton} R.,  1997, \mn@doi [\apj]
  {10.1086/303726}, \href {http://adsabs.harvard.edu/abs/1997ApJ...477..765C}
  {477, 765}

\bibitem[\protect\citeauthoryear{{Churchwell} et~al.,}{{Churchwell}
  et~al.}{2009}]{2009PASP..121..213C}
{Churchwell} E.,  et~al., 2009, \mn@doi [\pasp] {10.1086/597811}, \href
  {http://adsabs.harvard.edu/abs/2009PASP..121..213C} {121, 213}

\bibitem[\protect\citeauthoryear{{Cirasuolo} et~al.,}{{Cirasuolo}
  et~al.}{2012}]{2012SPIE.8446E..0SC}
{Cirasuolo} M.,  et~al., 2012, in Ground-based and Airborne Instrumentation for
  Astronomy IV. p. 84460S (\mn@eprint {arXiv} {1208.5780}),
  \mn@doi{10.1117/12.925871}

\bibitem[\protect\citeauthoryear{{Dalton} et~al.,}{{Dalton}
  et~al.}{2014}]{2014SPIE.9147E..0LD}
{Dalton} G.,  et~al., 2014, in Ground-based and Airborne Instrumentation for
  Astronomy V. p. 91470L (\mn@eprint {arXiv} {1412.0843}),
  \mn@doi{10.1117/12.2055132}

\bibitem[\protect\citeauthoryear{{Edvardsson}, {Andersen}, {Gustafsson},
  {Lambert}, {Nissen}  \& {Tomkin}}{{Edvardsson}
  et~al.}{1993}]{1993A&A...275..101E}
{Edvardsson} B.,  {Andersen} J.,  {Gustafsson} B.,  {Lambert} D.~L.,  {Nissen}
  P.~E.,   {Tomkin} J.,  1993, \aap, \href
  {http://adsabs.harvard.edu/abs/1993A%26A...275..101E} {275, 101}

\bibitem[\protect\citeauthoryear{{Eggen}, {Lynden-Bell}  \& {Sandage}}{{Eggen}
  et~al.}{1962}]{1962ApJ...136..748E}
{Eggen} O.~J.,  {Lynden-Bell} D.,   {Sandage} A.~R.,  1962, \mn@doi [\apj]
  {10.1086/147433}, \href {http://adsabs.harvard.edu/abs/1962ApJ...136..748E}
  {136, 748}

\bibitem[\protect\citeauthoryear{{Elmegreen} \& {Struck}}{{Elmegreen} \&
  {Struck}}{2013}]{2013ApJ...775L..35E}
{Elmegreen} B.~G.,  {Struck} C.,  2013, \mn@doi [\apjl]
  {10.1088/2041-8205/775/2/L35}, \href
  {http://adsabs.harvard.edu/abs/2013ApJ...775L..35E} {775, L35}

\bibitem[\protect\citeauthoryear{{Elmegreen} \& {Struck}}{{Elmegreen} \&
  {Struck}}{2016}]{2016ApJ...830..115E}
{Elmegreen} B.~G.,  {Struck} C.,  2016, \mn@doi [\apj]
  {10.3847/0004-637X/830/2/115}, \href
  {http://adsabs.harvard.edu/abs/2016ApJ...830..115E} {830, 115}

\bibitem[\protect\citeauthoryear{{Fall} \& {Efstathiou}}{{Fall} \&
  {Efstathiou}}{1980}]{1980MNRAS.193..189F}
{Fall} S.~M.,  {Efstathiou} G.,  1980, \mn@doi [\mnras]
  {10.1093/mnras/193.2.189}, \href
  {http://adsabs.harvard.edu/abs/1980MNRAS.193..189F} {193, 189}

\bibitem[\protect\citeauthoryear{{Fathi}, {Allen}, {Boch}, {Hatziminaoglou}  \&
  {Peletier}}{{Fathi} et~al.}{2010}]{2010MNRAS.406.1595F}
{Fathi} K.,  {Allen} M.,  {Boch} T.,  {Hatziminaoglou} E.,   {Peletier} R.~F.,
  2010, \mn@doi [\mnras] {10.1111/j.1365-2966.2010.16812.x}, \href
  {http://adsabs.harvard.edu/abs/2010MNRAS.406.1595F} {406, 1595}

\bibitem[\protect\citeauthoryear{{Feast}, {Menzies}, {Matsunaga}  \&
  {Whitelock}}{{Feast} et~al.}{2014}]{2014Natur.509..342F}
{Feast} M.~W.,  {Menzies} J.~W.,  {Matsunaga} N.,   {Whitelock} P.~A.,  2014,
  \mn@doi [\nat] {10.1038/nature13246}, \href
  {http://adsabs.harvard.edu/abs/2014Natur.509..342F} {509, 342}

\bibitem[\protect\citeauthoryear{{Flynn}, {Holmberg}, {Portinari}, {Fuchs}  \&
  {Jahrei{\ss}}}{{Flynn} et~al.}{2006}]{2006MNRAS.372.1149F}
{Flynn} C.,  {Holmberg} J.,  {Portinari} L.,  {Fuchs} B.,   {Jahrei{\ss}} H.,
  2006, \mn@doi [\mnras] {10.1111/j.1365-2966.2006.10911.x}, \href
  {http://adsabs.harvard.edu/abs/2006MNRAS.372.1149F} {372, 1149}

\bibitem[\protect\citeauthoryear{{Foreman-Mackey}, {Hogg}, {Lang}  \&
  {Goodman}}{{Foreman-Mackey} et~al.}{2013}]{2013PASP..125..306F}
{Foreman-Mackey} D.,  {Hogg} D.~W.,  {Lang} D.,   {Goodman} J.,  2013, \mn@doi
  [\pasp] {10.1086/670067}, \href
  {http://adsabs.harvard.edu/abs/2013PASP..125..306F} {125, 306}

\bibitem[\protect\citeauthoryear{{Freeman}}{{Freeman}}{1970}]{1970ApJ...160..811F}
{Freeman} K.~C.,  1970, \mn@doi [\apj] {10.1086/150474}, \href
  {http://adsabs.harvard.edu/abs/1970ApJ...160..811F} {160, 811}

\bibitem[\protect\citeauthoryear{{Fuhrmann}}{{Fuhrmann}}{1998}]{1998A&A...338..161F}
{Fuhrmann} K.,  1998, \aap, \href
  {http://adsabs.harvard.edu/abs/1998A%26A...338..161F} {338, 161}

\bibitem[\protect\citeauthoryear{{Gaia Collaboration} et~al.,}{{Gaia
  Collaboration} et~al.}{2016}]{2016A&A...595A...1G}
{Gaia Collaboration} et~al., 2016, \mn@doi [\aap]
  {10.1051/0004-6361/201629272}, \href
  {http://adsabs.harvard.edu/abs/2016A%26A...595A...1G} {595, A1}

\bibitem[\protect\citeauthoryear{{Garc{\'{\i}}a P{\'e}rez}
  et~al.,}{{Garc{\'{\i}}a P{\'e}rez} et~al.}{2016}]{2016AJ....151..144G}
{Garc{\'{\i}}a P{\'e}rez} A.~E.,  et~al., 2016, \mn@doi [\aj]
  {10.3847/0004-6256/151/6/144}, \href
  {http://adsabs.harvard.edu/abs/2016AJ....151..144G} {151, 144}

\bibitem[\protect\citeauthoryear{{Gilmore} \& {Reid}}{{Gilmore} \&
  {Reid}}{1983}]{1983MNRAS.202.1025G}
{Gilmore} G.,  {Reid} N.,  1983, \mn@doi [\mnras] {10.1093/mnras/202.4.1025},
  \href {http://adsabs.harvard.edu/abs/1983MNRAS.202.1025G} {202, 1025}

\bibitem[\protect\citeauthoryear{{Gilmore} et~al.,}{{Gilmore}
  et~al.}{2012}]{2012Msngr.147...25G}
{Gilmore} G.,  et~al., 2012, The Messenger, \href
  {http://adsabs.harvard.edu/abs/2012Msngr.147...25G} {147, 25}

\bibitem[\protect\citeauthoryear{Goodman \& Weare}{Goodman \&
  Weare}{2010}]{goodmanweare2010}
Goodman J.,  Weare J.,  2010, Comm. App. Math. and Comp. Sci., 65

\bibitem[\protect\citeauthoryear{{Grand} et~al.,}{{Grand}
  et~al.}{2016}]{2016MNRAS.460L..94G}
{Grand} R.~J.~J.,  et~al., 2016, \mn@doi [\mnras] {10.1093/mnrasl/slw086},
  \href {http://ukads.nottingham.ac.uk/abs/2016MNRAS.460L..94G} {460, L94}

\bibitem[\protect\citeauthoryear{{Green} et~al.,}{{Green}
  et~al.}{2015}]{2015ApJ...810...25G}
{Green} G.~M.,  et~al., 2015, \mn@doi [\apj] {10.1088/0004-637X/810/1/25},
  \href {http://adsabs.harvard.edu/abs/2015ApJ...810...25G} {810, 25}

\bibitem[\protect\citeauthoryear{{Gunn} et~al.,}{{Gunn}
  et~al.}{2006}]{2006AJ....131.2332G}
{Gunn} J.~E.,  et~al., 2006, \mn@doi [\aj] {10.1086/500975}, \href
  {http://adsabs.harvard.edu/abs/2006AJ....131.2332G} {131, 2332}

\bibitem[\protect\citeauthoryear{{Hayden} et~al.,}{{Hayden}
  et~al.}{2014}]{2014AJ....147..116H}
{Hayden} M.~R.,  et~al., 2014, \mn@doi [\aj] {10.1088/0004-6256/147/5/116},
  \href {http://adsabs.harvard.edu/abs/2014AJ....147..116H} {147, 116}

\bibitem[\protect\citeauthoryear{{Hayden} et~al.,}{{Hayden}
  et~al.}{2015}]{2015ApJ...808..132H}
{Hayden} M.~R.,  et~al., 2015, \mn@doi [\apj] {10.1088/0004-637X/808/2/132},
  \href {http://adsabs.harvard.edu/abs/2015ApJ...808..132H} {808, 132}

\bibitem[\protect\citeauthoryear{{Haywood}, {Di Matteo}, {Lehnert}, {Katz}  \&
  {G{\'o}mez}}{{Haywood} et~al.}{2013}]{2013A&A...560A.109H}
{Haywood} M.,  {Di Matteo} P.,  {Lehnert} M.~D.,  {Katz} D.,   {G{\'o}mez} A.,
  2013, \mn@doi [\aap] {10.1051/0004-6361/201321397}, \href
  {http://adsabs.harvard.edu/abs/2013A%26A...560A.109H} {560, A109}

\bibitem[\protect\citeauthoryear{{Herpich}, {Tremaine}  \& {Rix}}{{Herpich}
  et~al.}{2016}]{2016arXiv161203171H}
{Herpich} J.,  {Tremaine} S.,   {Rix} H.-W.,  2016, preprint, \href
  {http://adsabs.harvard.edu/abs/2016arXiv161203171H} {} (\mn@eprint {arXiv}
  {1612.03171})

\bibitem[\protect\citeauthoryear{{Holtzman} et~al.,}{{Holtzman}
  et~al.}{2015}]{2015AJ....150..148H}
{Holtzman} J.~A.,  et~al., 2015, \mn@doi [\aj] {10.1088/0004-6256/150/5/148},
  \href {http://adsabs.harvard.edu/abs/2015AJ....150..148H} {150, 148}

\bibitem[\protect\citeauthoryear{{Huertas-Company} et~al.,}{{Huertas-Company}
  et~al.}{2016}]{2016MNRAS.462.4495H}
{Huertas-Company} M.,  et~al., 2016, \mn@doi [\mnras] {10.1093/mnras/stw1866},
  \href {http://adsabs.harvard.edu/abs/2016MNRAS.462.4495H} {462, 4495}

\bibitem[\protect\citeauthoryear{Hunter}{Hunter}{2007}]{4160265}
Hunter J.~D.,  2007, \mn@doi [Computing in Science Engineering]
  {10.1109/MCSE.2007.55}, 9, 90

\bibitem[\protect\citeauthoryear{{Jayaraman}, {Gilmore}, {Wyse}, {Norris}  \&
  {Belokurov}}{{Jayaraman} et~al.}{2013}]{2013MNRAS.431..930J}
{Jayaraman} A.,  {Gilmore} G.,  {Wyse} R.~F.~G.,  {Norris} J.~E.,   {Belokurov}
  V.,  2013, \mn@doi [\mnras] {10.1093/mnras/stt221}, \href
  {http://adsabs.harvard.edu/abs/2013MNRAS.431..930J} {431, 930}

\bibitem[\protect\citeauthoryear{Jones, Oliphant, Peterson  et~al.}{Jones
  et~al.}{2001}]{Jones:2001aa}
Jones E.,  Oliphant T.,  Peterson P.,   et~al., 2001, {SciPy}: Open source
  scientific tools for {Python}, \url {http://www.scipy.org/}

\bibitem[\protect\citeauthoryear{{Juri{\'c}} et~al.,}{{Juri{\'c}}
  et~al.}{2008}]{2008ApJ...673..864J}
{Juri{\'c}} M.,  et~al., 2008, \mn@doi [\apj] {10.1086/523619}, \href
  {http://adsabs.harvard.edu/abs/2008ApJ...673..864J} {673, 864}

\bibitem[\protect\citeauthoryear{{Kalberla}, {Kerp}, {Dedes}  \&
  {Haud}}{{Kalberla} et~al.}{2014}]{2014ApJ...794...90K}
{Kalberla} P.~M.~W.,  {Kerp} J.,  {Dedes} L.,   {Haud} U.,  2014, \mn@doi
  [\apj] {10.1088/0004-637X/794/1/90}, \href
  {http://adsabs.harvard.edu/abs/2014ApJ...794...90K} {794, 90}

\bibitem[\protect\citeauthoryear{{Kazantzidis}, {Zentner}, {Kravtsov},
  {Bullock}  \& {Debattista}}{{Kazantzidis} et~al.}{2009}]{2009ApJ...700.1896K}
{Kazantzidis} S.,  {Zentner} A.~R.,  {Kravtsov} A.~V.,  {Bullock} J.~S.,
  {Debattista} V.~P.,  2009, \mn@doi [\apj] {10.1088/0004-637X/700/2/1896},
  \href {http://adsabs.harvard.edu/abs/2009ApJ...700.1896K} {700, 1896}

\bibitem[\protect\citeauthoryear{{Kobayashi} \& {Nakasato}}{{Kobayashi} \&
  {Nakasato}}{2011}]{2011ApJ...729...16K}
{Kobayashi} C.,  {Nakasato} N.,  2011, \mn@doi [\apj]
  {10.1088/0004-637X/729/1/16}, \href
  {http://adsabs.harvard.edu/abs/2011ApJ...729...16K} {729, 16}

\bibitem[\protect\citeauthoryear{{Kroupa}}{{Kroupa}}{2001}]{2001MNRAS.322..231K}
{Kroupa} P.,  2001, \mn@doi [\mnras] {10.1046/j.1365-8711.2001.04022.x}, \href
  {http://adsabs.harvard.edu/abs/2001MNRAS.322..231K} {322, 231}

\bibitem[\protect\citeauthoryear{{Kubryk}, {Prantzos}  \&
  {Athanassoula}}{{Kubryk} et~al.}{2015}]{2015A&A...580A.126K}
{Kubryk} M.,  {Prantzos} N.,   {Athanassoula} E.,  2015, \mn@doi [\aap]
  {10.1051/0004-6361/201424171}, \href
  {http://adsabs.harvard.edu/abs/2015A%26A...580A.126K} {580, A126}

\bibitem[\protect\citeauthoryear{{Larson}}{{Larson}}{1976}]{1976MNRAS.176...31L}
{Larson} R.~B.,  1976, \mn@doi [\mnras] {10.1093/mnras/176.1.31}, \href
  {http://adsabs.harvard.edu/abs/1976MNRAS.176...31L} {176, 31}

\bibitem[\protect\citeauthoryear{{Lindegren} et~al.,}{{Lindegren}
  et~al.}{2016}]{2016arXiv160904303L}
{Lindegren} L.,  et~al., 2016, preprint, \href
  {http://adsabs.harvard.edu/abs/2016arXiv160904303L} {} (\mn@eprint {arXiv}
  {1609.04303})

\bibitem[\protect\citeauthoryear{{Loebman}, {Ro{\v s}kar}, {Debattista},
  {Ivezi{\'c}}, {Quinn}  \& {Wadsley}}{{Loebman}
  et~al.}{2011}]{2011ApJ...737....8L}
{Loebman} S.~R.,  {Ro{\v s}kar} R.,  {Debattista} V.~P.,  {Ivezi{\'c}} {\v Z}.,
   {Quinn} T.~R.,   {Wadsley} J.,  2011, \mn@doi [\apj]
  {10.1088/0004-637X/737/1/8}, \href
  {http://adsabs.harvard.edu/abs/2011ApJ...737....8L} {737, 8}

\bibitem[\protect\citeauthoryear{{Lozinskaya} \& {Karadashev}}{{Lozinskaya} \&
  {Karadashev}}{1963}]{1963SvA.....6..658L}
{Lozinskaya} T.~A.,  {Karadashev} N.~S.,  1963, \sovast, \href
  {http://adsabs.harvard.edu/abs/1963SvA.....6..658L} {6, 658}

\bibitem[\protect\citeauthoryear{{Ma}, {Hopkins}, {Wetzel}, {Kirby},
  {Angles-Alcazar}, {Faucher-Giguere}, {Keres}  \& {Quataert}}{{Ma}
  et~al.}{2016}]{2016arXiv160804133M}
{Ma} X.,  {Hopkins} P.~F.,  {Wetzel} A.~R.,  {Kirby} E.~N.,  {Angles-Alcazar}
  D.,  {Faucher-Giguere} C.-A.,  {Keres} D.,   {Quataert} E.,  2016, preprint,
  \href {http://adsabs.harvard.edu/abs/2016arXiv160804133M} {} (\mn@eprint
  {arXiv} {1608.04133})

\bibitem[\protect\citeauthoryear{{Majewski}, {Zasowski}  \&
  {Nidever}}{{Majewski} et~al.}{2011}]{2011ApJ...739...25M}
{Majewski} S.~R.,  {Zasowski} G.,   {Nidever} D.~L.,  2011, \mn@doi [\apj]
  {10.1088/0004-637X/739/1/25}, \href
  {http://adsabs.harvard.edu/abs/2011ApJ...739...25M} {739, 25}

\bibitem[\protect\citeauthoryear{{Majewski} et~al.,}{{Majewski}
  et~al.}{2015}]{2015arXiv150905420M}
{Majewski} S.~R.,  et~al., 2015, preprint, \href
  {http://adsabs.harvard.edu/abs/2015arXiv150905420M} {} (\mn@eprint {arXiv}
  {1509.05420})

\bibitem[\protect\citeauthoryear{{Marshall}, {Robin}, {Reyl{\'e}}, {Schultheis}
   \& {Picaud}}{{Marshall} et~al.}{2006}]{2006A&A...453..635M}
{Marshall} D.~J.,  {Robin} A.~C.,  {Reyl{\'e}} C.,  {Schultheis} M.,   {Picaud}
  S.,  2006, \mn@doi [\aap] {10.1051/0004-6361:20053842}, \href
  {http://adsabs.harvard.edu/abs/2006A%26A...453..635M} {453, 635}

\bibitem[\protect\citeauthoryear{{Martell} et~al.,}{{Martell}
  et~al.}{2016}]{2016arXiv160902822M}
{Martell} S.,  et~al., 2016, preprint, \href
  {http://adsabs.harvard.edu/abs/2016arXiv160902822M} {} (\mn@eprint {arXiv}
  {1609.02822})

\bibitem[\protect\citeauthoryear{{Martig}, {Minchev}  \& {Flynn}}{{Martig}
  et~al.}{2014a}]{2014MNRAS.442.2474M}
{Martig} M.,  {Minchev} I.,   {Flynn} C.,  2014a, \mn@doi [\mnras]
  {10.1093/mnras/stu1003}, \href
  {http://adsabs.harvard.edu/abs/2014MNRAS.442.2474M} {442, 2474}

\bibitem[\protect\citeauthoryear{{Martig}, {Minchev}  \& {Flynn}}{{Martig}
  et~al.}{2014b}]{2014MNRAS.443.2452M}
{Martig} M.,  {Minchev} I.,   {Flynn} C.,  2014b, \mn@doi [\mnras]
  {10.1093/mnras/stu1322}, \href
  {http://adsabs.harvard.edu/abs/2014MNRAS.443.2452M} {443, 2452}

\bibitem[\protect\citeauthoryear{{Martig}, {Minchev}, {Ness}, {Fouesneau}  \&
  {Rix}}{{Martig} et~al.}{2016a}]{2016arXiv160901168M}
{Martig} M.,  {Minchev} I.,  {Ness} M.,  {Fouesneau} M.,   {Rix} H.-W.,  2016a,
  preprint, \href {http://ukads.nottingham.ac.uk/abs/2016arXiv160901168M} {}
  (\mn@eprint {arXiv} {1609.01168})

\bibitem[\protect\citeauthoryear{{Martig} et~al.,}{{Martig}
  et~al.}{2016b}]{2016MNRAS.456.3655M}
{Martig} M.,  et~al., 2016b, \mn@doi [\mnras] {10.1093/mnras/stv2830}, \href
  {http://adsabs.harvard.edu/abs/2016MNRAS.456.3655M} {456, 3655}

\bibitem[\protect\citeauthoryear{{Masseron} \& {Hawkins}}{{Masseron} \&
  {Hawkins}}{2017}]{2017A&A...597L...3M}
{Masseron} T.,  {Hawkins} K.,  2017, \mn@doi [\aap]
  {10.1051/0004-6361/201629938}, \href
  {http://adsabs.harvard.edu/abs/2017A%26A...597L...3M} {597, L3}

\bibitem[\protect\citeauthoryear{{Matteucci} \& {Francois}}{{Matteucci} \&
  {Francois}}{1989}]{1989MNRAS.239..885M}
{Matteucci} F.,  {Francois} P.,  1989, \mn@doi [\mnras]
  {10.1093/mnras/239.3.885}, \href
  {http://adsabs.harvard.edu/abs/1989MNRAS.239..885M} {239, 885}

\bibitem[\protect\citeauthoryear{{McKee}, {Parravano}  \& {Hollenbach}}{{McKee}
  et~al.}{2015}]{2015ApJ...814...13M}
{McKee} C.~F.,  {Parravano} A.,   {Hollenbach} D.~J.,  2015, \mn@doi [\apj]
  {10.1088/0004-637X/814/1/13}, \href
  {http://adsabs.harvard.edu/abs/2015ApJ...814...13M} {814, 13}

\bibitem[\protect\citeauthoryear{{Minchev}, {Famaey}, {Quillen}, {Dehnen},
  {Martig}  \& {Siebert}}{{Minchev} et~al.}{2012}]{2012A&A...548A.127M}
{Minchev} I.,  {Famaey} B.,  {Quillen} A.~C.,  {Dehnen} W.,  {Martig} M.,
  {Siebert} A.,  2012, \mn@doi [\aap] {10.1051/0004-6361/201219714}, \href
  {http://adsabs.harvard.edu/abs/2012A%26A...548A.127M} {548, A127}

\bibitem[\protect\citeauthoryear{{Minchev}, {Chiappini}  \& {Martig}}{{Minchev}
  et~al.}{2013}]{2013A&A...558A...9M}
{Minchev} I.,  {Chiappini} C.,   {Martig} M.,  2013, \mn@doi [\aap]
  {10.1051/0004-6361/201220189}, \href
  {http://adsabs.harvard.edu/abs/2013A%26A...558A...9M} {558, A9}

\bibitem[\protect\citeauthoryear{{Minchev}, {Chiappini}  \& {Martig}}{{Minchev}
  et~al.}{2014a}]{2014A&A...572A..92M}
{Minchev} I.,  {Chiappini} C.,   {Martig} M.,  2014a, \mn@doi [\aap]
  {10.1051/0004-6361/201423487}, \href
  {http://adsabs.harvard.edu/abs/2014A%26A...572A..92M} {572, A92}

\bibitem[\protect\citeauthoryear{{Minchev} et~al.,}{{Minchev}
  et~al.}{2014b}]{2014ApJ...781L..20M}
{Minchev} I.,  et~al., 2014b, \mn@doi [\apjl] {10.1088/2041-8205/781/1/L20},
  \href {http://adsabs.harvard.edu/abs/2014ApJ...781L..20M} {781, L20}

\bibitem[\protect\citeauthoryear{{Minchev}, {Martig}, {Streich}, {Scannapieco},
  {de Jong}  \& {Steinmetz}}{{Minchev} et~al.}{2015}]{2015ApJ...804L...9M}
{Minchev} I.,  {Martig} M.,  {Streich} D.,  {Scannapieco} C.,  {de Jong} R.~S.,
    {Steinmetz} M.,  2015, \mn@doi [\apjl] {10.1088/2041-8205/804/1/L9}, \href
  {http://adsabs.harvard.edu/abs/2015ApJ...804L...9M} {804, L9}

\bibitem[\protect\citeauthoryear{{Minchev}, {Steinmetz}, {Chiappini}, {Martig},
  {Anders}, {Matijevic}  \& {de Jong}}{{Minchev}
  et~al.}{2017}]{2017ApJ...834...27M}
{Minchev} I.,  {Steinmetz} M.,  {Chiappini} C.,  {Martig} M.,  {Anders} F.,
  {Matijevic} G.,   {de Jong} R.~S.,  2017, \mn@doi [\apj]
  {10.3847/1538-4357/834/1/27}, \href
  {http://adsabs.harvard.edu/abs/2017ApJ...834...27M} {834, 27}

\bibitem[\protect\citeauthoryear{{Nemec} \& {Nemec}}{{Nemec} \&
  {Nemec}}{1991}]{1991PASP..103...95N}
{Nemec} J.,  {Nemec} A.~F.~L.,  1991, \mn@doi [\pasp] {10.1086/132800}, \href
  {http://adsabs.harvard.edu/abs/1991PASP..103...95N} {103, 95}

\bibitem[\protect\citeauthoryear{{Nidever} et~al.,}{{Nidever}
  et~al.}{2014}]{2014ApJ...796...38N}
{Nidever} D.~L.,  et~al., 2014, \mn@doi [\apj] {10.1088/0004-637X/796/1/38},
  \href {http://adsabs.harvard.edu/abs/2014ApJ...796...38N} {796, 38}

\bibitem[\protect\citeauthoryear{{Nidever} et~al.,}{{Nidever}
  et~al.}{2015}]{2015AJ....150..173N}
{Nidever} D.~L.,  et~al., 2015, \mn@doi [\aj] {10.1088/0004-6256/150/6/173},
  \href {http://adsabs.harvard.edu/abs/2015AJ....150..173N} {150, 173}

\bibitem[\protect\citeauthoryear{{Nordstr{\"o}m} et~al.,}{{Nordstr{\"o}m}
  et~al.}{2004}]{2004A&A...418..989N}
{Nordstr{\"o}m} B.,  et~al., 2004, \mn@doi [\aap] {10.1051/0004-6361:20035959},
  \href {http://adsabs.harvard.edu/abs/2004A%26A...418..989N} {418, 989}

\bibitem[\protect\citeauthoryear{{Norris}}{{Norris}}{1987}]{1987ApJ...314L..39N}
{Norris} J.,  1987, \mn@doi [\apjl] {10.1086/184847}, \href
  {http://adsabs.harvard.edu/abs/1987ApJ...314L..39N} {314, L39}

\bibitem[\protect\citeauthoryear{{Ojha}}{{Ojha}}{2001}]{2001MNRAS.322..426O}
{Ojha} D.~K.,  2001, \mn@doi [\mnras] {10.1046/j.1365-8711.2001.04155.x}, \href
  {http://adsabs.harvard.edu/abs/2001MNRAS.322..426O} {322, 426}

\bibitem[\protect\citeauthoryear{{Papovich} et~al.,}{{Papovich}
  et~al.}{2015}]{2015ApJ...803...26P}
{Papovich} C.,  et~al., 2015, \mn@doi [\apj] {10.1088/0004-637X/803/1/26},
  \href {http://adsabs.harvard.edu/abs/2015ApJ...803...26P} {803, 26}

\bibitem[\protect\citeauthoryear{Perez \& Granger}{Perez \&
  Granger}{2007}]{4160251}
Perez F.,  Granger B.~E.,  2007, \mn@doi [Computing in Science Engineering]
  {10.1109/MCSE.2007.53}, 9, 21

\bibitem[\protect\citeauthoryear{{Pietrinferni}, {Cassisi}, {Salaris}  \&
  {Castelli}}{{Pietrinferni} et~al.}{2004}]{2004ApJ...612..168P}
{Pietrinferni} A.,  {Cassisi} S.,  {Salaris} M.,   {Castelli} F.,  2004,
  \mn@doi [\apj] {10.1086/422498}, \href
  {http://adsabs.harvard.edu/abs/2004ApJ...612..168P} {612, 168}

\bibitem[\protect\citeauthoryear{{Pietrinferni}, {Cassisi}, {Salaris}  \&
  {Castelli}}{{Pietrinferni} et~al.}{2006}]{2006ApJ...642..797P}
{Pietrinferni} A.,  {Cassisi} S.,  {Salaris} M.,   {Castelli} F.,  2006,
  \mn@doi [\apj] {10.1086/501344}, \href
  {http://adsabs.harvard.edu/abs/2006ApJ...642..797P} {642, 797}

\bibitem[\protect\citeauthoryear{{Pinsonneault} et~al.,}{{Pinsonneault}
  et~al.}{2014}]{2014ApJS..215...19P}
{Pinsonneault} M.~H.,  et~al., 2014, \mn@doi [\apjs]
  {10.1088/0067-0049/215/2/19}, \href
  {http://adsabs.harvard.edu/abs/2014ApJS..215...19P} {215, 19}

\bibitem[\protect\citeauthoryear{{Pohlen} \& {Trujillo}}{{Pohlen} \&
  {Trujillo}}{2006}]{2006A&A...454..759P}
{Pohlen} M.,  {Trujillo} I.,  2006, \mn@doi [\aap]
  {10.1051/0004-6361:20064883}, \href
  {http://adsabs.harvard.edu/abs/2006A%26A...454..759P} {454, 759}

\bibitem[\protect\citeauthoryear{{Portinari} \& {Chiosi}}{{Portinari} \&
  {Chiosi}}{2000}]{2000A&A...355..929P}
{Portinari} L.,  {Chiosi} C.,  2000, \aap, \href
  {http://adsabs.harvard.edu/abs/2000A%26A...355..929P} {355, 929}

\bibitem[\protect\citeauthoryear{{Recio-Blanco} et~al.,}{{Recio-Blanco}
  et~al.}{2014}]{2014A&A...567A...5R}
{Recio-Blanco} A.,  et~al., 2014, \mn@doi [\aap] {10.1051/0004-6361/201322944},
  \href {http://adsabs.harvard.edu/abs/2014A%26A...567A...5R} {567, A5}

\bibitem[\protect\citeauthoryear{{Reyl{\'e}}, {Marshall}, {Robin}  \&
  {Schultheis}}{{Reyl{\'e}} et~al.}{2009}]{2009A&A...495..819R}
{Reyl{\'e}} C.,  {Marshall} D.~J.,  {Robin} A.~C.,   {Schultheis} M.,  2009,
  \mn@doi [\aap] {10.1051/0004-6361/200811341}, \href
  {http://adsabs.harvard.edu/abs/2009A%26A...495..819R} {495, 819}

\bibitem[\protect\citeauthoryear{{Rix} \& {Bovy}}{{Rix} \&
  {Bovy}}{2013}]{2013A&ARv..21...61R}
{Rix} H.-W.,  {Bovy} J.,  2013, \mn@doi [\aapr] {10.1007/s00159-013-0061-8},
  \href {http://adsabs.harvard.edu/abs/2013A%26ARv..21...61R} {21, 61}

\bibitem[\protect\citeauthoryear{{Sale} et~al.,}{{Sale}
  et~al.}{2010}]{2010MNRAS.402..713S}
{Sale} S.~E.,  et~al., 2010, \mn@doi [\mnras]
  {10.1111/j.1365-2966.2009.15746.x}, \href
  {http://adsabs.harvard.edu/abs/2010MNRAS.402..713S} {402, 713}

\bibitem[\protect\citeauthoryear{{Schlafly} \& {Finkbeiner}}{{Schlafly} \&
  {Finkbeiner}}{2011}]{2011ApJ...737..103S}
{Schlafly} E.~F.,  {Finkbeiner} D.~P.,  2011, \mn@doi [\apj]
  {10.1088/0004-637X/737/2/103}, \href
  {http://adsabs.harvard.edu/abs/2011ApJ...737..103S} {737, 103}

\bibitem[\protect\citeauthoryear{{Sch{\"o}nrich} \& {Binney}}{{Sch{\"o}nrich}
  \& {Binney}}{2009a}]{2009MNRAS.396..203S}
{Sch{\"o}nrich} R.,  {Binney} J.,  2009a, \mn@doi [\mnras]
  {10.1111/j.1365-2966.2009.14750.x}, \href
  {http://adsabs.harvard.edu/abs/2009MNRAS.396..203S} {396, 203}

\bibitem[\protect\citeauthoryear{{Sch{\"o}nrich} \& {Binney}}{{Sch{\"o}nrich}
  \& {Binney}}{2009b}]{2009MNRAS.399.1145S}
{Sch{\"o}nrich} R.,  {Binney} J.,  2009b, \mn@doi [\mnras]
  {10.1111/j.1365-2966.2009.15365.x}, \href
  {http://adsabs.harvard.edu/abs/2009MNRAS.399.1145S} {399, 1145}

\bibitem[\protect\citeauthoryear{{Sellwood} \& {Binney}}{{Sellwood} \&
  {Binney}}{2002}]{2002MNRAS.336..785S}
{Sellwood} J.~A.,  {Binney} J.~J.,  2002, \mn@doi [\mnras]
  {10.1046/j.1365-8711.2002.05806.x}, \href
  {http://adsabs.harvard.edu/abs/2002MNRAS.336..785S} {336, 785}

\bibitem[\protect\citeauthoryear{{Shetrone} et~al.,}{{Shetrone}
  et~al.}{2015}]{2015ApJS..221...24S}
{Shetrone} M.,  et~al., 2015, \mn@doi [\apjs] {10.1088/0067-0049/221/2/24},
  \href {http://adsabs.harvard.edu/abs/2015ApJS..221...24S} {221, 24}

\bibitem[\protect\citeauthoryear{{Skrutskie} et~al.,}{{Skrutskie}
  et~al.}{2006}]{2006AJ....131.1163S}
{Skrutskie} M.~F.,  et~al., 2006, \mn@doi [\aj] {10.1086/498708}, \href
  {http://adsabs.harvard.edu/abs/2006AJ....131.1163S} {131, 1163}

\bibitem[\protect\citeauthoryear{{Sobeck} et~al.,}{{Sobeck}
  et~al.}{2014}]{2014AAS...22344006S}
{Sobeck} J.,  et~al., 2014, in American Astronomical Society Meeting Abstracts
  \#223. p. 440.06

\bibitem[\protect\citeauthoryear{{Spitoni}, {Romano}, {Matteucci}  \&
  {Ciotti}}{{Spitoni} et~al.}{2015}]{2015ApJ...802..129S}
{Spitoni} E.,  {Romano} D.,  {Matteucci} F.,   {Ciotti} L.,  2015, \mn@doi
  [\apj] {10.1088/0004-637X/802/2/129}, \href
  {http://adsabs.harvard.edu/abs/2015ApJ...802..129S} {802, 129}

\bibitem[\protect\citeauthoryear{{Stinson} et~al.,}{{Stinson}
  et~al.}{2013}]{2013MNRAS.436..625S}
{Stinson} G.~S.,  et~al., 2013, \mn@doi [\mnras] {10.1093/mnras/stt1600}, \href
  {http://adsabs.harvard.edu/abs/2013MNRAS.436..625S} {436, 625}

\bibitem[\protect\citeauthoryear{{Toyouchi} \& {Chiba}}{{Toyouchi} \&
  {Chiba}}{2016}]{2016arXiv161009869T}
{Toyouchi} D.,  {Chiba} M.,  2016, preprint, \href
  {http://adsabs.harvard.edu/abs/2016arXiv161009869T} {} (\mn@eprint {arXiv}
  {1610.09869})

\bibitem[\protect\citeauthoryear{{Tsikoudi}}{{Tsikoudi}}{1979}]{1979ApJ...234..842T}
{Tsikoudi} V.,  1979, \mn@doi [\apj] {10.1086/157565}, \href
  {http://adsabs.harvard.edu/abs/1979ApJ...234..842T} {234, 842}

\bibitem[\protect\citeauthoryear{{Villalobos} \& {Helmi}}{{Villalobos} \&
  {Helmi}}{2008}]{2008MNRAS.391.1806V}
{Villalobos} {\'A}.,  {Helmi} A.,  2008, \mn@doi [\mnras]
  {10.1111/j.1365-2966.2008.13979.x}, \href
  {http://adsabs.harvard.edu/abs/2008MNRAS.391.1806V} {391, 1806}

\bibitem[\protect\citeauthoryear{{Weinberg}, {Andrews}  \&
  {Freudenburg}}{{Weinberg} et~al.}{2017}]{2016arXiv160407435W}
{Weinberg} D.~H.,  {Andrews} B.~H.,   {Freudenburg} J.,  2017, \mn@doi [\apj]
  {10.3847/1538-4357/837/2/183}, \href
  {http://adsabs.harvard.edu/abs/2017ApJ...837..183W} {837, 183}

\bibitem[\protect\citeauthoryear{{Wilson} et~al.,}{{Wilson}
  et~al.}{2010}]{2010SPIE.7735E..1CW}
{Wilson} J.~C.,  et~al., 2010, in Ground-based and Airborne Instrumentation for
  Astronomy III. p. 77351C, \mn@doi{10.1117/12.856708}

\bibitem[\protect\citeauthoryear{{Wright} et~al.,}{{Wright}
  et~al.}{2010}]{2010AJ....140.1868W}
{Wright} E.~L.,  et~al., 2010, \mn@doi [\aj] {10.1088/0004-6256/140/6/1868},
  \href {http://adsabs.harvard.edu/abs/2010AJ....140.1868W} {140, 1868}

\bibitem[\protect\citeauthoryear{{Yoshii}}{{Yoshii}}{1982}]{1982PASJ...34..365Y}
{Yoshii} Y.,  1982, \pasj, \href
  {http://adsabs.harvard.edu/abs/1982PASJ...34..365Y} {34, 365}

\bibitem[\protect\citeauthoryear{{Yuan}, {Liu}  \& {Xiang}}{{Yuan}
  et~al.}{2013}]{2013MNRAS.430.2188Y}
{Yuan} H.~B.,  {Liu} X.~W.,   {Xiang} M.~S.,  2013, \mn@doi [\mnras]
  {10.1093/mnras/stt039}, \href
  {http://adsabs.harvard.edu/abs/2013MNRAS.430.2188Y} {430, 2188}

\bibitem[\protect\citeauthoryear{{Zamora} et~al.,}{{Zamora}
  et~al.}{2015}]{2015AJ....149..181Z}
{Zamora} O.,  et~al., 2015, \mn@doi [\aj] {10.1088/0004-6256/149/6/181}, \href
  {http://adsabs.harvard.edu/abs/2015AJ....149..181Z} {149, 181}

\bibitem[\protect\citeauthoryear{{Zasowski} et~al.,}{{Zasowski}
  et~al.}{2013}]{2013AJ....146...81Z}
{Zasowski} G.,  et~al., 2013, \mn@doi [\aj] {10.1088/0004-6256/146/4/81}, \href
  {http://adsabs.harvard.edu/abs/2013AJ....146...81Z} {146, 81}

\bibitem[\protect\citeauthoryear{{de Vaucouleurs}}{{de
  Vaucouleurs}}{1959}]{1959HDP....53..311D}
{de Vaucouleurs} G.,  1959, Handbuch der Physik, \href
  {http://adsabs.harvard.edu/abs/1959HDP....53..311D} {53, 311}

\bibitem[\protect\citeauthoryear{{van Dokkum} et~al.,}{{van Dokkum}
  et~al.}{2013}]{2013ApJ...771L..35V}
{van Dokkum} P.~G.,  et~al., 2013, \mn@doi [\apjl]
  {10.1088/2041-8205/771/2/L35}, \href
  {http://adsabs.harvard.edu/abs/2013ApJ...771L..35V} {771, L35}

\bibitem[\protect\citeauthoryear{van~der Walt, Colbert  \& Varoquaux}{van~der
  Walt et~al.}{2011}]{5725236}
van~der Walt S.,  Colbert S.~C.,   Varoquaux G.,  2011, \mn@doi [Computing in
  Science Engineering] {10.1109/MCSE.2011.37}, 13, 22

\makeatother
\end{thebibliography}




\appendix

\section{Density Fits}
\label{sec:densityfits}
\begin{figure*}
     \includegraphics[width=\textwidth]{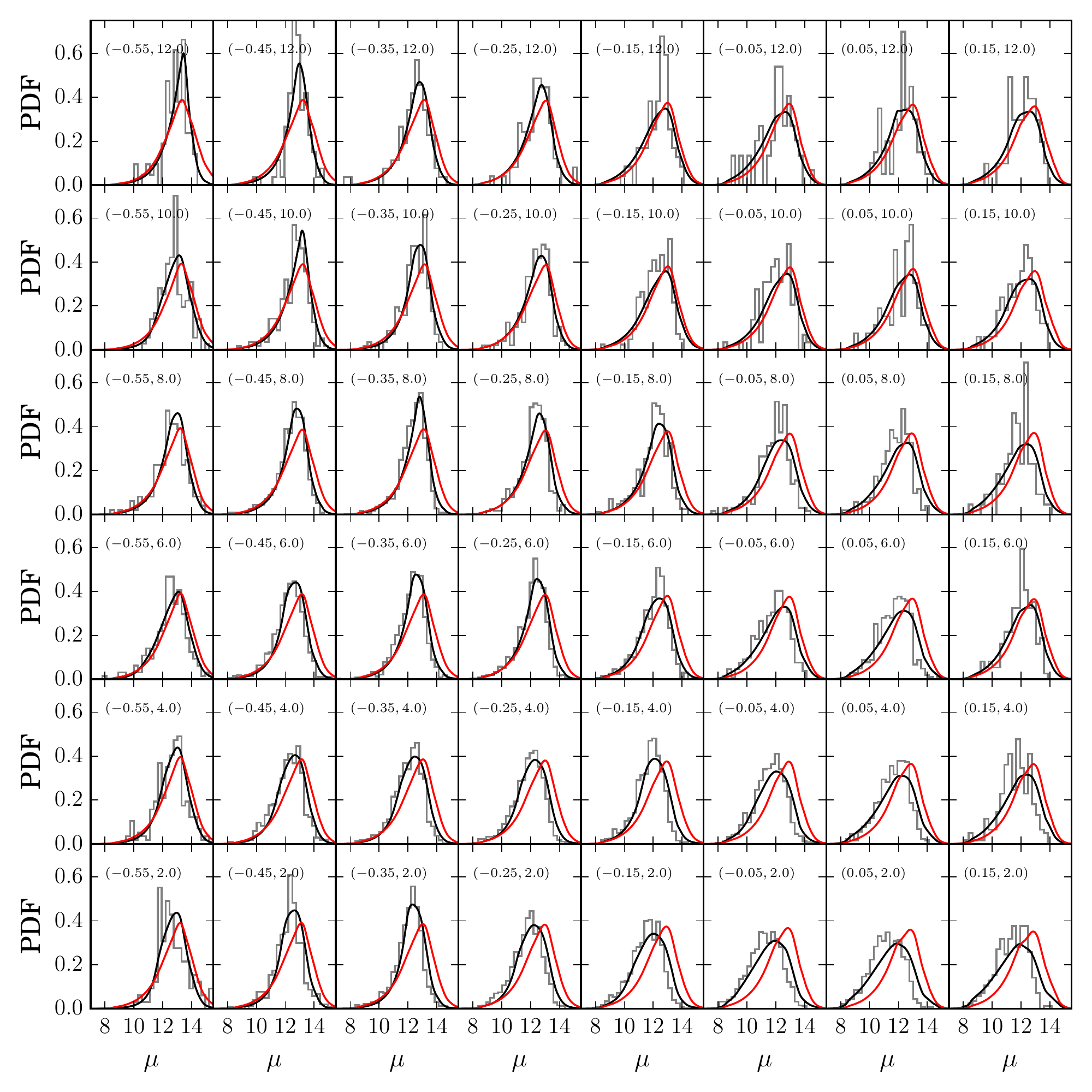}
   \caption{ Comparison between the best fit models and the APOGEE data for mono-age, mono-\feh{} populations in the low \afe{} sub-sample. The grey histogram shows the distance modulus distribution of the APOGEE data for the mono-age, mono-\feh{} bin indicated by the (\feh{} [dex], age [Gyr]) coordinate given in each panel. The coloured curves show the distance modulus distribution found when the best fit broken exponential (black) and single exponential (red) density model is run through the effective selection function. It is clear that the broken exponential density model provides a qualitatively better fit to the data in all cases.}
    \label{fig:low_fitcomp}
\end{figure*}

\begin{figure*}
     \includegraphics[width=\textwidth]{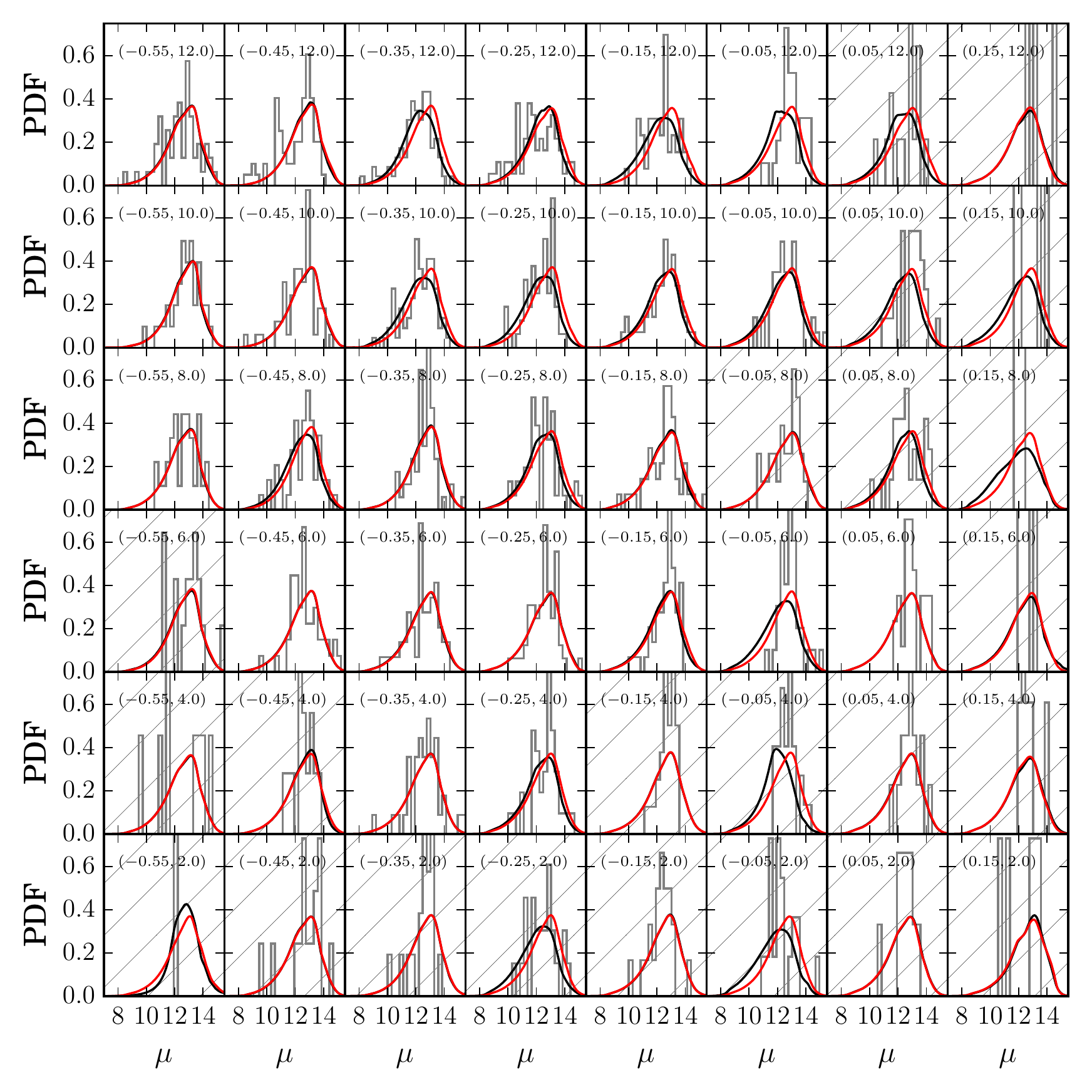}
   \caption{Comparison between the best fit models and the APOGEE data for mono-age, mono-\feh{} populations in the high \afe{} sub-sample. The grey histogram shows the distance modulus distribution of the APOGEE data for the mono-age, mono-\feh{} bin indicated by the (\feh{} [dex], age [Gyr]) coordinate given in each panel. The coloured curves show the distance modulus distribution found when the best fit broken exponential (black) and single exponential (red) density model is run through the effective selection function. In many cases the red and black curves are indistinguishable (only red is seen), or very similar. In cases where the black and red curves are different, the red provides a qualitatively better fit. Bins with less than 30 stars (which we disregard for the majority of the analysis) are hatched out. }
    \label{fig:high_fitcomp}
\end{figure*}

For completeness, we briefly discuss the quality of the fits performed with the method outlined in section \ref{sec:methoda}. Figures \ref{fig:low_fitcomp} and \ref{fig:high_fitcomp} show the distance modulus distribution of the APOGEE data in each of our mono-age and mono-\feh{} bins (grey histograms) and the resulting distance modulus distribution when the best fit density model for each bin is run through the calculated effective selection function (which is the space in which models are fit in our procedure). The red line represents a a single-exponential fit to the radial and vertical spatial distribution and the black lines give the best fit broken-exponential density model (upon which we base our results). We show the single-exponential fit in order to demonstrate that, in most cases, this does not provide a good fit to the data, and that when a single exponential is a better fit, the broken-exponential density fit matches it.

Regarding Figure \ref{fig:low_fitcomp}, which shows the low \afe{} sub-populations, it is clear that the black curve (broken exponential) represents a far better model for the data than the red curve (single exponential), in all mono-age, mono-\feh{} bins. While the black curve is not perfect in all cases, the peak of the distribution tends to lie at the correct $\mu$, whereas the red curve finds a peak at higher $\mu$ in most cases (due to the higher than necessary density at low Galactocentric radius in this model).

Figure \ref{fig:high_fitcomp} demonstrates the fits for the high \afe{} sub-populations. The greyed out panels reflect those with less than 30 stars, which we deem too noisy to render reliable fits. In many of the remaining panels, the red curve is similar or identical to the black, due to the fact that many of the high \afe{} populations are better described by single exponentials, and the broken exponential generally recovers this result. In most of the cases where the curves differ greatly, the red curve recovers the peak of the distribution better than the black - suggesting that breaks which were fit in the radial range we consider are artificial, and due to the noisy data in this regime. We discuss the broken exponential fits in the main text in order to make proper comparison with the low \afe{} sample, although it seems plausible that the single exponential model provides a better explanation of the data. 

\section{The effect of uncertainties on trends with age}
\label{sec:ageerror}

\begin{figure}
     \includegraphics[width=\columnwidth]{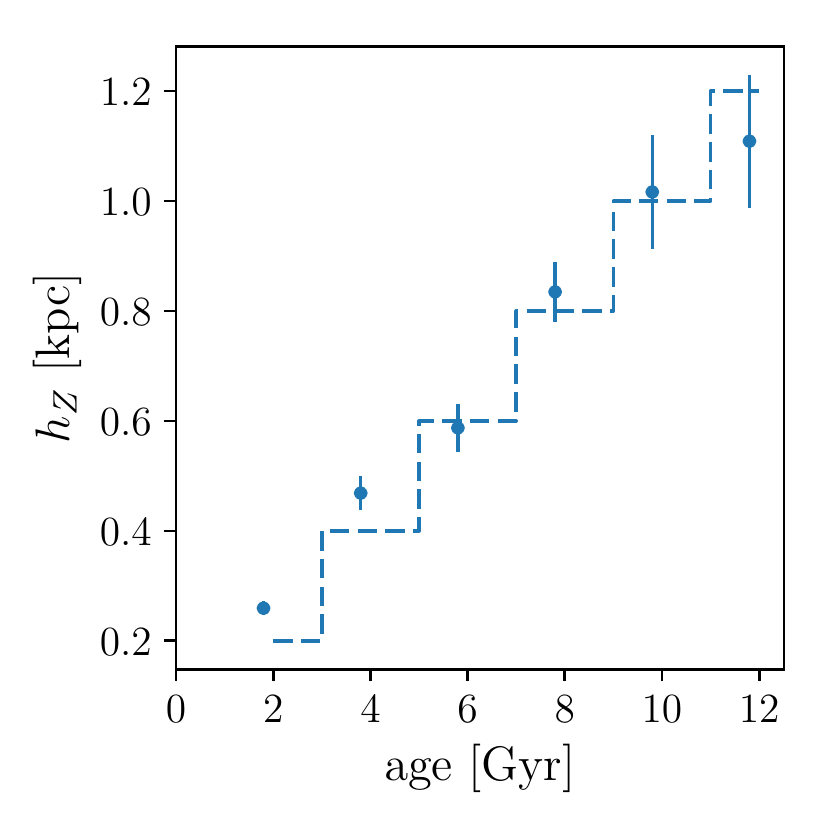}
   \caption{\added[id=TM]{The resulting age-$h_Z$ trend from the monte carlo sampling of a set of mock density distributions. The input  density models had $h_Z$ increasing monotonically with age (in bins of $\Delta\mathrm{age}= 2$ Gyr) from 0.2 to 1.2 kpc (shown by the \emph{blue} dashed line). After sampling of the density distribution, we applied random errors of $40\%$ to the mock ages, and measured the structural parameters using the exact density fitting method applied to the APOGEE data. The method is able to approximately recover the general shape of the input age-$h_Z$ relation, showing a clear trend with age. The age errors increase the error bar sizes significantly where mixing does occur, but the results are consistent with the input in most cases.}}
    \label{fig:montecarlo_hz}
\end{figure}

\added[id=TM]{In order to demonstrate and characterise the effect that the age errors have on our interpretation of the trends between structural parameters and age, we created a mock data set from a set of randomly sampled density distributions. We created a mock data set with an input trend of $h_Z$ with age which increased monotonically with age from 0.2 to 1.2 kpc. Ages were assigned to each $h_Z$ population, sampling uniformly in bins of width 2 Gyr, to which we then added a random gaussian error of $40\%$, replicating the shifting of stars with different $h_Z$ into each age bin. In each bin, we sample a single exponential (a broken exponential with $R_\mathrm{peak}=0$) with scale length 8 kpc. This higher scale length is required to make the test computationally efficient, to produce realistic numbers of stars when selecting stars in APOGEE fields and does not impact on the results of this test. We assume no error on the stellar positions, in order to isolate the effect of the age errors.} 

\added[id=TM]{While this test is a somewhat simplistic representation of the underlying processes, it serves as a good example of the effect of the age uncertainties which are expected in the present data. One example of its simplicity is the assignment of a single $h_Z$ to relatively wide bins in age. It seems logical to assume that if there is an age-$h_Z$ relation, then the change in $h_Z$ should be somewhat continuous with age. Our test assigns the same $h_Z$ to stars at bin edges (which should have $h_Z$ close to that of the bin-edge stars in the neighbouring bin). This may artificially increase the amount of blurring of the age-$h_Z$ trend. We also simplify the test by assuming that the only changing parameter is the scale height. Realistic structural parameters would change the relative number of stars within each bin observed by APOGEE (and considered in our test), and may change the level of contamination between bins. However, we assume that our simple approximation represents a `worst case' scenario, where the mixing between bins is maximal.}

\added[id=TM]{We restrict the mock data to the APOGEE fields, simplifying the selection to a distance cut (assuming the selection fraction is 1 out to a distance which corresponds to $M_\mathrm{H}= -1.5$, assuming no extinction).  We apply the method described in Section \ref{sec:methoda} to our mock data, fitting a broken exponential profile, and using the best fit solution to initiate an MCMC sampling of the posterior probability distribution. As in the main body of the paper, the reported parameter values reflect the median and $\sigma$ of one dimensional projections of the MCMC chain. The resulting age-$h_Z$ relation is shown in Figure \ref{fig:montecarlo_hz}.  A clear trend is recovered between age and $h_Z$. The trend is still recovered at high age, regardless of the high level of mixing between bins, which increases the size of the error bars. The higher-scale height components are recovered by the analysis, but results are scattered around the input values, with large error bars. This serves to show that even in the face of large age uncertainties causing mixing between the adopted bins, our method is still able to recover the underlying trends of parameters with age.}


\bsp	
\label{lastpage}
\end{document}